\documentclass[fleqn,usenatbib]{mnras}
\usepackage[T1]{fontenc}
\usepackage{ae,aecompl}
\usepackage{graphicx}	% Including figure files
\usepackage{amsmath}	% Advanced maths commands
\usepackage{amssymb}	% Extra maths symbols
\usepackage{xcolor}
\usepackage{url}
\usepackage{soul}
\usepackage{newtxtext,newtxmath}
\title[Cold clouds in a low metallicity starburst]{The masses, structure and lifetimes of cold clouds in a high-resolution simulation of a low metallicity starburst}
\author[C. M. Fotopoulou et al.]
{Constantina M. Fotopoulou$^{1}$\thanks{E-mail: constant@mpa-garching.mpg.de},
Thorsten Naab$^{1}$,
Natalia Lah\'{e}n$^{1}$,
Miha Cernetic$^{1}$,
\newauthor
Tim-Eric Rathjen$^{2}$,
Ulrich P. Steinwandel$^{3}$,
Jessica M. Hislop$^{4}$,
Stefanie Walch$^{2}$,
\newauthor
Peter H. Johansson$^{4}$
\\
$^{1}$Max Planck Institute for Astrophysics, Karl-Schwarzschild-Str. 1, D-85740, Garching, Germany\\
$^{2}$I. Physikalisches Institut, Universitat zu K\"{o}ln, Z\"{u}lpicher Str. 77, D-50937 K\"{o}ln, Germany\\
$^{3}$Center for Computational Astrophysics, Flatiron Institute, 162 5th Avenue, New York, NY 10010, USA\\
$^{4}$Department of Physics, University of Helsinki, Gustaf H\"{a}llstr\"{o}min katu 2, FI-00014 Helsinki, Finland\\
}
\date{Accepted XXX. Received YYY; in original form ZZZ}
\pubyear{2024}
\begin{document}
\label{firstpage}
\pagerange{\pageref{firstpage}--\pageref{lastpage}}
\maketitle
\begin{abstract}
We present an analysis of the cold gas phase in a low metallicity starburst generated in a high-resolution hydrodynamical simulation of a gas-rich dwarf galaxy merger as part of the {\textsc{griffin}} project. The simulations resolve (4 M$_\odot$ gas phase mass resolution, $\sim$ 0.1 pc spatial resolution) the multi-phase interstellar medium with a non-equilibrium chemical heating/cooling network at temperatures below \mbox{10$^4$ K}. Massive stars are sampled individually and interact with the ISM through the formation of HII regions and supernova explosions. In the extended starburst phase, the ISM is dominated by cold ($T_\mathrm{gas} < 300$ K) filamentary clouds with self-similar internal structures. The clouds have masses of $10^{2.6}$ – $10^{5.6}$ M$_\odot$ with a power law mass function, $dN/dM \propto M^\alpha$ with $\alpha = -1.78 (\pm 0.08) $. They also follow the Larson relations, in good agreement with observations. We trace the lifecycle of the cold clouds and find that they follow an exponential lifetime distribution and an e-folding time of $\sim$ 3.5 Myr. Clouds with peak masses below $10^4$ M$_\odot$ follow a power law relation with their average lifetime $\tau_\mathrm{life} \propto M^{0.3}_\mathrm{max}$ which flattens out for higher cloud masses at $ < 10$ Myr. A similar relation exists between cloud size at peak mass and lifetime. This simulation of the evolution of a realistic galactic cold cloud population supports the rapid formation and disruption of star-forming clouds by stellar radiation and supernovae on a timescale less than 10 Myr. 
\end{abstract}
%%%%%%%%%%%%%%%%%%%%%%%%%%%%%%%%%%%%%%%%%%%%%%%%%%
%%%%%%%%%%%%%%%%% BODY OF PAPER %%%%%%%%%%%%%%%%%%
\section{Introduction}\label{sec:Introduction}
There is clear observational evidence that galactic star formation is closely connected to the galactic gas content and in particular to molecular gas \citep[see e.
g.][]{1998ARA&A..36..189K,2008AJ....136.2846B,2008AJ....136.2782L,Kennicutt2012}. Most molecular gas is organised in giant clouds with sizes of a few to $\sim$ 100 pc and masses of 10 to $\sim 10^6$ M$_\odot$ \citep[see e.g.][]{Heyer2015} which are observed to follow several scaling relations. The cloud mass functions (CMFs) follow a power law with slope $\alpha$, i.e. $dN/dM \propto M^\alpha$. The slope varies between $\alpha = -1.5$ and $\alpha = -2.1$ \citep{Rosolowsky2005} with typical values of $\alpha \sim -1.8$ \citep{Heyer2001} for the Milky Way. The formation of dense cold clouds and star clusters therein can be triggered and enhanced in extreme environments \citep{Elmegreen1997} such as starburst galaxies \citep{Holtzman1992, Leroy2018}, interacting massive galaxies \citep{Whitmore2010, Herrera2012, Johnson2015} or merging dwarf galaxies \citep{Johnson2000, Kepley2016, Ochsendorf2017}.

Molecular clouds also follow relations between their kinematic and structural properties \citep{Larson1981}. The so-called ``Larson relations'' connect the cloud velocity dispersion along the line of sight, $\sigma_\mathrm{v}$, (from $^{12}$CO line-width observations) to cloud size, $R_\mathrm{c}$ as $\sigma_v \, [\mathrm{km\, s}^{-1}] \propto R^{0.38}_\mathrm{c} \, [\mathrm{pc}]$, the velocity dispersion to cloud mass as $\sigma_\mathrm{v} \, [\mathrm{km\, s}^{-1}] \propto M_\mathrm{c}^{0.2}$, and the cloud density $n_\mathrm{c}$ to cloud size as $n_\mathrm{c} \, [\mathrm{cm^{-3}}] \propto R_\mathrm{c}^{-1.1} \, [\mathrm{pc}]$. More recent estimates of the slope of the first Larson relation indicate values close to $\sim 0.5$ \citep[see e.g.][]{Solomon1987,Heyer2001,Heyer2015}. However, there seem to be considerable variations in the observed slope ranging between 0.15 - 0.91, depending on the respective study \citep[e.g.][]{2008ApJ...686..948B,2020ApJ...895...21I,Duarte-Cabral2021}. The third Larson relation implies approximately constant cloud surface densities \citep[see e.g.][]{2010A&A...519L...7L}. Several other studies indicate surface densities varying from $\sim 10\, \mathrm{M_\odot\, pc}^{-2}$ to $1000\, \mathrm{M_\odot\, pc}^{-2}$ \citep{Heyer2009,2013ApJ...779...46H,Duarte-Cabral2021}. It seems, therefore, unclear how universal observed molecular/cold cloud scaling relations behave. 

The very existence of a physical scaling relation between the density and the size of molecular clouds has been challenged by simulations \citep{MacLow1999, Ballesteros-Paredes2002, MacLow2004}, attributing its existence to observational artefacts, namely the unavoidable existence of a minimum threshold in column density. 
On the other hand, the size-velocity dispersion relation is not disputed since it is reliably reproduced by both observations and simulations. The debate lies in whether its existence confirms the presence of virial equilibrium or even energy equipartition, as argued by previous studies \citep{Ballesteros-Paredes2002, MacLow2004}.

The lifetimes of molecular clouds are even more difficult to determine observationally \citep[see][for a discussion]{Heyer2015}. Already early studies have estimated either short lifetimes $\sim$10 Myr or long lifetimes $\sim$100 Myr \citep{1975ApJ...199L.105S,1979ApJ...229..578S}. More recent work seems to favour maximum lifetimes of several 10 Myr \citep[e.g.][]{2007prpl.conf...81B,2015ApJ...806...72M,2019Natur.569..519K}. Knowledge about the cloud lifetimes is of fundamental astrophysical interest as it can reveal information about the formation (gravity, turbulence, galactic gas thermodynamics) and destruction (galactic shear, star formation and stellar feedback) mechanisms of the cold and dense phase of the interstellar medium (ISM) which is the site of star formation in all galaxies. 

A large number of numerical studies have investigated the formation and disruption of the cold cloud population in massive, Milky Way-like, galaxies. The different simulation studies have been carried out with all major available (magneto-)hydrodynamical galaxy evolution codes \citep[see e.g.][]{Naab2017}. The studies have used different mass and spatial resolutions as well as different sub-grid models for ISM heating and cooling processes, chemistry, star formation and stellar feedback when included. In some studies the stellar disk (with/without spiral structure) and/or the galactic halo potential is analytic \citep[][]{2009ApJ...700..358T,2011MNRAS.413.2935D,2020MNRAS.492.1594S}, in some the halo is represented by particles and can dynamically respond \citep[e.g.][]{2012MNRAS.421.3488H}. A general conclusion from these studies is that the formation and disruption of cold clouds is complex and deeply connected to the larger scale galactic ISM properties \citep[see e.g.][for a review]{2020SSRv..216...50C}. Clouds simulated in the galactic environment can have highly irregular structure \citep[e.g.][]{2013ApJ...776...23B,2020MNRAS.492.1594S,2021MNRAS.505.5438T} and in particular massive clouds can rapidly grow by cloud collisions \citep[][]{2018PASJ...70S..56L,2023MNRAS.519.1887S}. Isolated cloud formation/disruption studies starting with more or less idealised initial conditions are well suited for understanding the cloud internal processes but cannot deliver the full picture. 

Most simulations, even without stellar feedback, indicated that cold clouds form with a declining CMF above the resolution limit with power law slopes of $\alpha \sim -1.8$ \citep[e.g.][]{2009ApJ...700..358T, 2011MNRAS.413.2935D,2012MNRAS.421.3488H, 2018MNRAS.479.3167G, 2020MNRAS.499.5862L, Jeffreson2021}. 
It has been noticed \citep{2012MNRAS.421.3488H} that this slope is similar to analytical expectations from gravitational collapse \citep{1974ApJ...187..425P,1991ApJ...379..440B}. However, other factors could affect the mass function slope, e.g. \citet{2014MNRAS.442.3674W,2020MNRAS.499.5862L} find a slope increase for models with stronger stellar feedback.

Many studies indicate that simulated cold and dense clouds are marginally gravitationally bound or unbound \citep[e.g.][]{2012MNRAS.421.3488H,2021MNRAS.505.5438T} and only the densest regions might collapse \citep{2011MNRAS.417.1318D}. Galaxy interactions are reported to produce clouds which are less bound \citep{2018MNRAS.480.3356P}. Whether the simulated clouds "automatically" follow the Larson relations remains uncertain. In some high-resolution galaxy evolution simulations, the forming clouds seem to follow the linewidth-size relation \citep[e.g.][]{2012MNRAS.421.3488H,Jeffreson2021}. \citet{2019MNRAS.484.1238N} use one of the most complete modelling approaches including radiation transfer and molecular chemistry. Their linewidth-size relation appears to be steeper than observed \citep[see also][for the same effect]{2021MNRAS.505.5438T}. \citet{2018MNRAS.479.3167G} argue that the slope of the linewidth-size relation depends on the implementation of feedback (steeper without feedback) and on whether the analysis is performed in projection (2D) or 3D. 

All galaxy scale simulation studies indicate short cloud lifetimes of several Myr not exceeding a few 10 Myr \citep[e.g][]{2011ApJ...730...11T,2012MNRAS.421.3488H,2013MNRAS.432..653D,2020MNRAS.497.3993B,Jeffreson2021}. For example, \citet{Jeffreson2021} studied the time evolution of molecular clouds in simulated Milky Way-like galaxies and found self-gravitating clouds with lifetimes that scale with their sizes as $\tau_\mathrm{life} \propto \ell^{-0.3}$. In this context, it is worth noting that most if not all simulation based cold (molecular) cloud studies suffer from two major uncertainties - the cloud identification algorithm \citep[e.g.][]{2019MNRAS.484.1238N} and the (sub-grid) modelling of the ISM \citep[e.g.][]{2020MNRAS.499.5862L}, in particular the modelling of molecular gas. For cloud identification different methods have been used like group-finding algorithms \citep[][]{2008MNRAS.391..844D,2012MNRAS.421.3488H,2019MNRAS.484.1238N}, coherent peaks of high-density gas \citep{2009ApJ...700..358T} or dendrogram-based algorithms \citep[][]{2020MNRAS.499.5862L,2021MNRAS.505.5438T,Jeffreson2021}. 
To identify `molecular clouds' different studies have used density thresholds \citep[e.g.][]{2009ApJ...700..358T}, various post-processing methods for molecular hydrogen \citep[e.g.][]{2009ApJ...700..358T}, or direct chemical networks \citep[e.g.][]{2021MNRAS.505.5438T}. The simulations presented here also use a chemical network to follow molecular gas directly. However, we use a clump finding base algorithm as we are simulating a low metallicity system where the main reservoir for star formation is the cold and dense neutral gas phase \citep{Hu2016}. 

In this work, we investigate the formation, the evolution and the properties of simulated cold clouds. For this purpose we analyse the star-forming cold dense gas clouds emerging after a massive starburst caused by the merger of two gas-rich and metal-poor dwarf galaxies in the smoothed particle hydrodynamics (SPH) simulation carried out by \citet{Lahen2019, Lahen2020a, Lahen2020b}. The simulation is part of the \textsc{griffin} project\footnote{\url{https://wwwmpa.mpa-garching.mpg.de/~naab/griffin-project/}} (Galaxy Realizations Including Feedback From INdividual massive stars), which aims to resolve current challenges in galaxy formation \citep[see e.g. review by][]{Naab2017} implementing a detailed model for the ISM \citep{Hu2014, Hu2016, Hu2017, Lahen2019, Lahen2020a, Lahen2020b, Steinwandel2020, Bisbas2022, Hislop2022, Szakacs2022}. The high-resolution simulations have solar mass gas phase resolution and sub-parsec spatial resolution and include a non-equilibrium chemical network/heating and cooling, individually realised massive stars and stellar lifetimes as well as a model for HII regions and resolved supernova explosions. We focus especially on the evolution of the emerging cold clouds and the comparison to observed scaling relations.

This paper is structured as follows. In Sec. \ref{sec:Methods}, we introduce the underlying simulation with its numerical realisation and initial conditions, as well as the developed analysis tools. A description of the cold interstellar medium (ISM) in our numerical experiment follows in Sec. \ref{sec:EvolutionOfTheColdISM}. We analyse the mass and size functions of the detected dense cold gas clouds in Sec. \ref{sec:ColdCloudScalingRelations} and test them for the aforementioned scaling relations in Sec. \ref{sec:ColdCloudScalingRelations}. The life cycle of the cold clouds is studied in Sec. \ref{sec:TheColdCloudlifecycle}. We discuss our findings in context, as well as possible caveats of our methods, in Sec. \ref{sec:Discussion} and conclude this work with a summary in Sec. \ref{sec:Conclusions}. 

%%%%%%%%%%%%%%%%%%%%%%%%%%%%%%%%%%%%%%%%%%%%%%%%%%
\begin{figure*}
\includegraphics[width=1\textwidth]{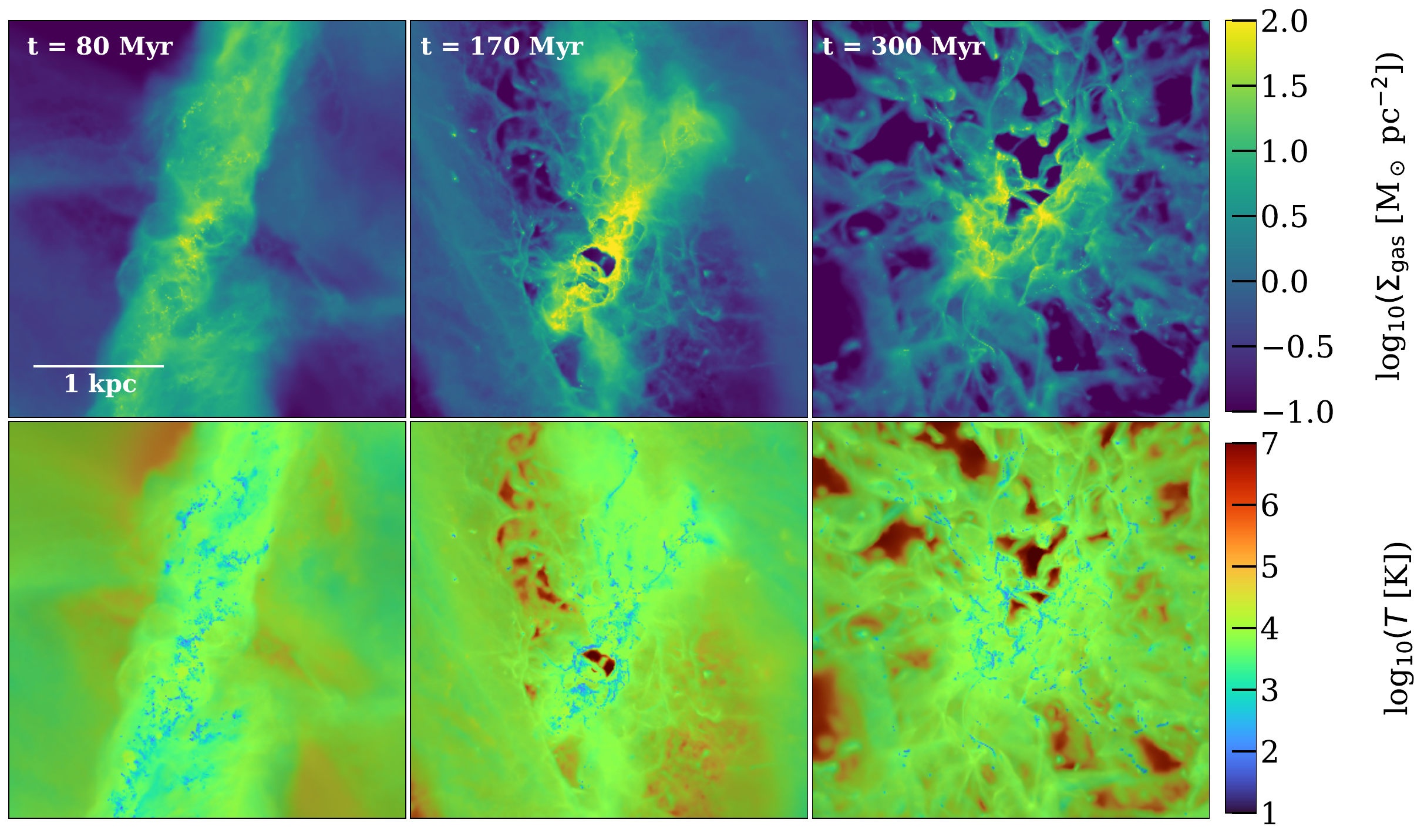}
\caption{Gas surface density maps (top panels) and mass-weighted gas temperature maps (bottom panels) of the central 3 kpc of the dwarf galaxy merger at three different stages: first encounter at 80 Myr (left), the second encounter and peak of star formation (middle panel), and at the new equilibrium configuration at 300 Myr (right panel). Dense (yellow) and cold (blue) gas is structured in clouds and filaments. The hot (red, bottom panel) gas is generated by supernova explosions. }
\label{fig:cold_gas_zoomed}
\end{figure*}
%%%%%%%%%%%%%%%%%%%%%%%%%%%%%%%%%%%%%%%%%%%%%%%%%%

\section{Simulation and methods}\label{sec:Methods}
The dwarf galaxy merger simulation analysed in this paper is presented in \citet{Lahen2019} as part of the \textsc{Griffin} project$^1$. It utilises the \textsc{SPHGal} implementation \citep{Hu2014, Hu2016, Hu2017} of the smoothed particle hydrodynamics (SPH) code \textsc{gadget}-3 \citep{Springel2005} and accounts for artificial viscosity and thermal conduction as described in \citet{Hu2014}. For full details about the simulation, we refer to \citet{Lahen2019, Lahen2020a}. In the following, we briefly describe the setup and star formation, stellar feedback, and the chemistry model.

\subsection{Initial conditions} 
The simulation computes the merger of two identical gas-rich low-metallicity isolated dwarf galaxies \citep[see e.g.][]{Hu2016,Hu2017}. Each dwarf galaxy has a dark matter halo following a \citet{Hernquist1990} profile with a virial mass of $M_\mathrm{vir} = 2 \times 10^{10}\, \mathrm{M}_\odot$, a virial radius of $r_\mathrm{vir} = 44\, \mathrm{kpc}$, a spin parameter of $\lambda = 0.03$, and an equivalent \citet{1997ApJ...490..493N} concentration parameter of $c = 10$.
The embedded rotationally supported gas disk has a mass of $M_\mathrm{gas} = 4 \times 10^7\, \mathrm{M_\odot}$, a scale radius of $r_\mathrm{gas} = 1.46\, \mathrm{kpc}$ and an initial metallicity of $Z_\mathrm{gas} \approx 0.1\, \mathrm{Z_\odot}$, which is a typical value for low-mass dwarf galaxies. The stellar disk has a mass of $M_* = 2 \times 10^7\, \mathrm{M_\odot}$, a scale radius of $r_* = 0.73\, \mathrm{kpc}$ and a scale height of $h_* = 0.35\, \mathrm{kpc}$.
The dwarf galaxies are set up on an initially parabolic orbit with a pericentric distance of $d_\mathrm{peri} = 1.46\,\mathrm{kpc}$ and separation of $d_\mathrm{init} = 5\,\mathrm{kpc}$.

The mass resolution is $m_\mathrm{DM} = 7 \times 10^3\,\mathrm{M}_\odot$ for the dark matter particles and $m_\mathrm{baryonic} = 4\,\mathrm{M}_\odot$ for the baryonic matter (gas and stars) resulting in $5 \times 10^6$ dark matter, $10^7$ gas and $4\times10^6$ stellar particles per dwarf galaxy. The gravitational softening length is $\epsilon_\mathrm{DM} = 62\,\mathrm{pc}$ and $\epsilon_\mathrm{baryonic} = 0.1\,\mathrm{pc}$ for the dark matter and baryonic matter, respectively. 

\subsection{Star formation and stellar feedback}
A gas particle $i$ becomes eligible for star formation once its associated Jeans mass, 
\begin{equation}
 M_\mathrm{J, i} = \dfrac{\pi^{5/2}c_\mathrm{s}^3}{6G^{3/2}\rho_\mathrm{i}^{1/2}},
\end{equation} 
crosses a threshold density of 8 SPH kernel masses ($M_{J, i} < 3200\,\mathrm{M_\odot}$) \citep{Hu2017}, where $c_\mathrm{s}$ is the local sound speed, $G$ is the gravitational constant, and $\rho_\mathrm{i}$ is the local gas density. Each SPH kernel contains approximately 100 neighbour particles.
For gas particles with Jeans masses between 0.5 and 8 SPH kernel masses, star formation is modelled following \citet{Schmidt1959}, such that
\begin{equation}
 \dfrac{d\rho_*}{dt} = \epsilon_\mathrm{ff}\dfrac{\rho_\mathrm{gas}}{t_\mathrm{ff}},
\end{equation}
where $\rho_*$ and $\rho_\mathrm{gas}$ are the stellar and gas densities, respectively, and $t_\mathrm{ff}$ is the free-fall time of the star-forming gas. For the efficiency per free-fall time, $\epsilon_\mathrm{ff}$, we assume a typical value of 0.02 \citep[see][for a discussion on the impact of different efficiencies]{Hislop2022}. For Jeans masses below 0.5 kernel masses ($M_{J, i} < 200\,\mathrm{M_\odot}$), corresponding to number densities of $n_\mathrm{H} > 10^{3.5}\,\mathrm{cm^{-3}}$ at temperatures between $T = 10 - 100\,\mathrm{K}$, we enforce instantaneous star formation. Once a gas particle is labelled as star-forming, we sample stars from a \citet{Kroupa2001} IMF in the mass range of 0.08 $\mathrm{M}_\odot$ to 50 $\mathrm{M}_\odot$. Stars with masses exceeding 4 $\mathrm{M}_\odot$ are represented individually as distinct star particles, while stars with lower masses are aggregated into stellar particles of approximately 4 $\mathrm{M}_\odot$. Each massive star ($M > 4\, \mathrm{M}_\odot$) injects stellar feedback and contributes via far-ultraviolet (FUV) emission to the interstellar radiation field (ISRF). The FUV from the stars is integrated from the \textsc{BaSeL} library \citep{Westera2002} and propagated into the medium considering dust shielding calculated with \textsc{TreeCol} \citep{Clark2012}.  

Additionally, young massive stars ($> 8 \,\mathrm{M_\odot}$) also contribute with hydrogen ionising radiation and the resulting HII regions are modelled as Str\"omgren spheres. The gas within these regions is heated to 10$^4$ K and the size of the regions is iterated until ionisation equilibrium is reached. All details about the modelling of FUV and hydrogen ionising radiation are presented in \citet{Hu2017}.

Massive stars explode as core-collapse supernovae using the lifetimes of \citet{Georgy2013}. We realise Type II supernovae (SNII) by injecting thermal energy equivalent to $E_\mathrm{SN} = 10^{51}\,\mathrm{erg}$ into the surrounding gas particles alongside with ejecta of mass and metals obtained from \citet{Chieffi2004}. The environmental densities at SN explosion sites are pre-processed by the ionising radiation and previous SN explosions which enables a stronger coupling of the injected SN energy to the gas dynamics \citep[see e.g.][]{Walch2015a,Hu2016,Hu2017,Lahen2020b,Rathjen2021,Smith2021,Hislop2022}. We also account for stellar winds from asymptotic giant branch (AGB) stars with metal yields from \citet{Karakas2010}.

\subsection{Chemistry} 
%%%%%%%%%%%%%%%%%%%%%%%%%%%%%%%%%%%%%%%%%%%%%%%%%%
\begin{figure}
\includegraphics[width=1.0\columnwidth]{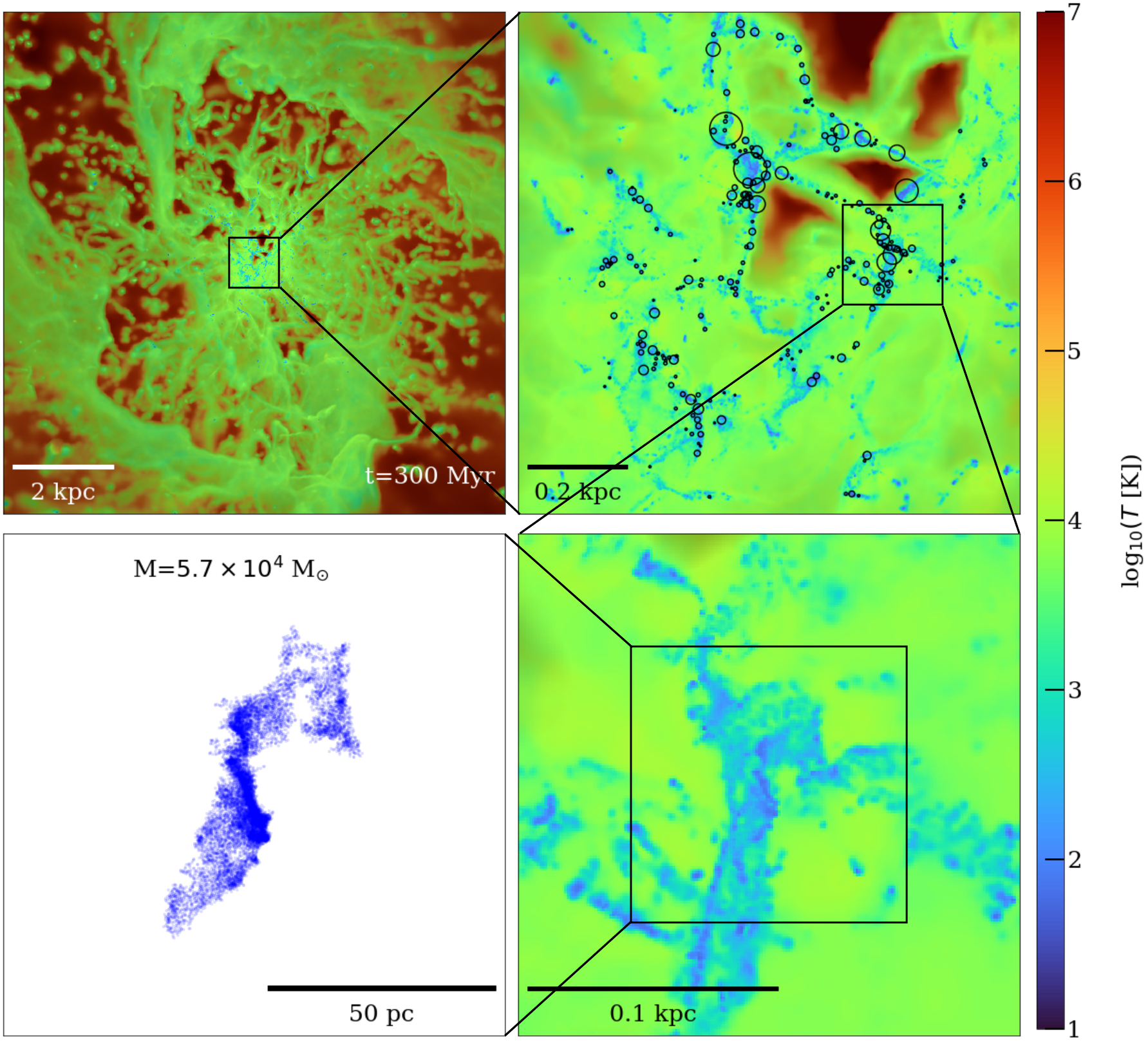}
\caption{Zoom sequence (top left, top right, bottom right) of the gas distribution colour-coded by the mass weighted temperature (first three panels) of the simulation at \mbox{300 Myr} (right panel in Fig. \ref{fig:cold_gas_zoomed}) from \mbox{10 kpc} down to \mbox{100 pc} with one of the most massive cold clouds in the system ($5.7 \times 10^4\, \mathrm{M_{\odot}}$). The black circles in the top right panel indicate the positions and half-mass radii of individual cold clouds identified with the FoF algorithm. We focus solely on the cold ISM gas phase targeting gas with $n_\mathrm{gas}>100\,\mathrm{cm}^{-3}$ and $T_\mathrm{gas}<300\,\mathrm{K}$.
In the bottom right panel, we show a zoom onto a massive cloud with its environment and the raw particle distribution identified by FoF is shown in the bottom left panel. The highly irregular shape is typical for the massive cold clouds in the simulation.}
\label{fig:one_cloud_zoomed}
\end{figure}
%%%%%%%%%%%%%%%%%%%%%%%%%%%%%%%%%%%%%%%%%%%%%%%%%%
The code models non-equilibrium chemistry with heating and cooling while explicitly following the species H, H$^+$, H$_2$, C, C$^+$, O, CO and e$^-$, based on the chemical network by \citet{Nelson1997} and adapted by \citet{Glover2007}. We account for photoelectric heating on dust grains, cosmic ray ionisation and photo-dissociation of H$_2$, as well as fine-structure and metal cooling. However, in the high-temperature regime ($T > 3 \times 10^4\,\mathrm{K}$), we switch to metallicity-dependent equilibrium cooling tables by \citet{Wiersma2009} under the assumption of an optically thin ISM subjected to an ionising UV field from the galactic background \citep{Haardt1996}.

\subsection{Cold cloud identification} 
\label{sec:cloud-identification}
In metal rich galaxies, stars typically form within molecular clouds \citep[see e.g.][]{Lada2003} that consist mainly of H$_2$. Observational data from nearby galaxies have revealed a correlation between the surface density of the SFR ($\Sigma_\mathrm{SFR}$) and the molecular gas surface
density ($\Sigma_\mathrm{H_2}$) \citep{Bigiel2008, Bigiel2011}. Dwarf galaxies, however, are typically metal-poor with a shallow H$_2$ content \citep{Tremonti2004, Hunter2012}. For many star-forming low mass dwarf galaxies very little or no H$_2$ is detected \citep{Schruba2012, Michalowski2015}. Therefore, the larger scale reservoir for star formation in dwarf galaxies might be cold atomic gas, which is also supported by the results from numerical simulations \citep{Glover2012, Krumholz2012, Hu2014, Hu2016}. In our simulated system only small amounts of H$_2$ form, which is typical for metal-poor systems, such as dwarf galaxies \citep[see e.g.][]{2013ApJ...779..102K}. To be more specific, the formation of H$_2$ peaks around the time of the major starburst, i.e. between 150-180 Myr and even during this peak period, H$_2$ accounts for -- on average -- 0.04 per cent of the total gas mass of the system. Therefore, we do not use H$_2$ to identify the sites of star forming clouds, but instead we use gas density and temperature criteria, i.e. gas which is sufficiently dense and cold to form stars. 

To identify the cold clouds in each simulation snapshot we use the friends-of-friends algorithm (FoF) implemented in the Python based \textsc{pygad} analysis package \citep{Rottgers2020}. For the cold and dense ISM gas phase we only consider gas above a defined density and below a defined temperature threshold of $n_\mathrm{gas}>100\,\mathrm{cm}^{-3}$ and $T_\mathrm{gas}<300\,\mathrm{K}$ \citep[e.g.][]{Walch2015, Rathjen2021}. We varied the FoF linking length over an order of magnitude and found that a linking length of 1 pc results in stable cloud masses (see Appendix \ref{appendixA}). 

We only consider clouds (FoF groups) more massive than 100 SPH gas particles (approximately 400 $M_\odot$). We track the individual clouds through the simulation snapshots (separated by 1 Myr) via backward-tracing. A cloud in any given simulation snapshot is considered to be the descendant of a cloud in a previous snapshot if it contains at least 20 per cent of the particles of the parental cloud. We tested a wide range of thresholds between 10 and 90 per cent and found that 20 per cent gives the most stable results, mapping both the growth and the destruction phases of the cold cloud life-cycles. 
%%%%%%%%%%%%%%%%%%%%%%%%%%%%%%%%%%%%%%%%%%%%%%%%%%
\begin{figure} 
\includegraphics[width=1\columnwidth]{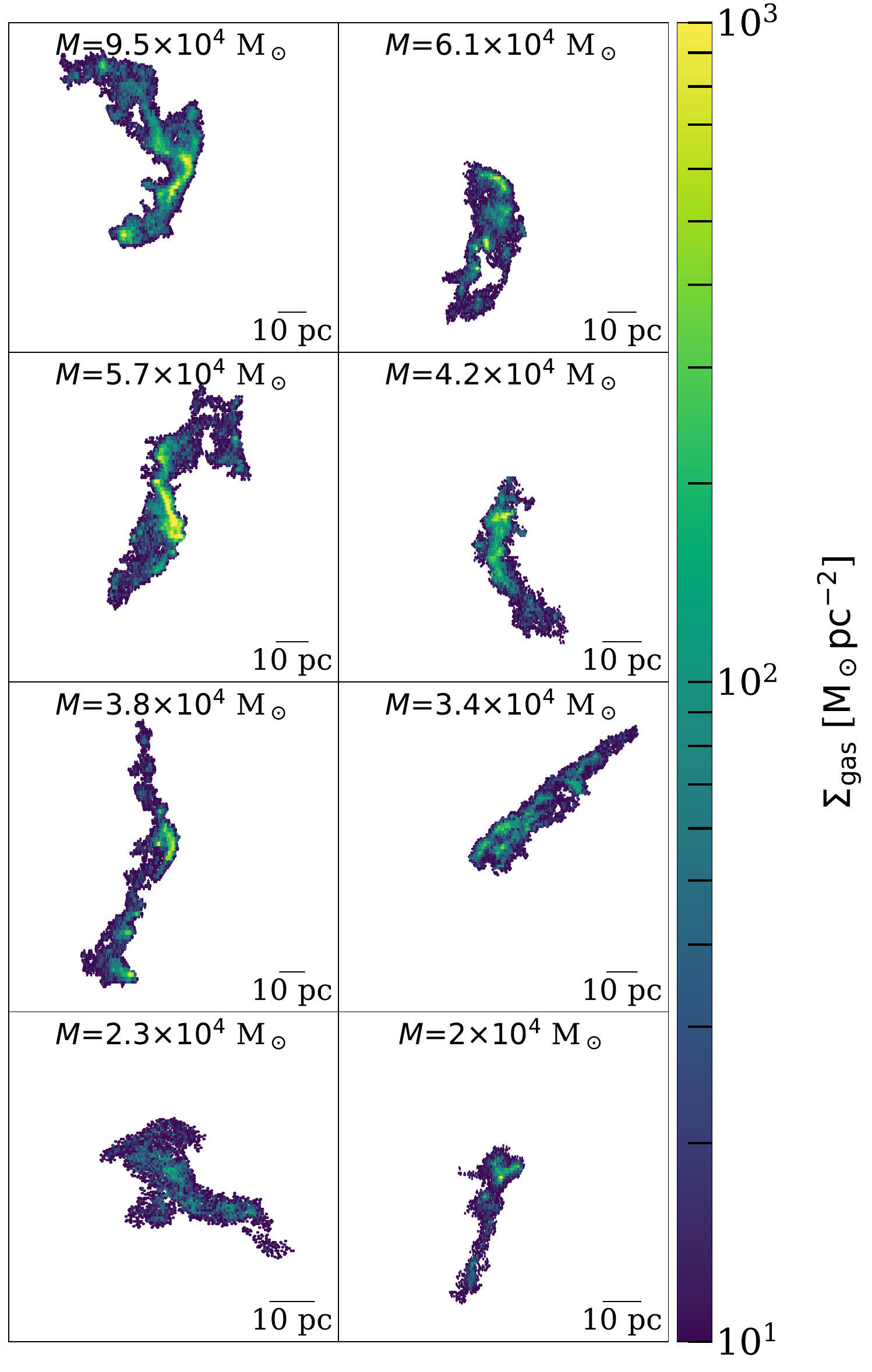}
\caption{Visualisation of the FoF identified SPH gas particle distribution of the eight most massive cold clouds at 300 Myr. Each cloud is centred to its respective densest region in 3D projection. The massive cold clouds typically have filamentary and irregular shapes with density substructure. The colourmap indicates the surface density of each cold gas cloud, from low density areas in dark blue to high density areas in yellow.} 
\label{fig:hateful8_300Myr}
\end{figure}
%%%%%%%%%%%%%%%%%%%%%%%%%%%%%%%%%%%%%%%%%%%%%%%%%%
%%%%%%%%%%%%%%%%%%%%%%%%%%%%%%%%%%%%%%%%%%%%%%%%%%
\begin{figure} 
\includegraphics[width=1\columnwidth]{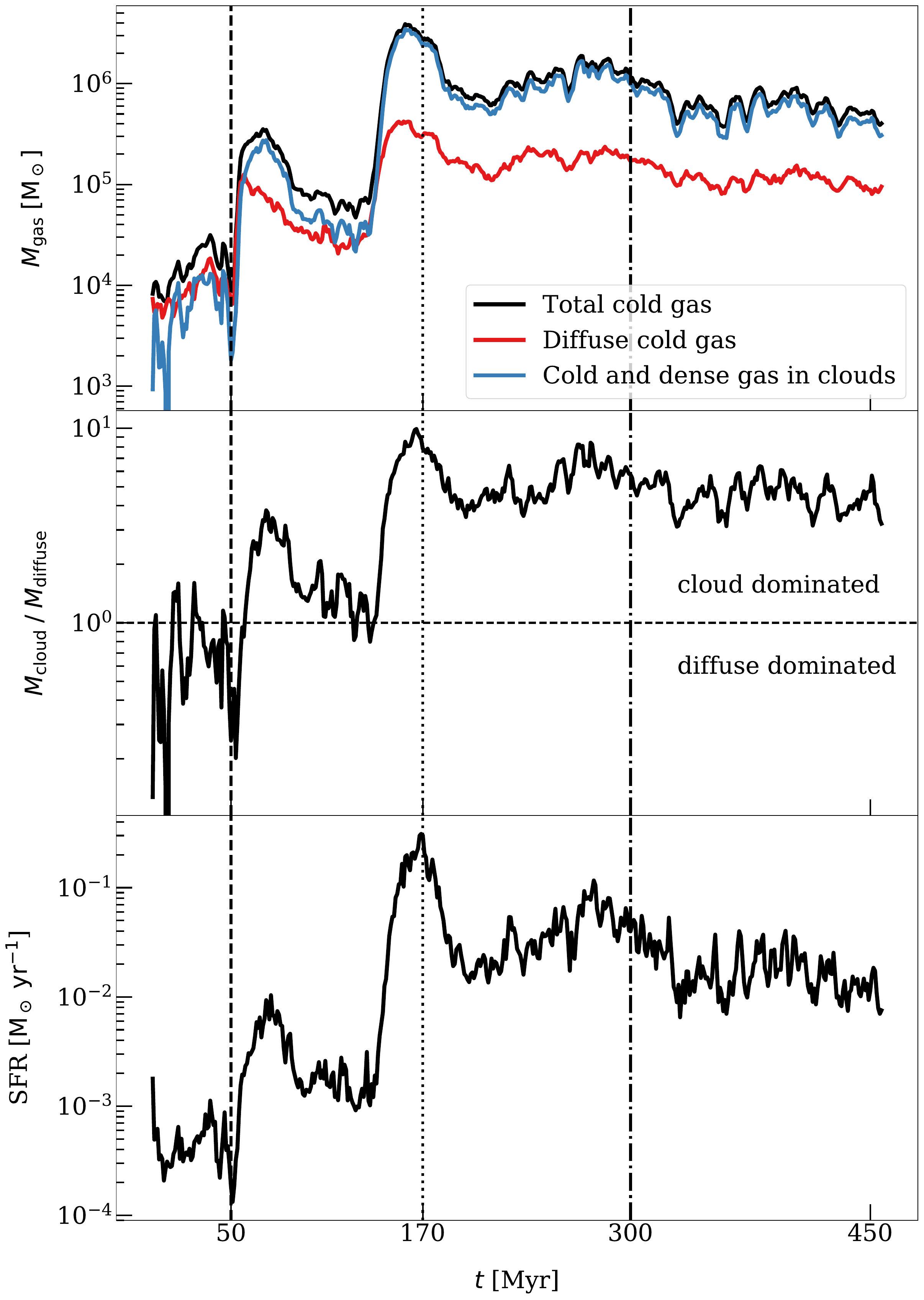}
\caption{{\it Top panel:} Time evolution of the total mass in dense and cold gas ($n_{\mathrm{gas}}>100\,\mathrm{cm^{-3}}$ and $T_{\mathrm{gas}}<300\,\mathrm{K}$, black) split up into cold clouds (blue) and the component not bound in clouds, i.e. the diffuse gas (red). The time of the first encounter at 50 Myr, the second encounter at 170 Myr and the post-merger state at 300 Myr are indicated by dashed, dotted and dashed-dotted lines, respectively. The mass in cold gas strongly increases around the peak starburst. {\it Middle panel:} The mass ratio of cold clouds and cold diffuse gas. Unity is indicated by the horizontal line. The starburst drives the cold medium, which is initially dominated by diffuse gas, into a new equilibrium where the mass of the cold gas bound in clouds is now about four times the mass of the diffuse gas. {\it Bottom panel:} Star formation rate (SFR) as a function of time. The interaction increases the SFR by about two orders of magnitude and drives the system into an extended starburst. The SFR evolution is qualitatively similar to the mass evolution of cold gas in clouds (see top and middle panel).}
\label{fig:FoF_cold_mass_evolution}
\end{figure}
%%%%%%%%%%%%%%%%%%%%%%%%%%%%%%%%%%%%%%%%%%%%%%%%%%
\section{Evolution of the cold ISM}\label{sec:EvolutionOfTheColdISM}

In Fig. \ref{fig:cold_gas_zoomed}, we show an overview of the gas distribution (top panels) and the density-weighted temperature distribution (bottom panels) of the central 3 kpc of the dwarf galaxy merger at the first encounter at 80 Myr, at the peak of star formation during the second encounter after 170 Myr and at 300 Myr, when the system is more relaxed. At the first encounter (left panels) the gas gets compressed and reaches surface densities in excess of $\Sigma_\mathrm{gas} \gtrsim 1000\, \mathrm{M_\odot\, pc^{-2}}$ (top panel), more than two orders of magnitude higher than for isolated systems \citep{Hu2016,Hu2017,Hislop2022}.
At these high surface densities, cold clouds (the blue regions in the bottom panel) are already forming in the turbulent ISM. At the second encounter (middle panels), the merger has the highest SFR and the effect of supernova feedback is visible in the form of filaments and shells at high gas surface density and SN driven bubbles at low surface densities. These bubbles are filled by hot gas (red-coloured, middle bottom panel) generated by SN explosions. The central super-bubble originates from a massive star cluster as discussed in \citet{Lahen2019, Lahen2020a}. Gas surface densities here can reach extreme values of $\Sigma_\mathrm{gas} \gtrsim 10^3\, \mathrm{M_\odot\, pc^{-2}}$ when the system reaches its highest SFR \citep[see e.g.][]{Lahen2020a}. The ISM now shows a clear multi-phase structure with clumpy cold (blue), volume-filling warm (green) and hot (red) gas covering a temperature range of $10 \,\mathrm{K} \lesssim T \lesssim 10^7\, \mathrm{K}$. 
The characteristic filamentary multi-phase structure of the cold ISM in the post-merger state is depicted in the right panels of Fig. \ref{fig:cold_gas_zoomed}. After the peak starburst at $\sim 170$ Myr, the remnant remains in a turbulent state, with mass-weighted velocity dispersion of the cold gas exceeding 10 km s$^{-1}$, for several dynamical times until the end of the simulation at $\sim$ 450 Myr. 

We identify coherent structures in the cold ISM with the FoF identification procedure described in Sec. \ref{sec:cloud-identification}. We term these structures (FoF groups) as "cold clouds" and we identify hundreds of cold clouds at every simulation snapshot. An example of the cloud identification process is shown in Fig. \ref{fig:one_cloud_zoomed} for the merger system at $t_\mathrm{sim}=300\,\mathrm{Myr}$ (e.g. the right panels in Fig. \ref{fig:cold_gas_zoomed}). The top left panel shows the large scale gas distribution colour-coded by temperature on a 10 kpc scale. The hot gas (red) is generated by SNII explosions. A zoom into the central kiloparsec makes the filamentary cold cloud structure visible (blue regions, top right panel). Each black circle represents an identified cold cloud and the maximum spatial extent of the clouds is indicated by the size of the respective circle. A zoom into the region of one of the most massive clouds in this snapshot (bottom right panel) on a 200 pc scale depicts the typical highly irregular structure of the cold gas (blue). In the bottom left panel, we show the underlying SPH gas particle distribution of one identified cloud with a mass of $M=5.7\times10^4\,\mathrm{M_\odot}$. The substructure within the clouds with several dense sub-clumps becomes visible in this representation. This zoom sequence from \mbox{10 kpc} down to \mbox{50 pc} showcases the large dynamic range covered by the simulation. 
%%%%%%%%%%%%%%%%%%%%%%%%%%%%%%%%%%%%%%%%%%%%%%%%%%
\begin{figure} 
\includegraphics[width=1\columnwidth]{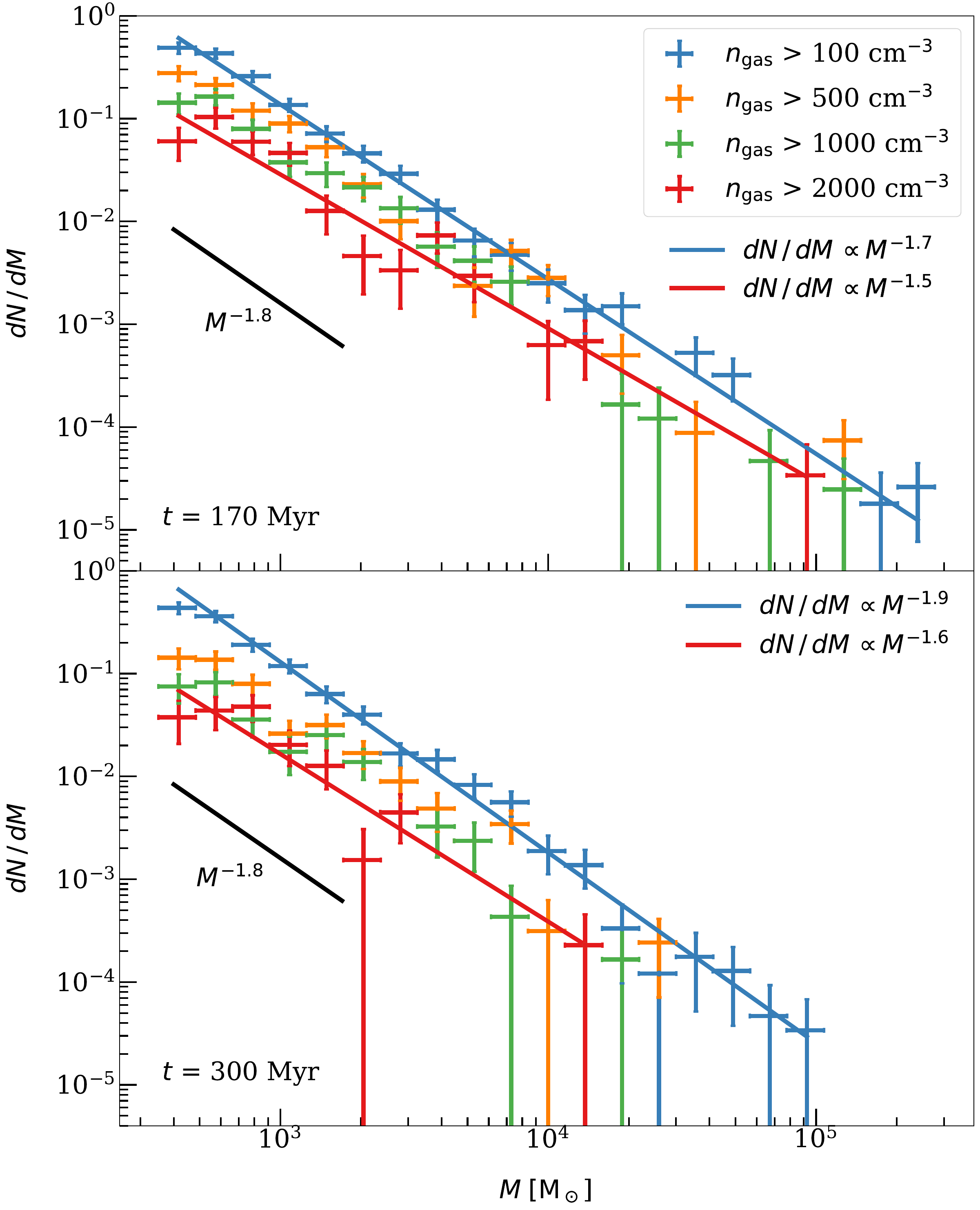}
\caption{Mass function of cold ($T < 300 \mathrm{K}$) clouds at 170 Myr (top panel) and 300 Myr (bottom panel) and increasing density cuts from $n_\mathrm{gas} > 100\, \mathrm{cm}^{-3}$ to $n_\mathrm{gas} > 2000\, \mathrm{cm}^{-3}$ shown in blue, orange, green, and red, respectively. The CMFs follow a power law $dN/dM \propto M^{-\alpha}$ for all density thresholds, indicating self-similar structure. The power law slope for these two snapshots ranges between $-1.6 \lesssim \alpha \lesssim -1.8$. A typical observational estimate of $dN/dM \propto M^{-1.8}$ from CO is shown by the black line \citep{Heyer2001}. The error bars represent the errors in the x- and y- axis. As errors in the x-axis $\delta x$, we use the widths of the respective mass bins: $\delta x = dM$. As errors in the y-axis $\delta y$, we calculate the standard errors of the mass distribution: $\delta y = \sqrt{dN/dM} / dM$. 
}
\label{fig:CMF_diff_dens}
\end{figure}
%%%%%%%%%%%%%%%%%%%%%%%%%%%%%%%%%%%%%%%%%%%%%%%%%%

A gallery of the morphology of the eight most massive cold clouds identified at $t$ = 300 Myr is presented in Fig. \ref{fig:hateful8_300Myr}. Each cloud is centred on its densest region. The visualisation of the SPH gas particle distribution using the gas surface density as a colour map highlights the variations in shape and substructure. At this snapshot the clouds can become as massive as $\mathrm{9.5 \, \times 10^4 \, M_\odot}$. The most massive cloud identified in the entire simulation has a mass of $\mathrm{3.9 \, \times 10^5 \, M_\odot}$ at 156 Myr. The cloud shown in Fig. \ref{fig:one_cloud_zoomed} is the third most massive cloud in this simulation snapshot (second row, left panel of Fig. \ref{fig:hateful8_300Myr}). At the more massive end, the clouds are typically elongated with sizes up to several tens or hundred parsecs and show density substructure. 

With the results of the cloud analysis, we separate the dense and cold ($n_{\mathrm{gas}}>100\,\mathrm{cm^{-3}}$ and $T_{\mathrm{gas}}<300\,\mathrm{K}$) gas phase in two categories: a ``cloud component'', i.e. the gas associated in clouds which are identified using the FoF algorithm (see Sec. \ref{sec:cloud-identification}) and a ``diffuse component'', i.e. the gas that is not identified as part of clouds. 
We present the time evolution of the cold gas for the full simulation in Fig. \ref{fig:FoF_cold_mass_evolution}. In the top panel, the cold gas (black) is divided into a cloud (blue) and a diffuse (no-cloud, red) component. The mass in cold gas is low ($\sim 10^4\, \mathrm{M}_\odot$) and dominated by diffuse gas at the beginning of the simulation when the galaxies are entirely dominated by warm gas. At this time, cold gas constitutes only $\sim 0.01\,\mathrm{per\,cent}$ of the total $\sim 10^7\, \mathrm{M}_\odot$ of gas in the two merging galaxies. Already after the first encounter (vertical dashed line) the gas is compressed and cools so that the cold gas mass increases by about one order of magnitude. Now the cold phase ISM becomes dominated by clouds as can also be seen in the middle panel of Fig. \ref{fig:FoF_cold_mass_evolution}. The effect becomes even stronger after the second encounter (vertical dotted line). The cold gas mass reaches $\sim 10^6\, \mathrm{M}_\odot$ and the cold ISM is now predominantly in a clumpy phase and stays in such a configuration until the end of the simulation. The SFR of the system closely follows the evolution of the cold gas mass as well as the mass fraction of gas in clouds (bottom panel in Fig. \ref{fig:FoF_cold_mass_evolution}).

\section{Mass and size function}\label{sec:MassAndSizeFunction}

In this Section and Section \ref{sec:ColdCloudScalingRelations} we compare the two-dimensional projected properties of the cold clouds identified from the simulation with a three-dimensional friends-of-friends finder as explained in Section \ref{sec:cloud-identification}. It has been indicated in the literature that two-dimensional and three-dimensional identification methods from simulation might result in differences in scaling relations. We do not address this issue here as we would like to keep the cloud definition consistent for all sections of the paper. We refer the reader to \citet{2018MNRAS.479.3167G} for a detailed discussion on cloud identification.
%%%%%%%%%%%%%%%%%%%%%%%%%%%%%%%%%%%%%%%%%%%%%%%%%%
\begin{figure}
\includegraphics[width=1\columnwidth]{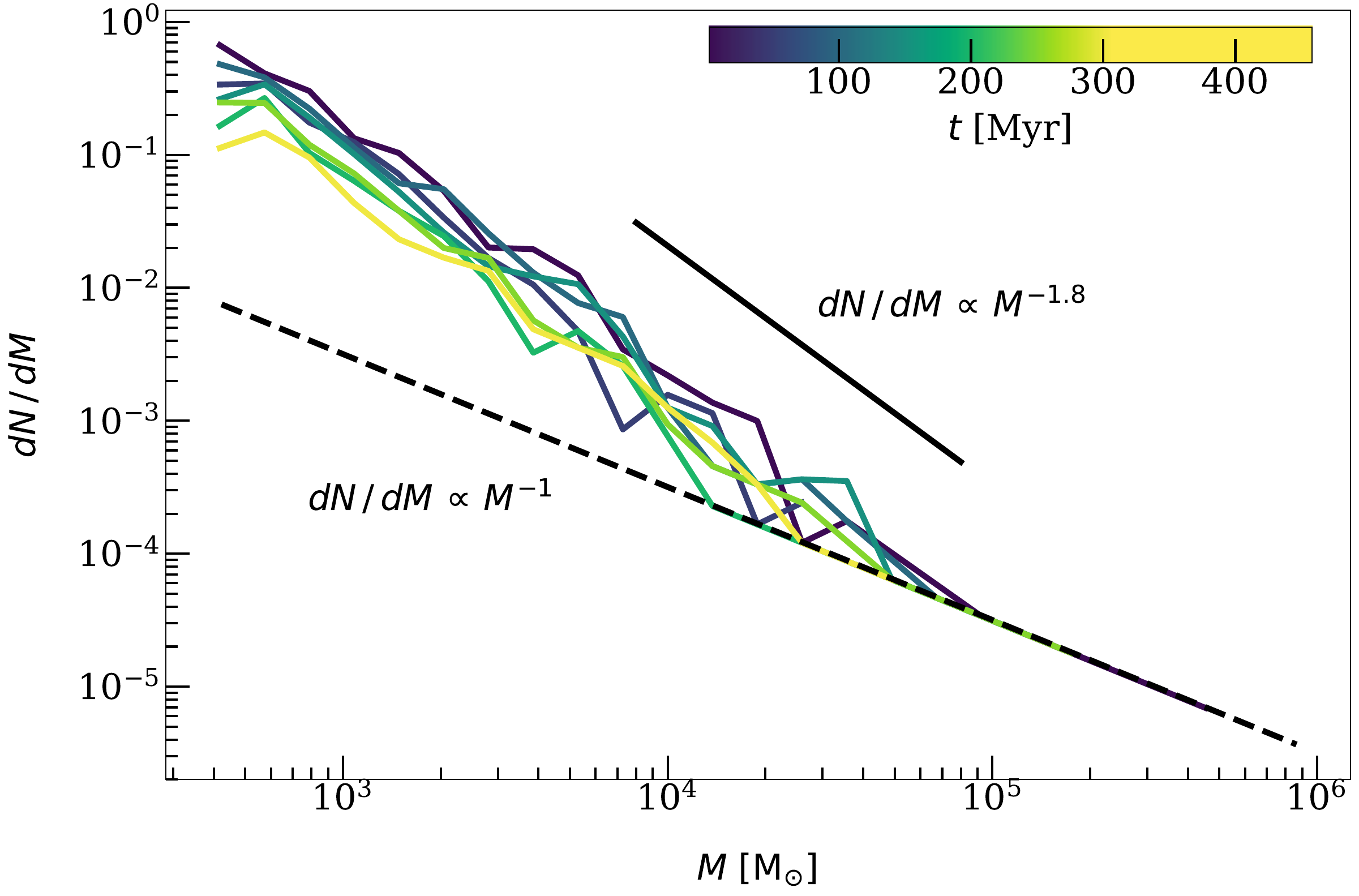}
\caption{Time evolution of the cold CMF from pre-merger times (dark blue) towards the new post-merger equilibrium (yellow) plotted every \mbox{50 Myr}. At all phases, the mass function follows a power-law distribution. For comparison, we indicate by the black line ($dN/dM \propto M^{-1.8}$) the observed slope from CO from \citep{Heyer2001}. The dotted line ($dN/dM \propto M^{-1}$) indicates the one cloud per mass bin threshold.}
\label{fig:CMF_time_evolution}
\end{figure}
%%%%%%%%%%%%%%%%%%%%%%%%%%%%%%%%%%%%%%%%%%%%%%%%%%

%%%%%%%%%%%%%%%%%%%%%%%%%%%%%%%%%%%%%%%%%%%%%%%%%%
\begin{figure} 
\includegraphics[width=1\columnwidth]{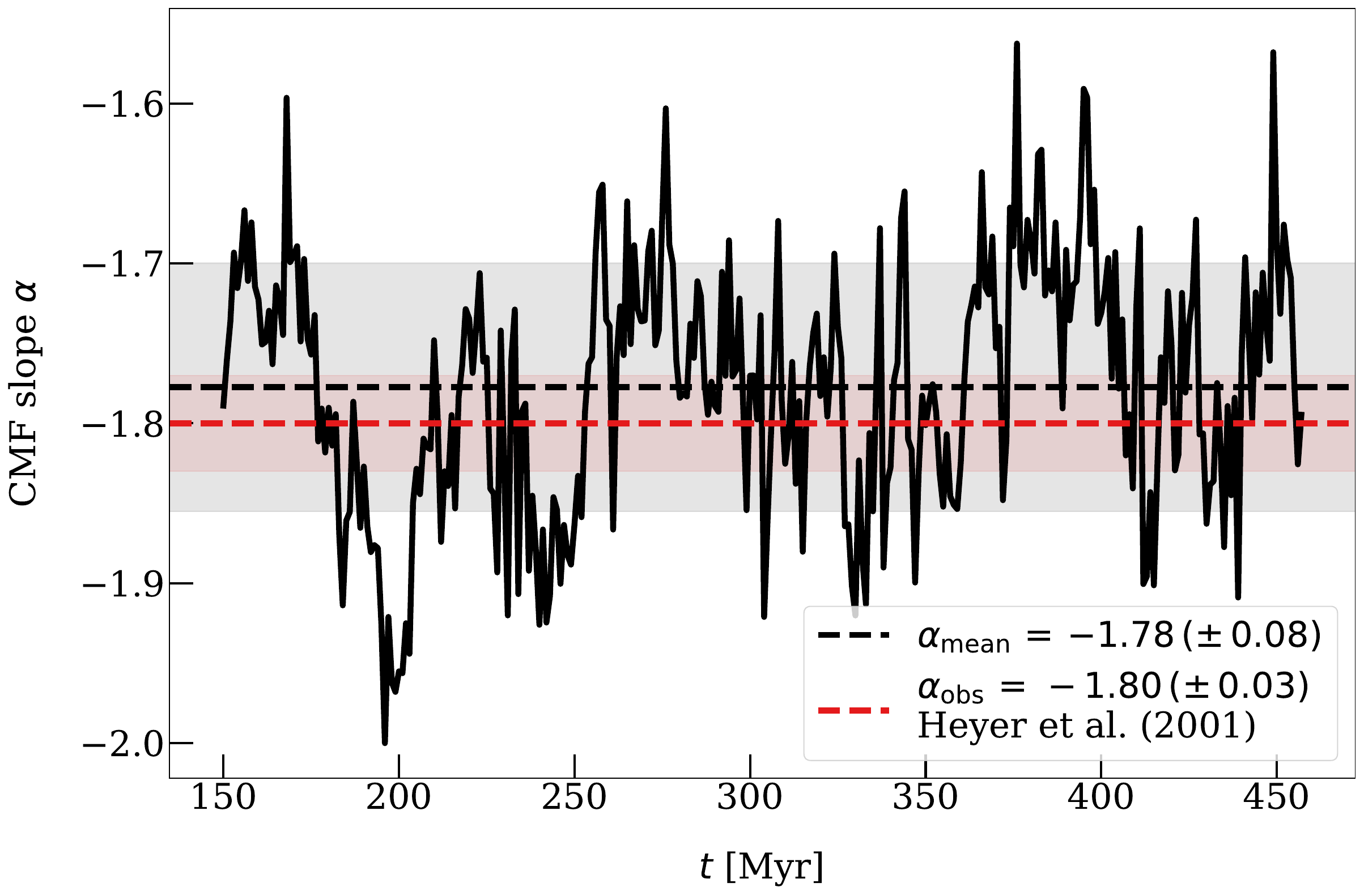}
\caption{Time evolution of the power law slope $\mathrm{\alpha}$ of the cold CMFs shown in Fig. \ref{fig:CMF_time_evolution} starting at 150 Myr (i.e. shortly before the starburst). The mean value of the slope is indicated by the black horizontal dashed line ($\alpha_\mathrm{mean} = -1.78$) and the scatter of 1 $\sigma (\pm 0.08)$ is indicated by the grey area. The value of the observational slope of the CO luminosity function ($\Delta N/\Delta L_\mathrm{CO} \propto L^\mathrm{\alpha}_\mathrm{CO}$) is indicated by the red horizontal dashed line ($\alpha_\mathrm{obs} = -1.8$) and the observational scatter ($\pm 0.03$) is indicated by the red area \citep{Heyer2001}. The mass function slope is similar to the observations at all stages of the simulation.}
\label{fig:CMF_slope_time_evolution}
\end{figure}
%%%%%%%%%%%%%%%%%%%%%%%%%%%%%%%%%%%%%%%%%%%%%%%%%%
In Fig. \ref{fig:CMF_diff_dens}, we present the cold CMF from the minimum mass of $\sim 400\, \mathrm{M}_\odot$ to $\sim 4 \times 10^5\, \mathrm{M}_\odot$ at the peak of star formation at \mbox{170 Myr} (top panel) and at a later phase of the system at \mbox{300 Myr} (bottom panel, see Fig. \ref{fig:cold_gas_zoomed} for the corresponding global gas distributions). We compare the CMFs for four increasing upper limits for the cold gas density. In addition to our fiducial value of $n_\mathrm{gas}= 100\, \mathrm{cm}^{-3}$ (blue) we also perform the FoF analysis for cold gas denser than $n_\mathrm{gas} = 500\, \mathrm{cm}^{-3}$ (orange), $n_\mathrm{gas} = 1000\, \mathrm{cm}^{-3}$ (green), and $n_\mathrm{gas} = 2000\, \mathrm{cm}^{-3}$ (red). In all cases, we study the cold gas, i.e. the temperature limit of 300 K remains constant.
For all density thresholds, the cold CMFs follow similar $dN/dM \propto M^\alpha$ power law distributions with slopes in the range of $-1.9 \lesssim \alpha \lesssim -1.5$. As expected, the normalisation decreases with increasing density thresholds. The slopes of the cold CMFs are comparable to observational values $\alpha \sim -1.8$ for molecular CMFs inferred from the CO luminosity observations \citep[e.g.][]{Heyer2001}. The similar slopes for denser substructures indicate a self-similar density structure within the clouds on the spatial and density scales investigated here. Such self-similarity is typically attributed to the presence of turbulence \citep[see e.g.][]{Elmegreen1996,1998A&A...336..697S,2004ARA&A..42..211E}

The time evolution of the cold CMF for the full simulation is shown in Fig. \ref{fig:CMF_time_evolution} in steps of 50 Myr. The time evolution of the best-fitting slope is shown in Fig. \ref{fig:CMF_slope_time_evolution}. From the early pre-merger stages of the simulation (dark blue) towards the late post-merger equilibrium (yellow) the mass function approximately follows a power-law distribution $dN/dM \propto M^\alpha$ at every stage. The limiting minimum slope $\alpha = -1$ for one cloud per log mass bin is indicated by the black dashed line in Fig. \ref{fig:CMF_time_evolution}. The slope of the cold CMF is in good agreement with the observed CO luminosity function \citep{Heyer2001} (solid black line) at all evolutionary stages (see Fig. \ref{fig:CMF_slope_time_evolution}).

For more than 300 Myr, the best-fitting slope of the CMF varies little (see Fig. \ref{fig:CMF_slope_time_evolution}). The mean value is $\alpha \sim -1.8$ with a scatter of $\sim 0.1$ dex. To test the correlation between the mass function slope and normalisation and the SFR we use the Pearson correlation coefficient $\rho$ that is defined as the ratio between the covariance of two variables $X, \, Y$ and the product of their standard deviations $\sigma_\mathrm{X}, \, \sigma_\mathrm{Y}$, i.e. $\rho(X, Y) = \mathrm{cov}(X, Y)/ (\sigma_\mathrm{X}\sigma_\mathrm{Y})$. The Pearson correlation coefficient is $\rho(\alpha, \mathrm{SFR}) = 0.3$ suggesting that there is only a weak correlation between the slope of the CMF $\alpha$ and the actual SFR of the system. The correlation coefficient for the normalisation of the CMF $\beta$ and the SFR is $\rho(\beta, \mathrm{SFR}) = 0.2$, also suggesting a weak correlation. Therefore, the star formation activity of the system does not appear to have a strong impact on cold cloud mass function. However, the mass of the most massive cloud identified in each snapshot between 150 Myr and 458 Myr correlates strongly with the SFR with $\rho(M_\mathrm{max}, \mathrm{SFR}) = 0.8$. This is in agreement with a similar correlation between the most massive forming star cluster and the SFR in the same simulation \citep{Lahen2020a}.

%%%%%%%%%%%%%%%%%%%%%%%%%%%%%%%%%%%%%%%%%%%%%%%%%%
\begin{figure} 
\includegraphics[width=1\columnwidth]{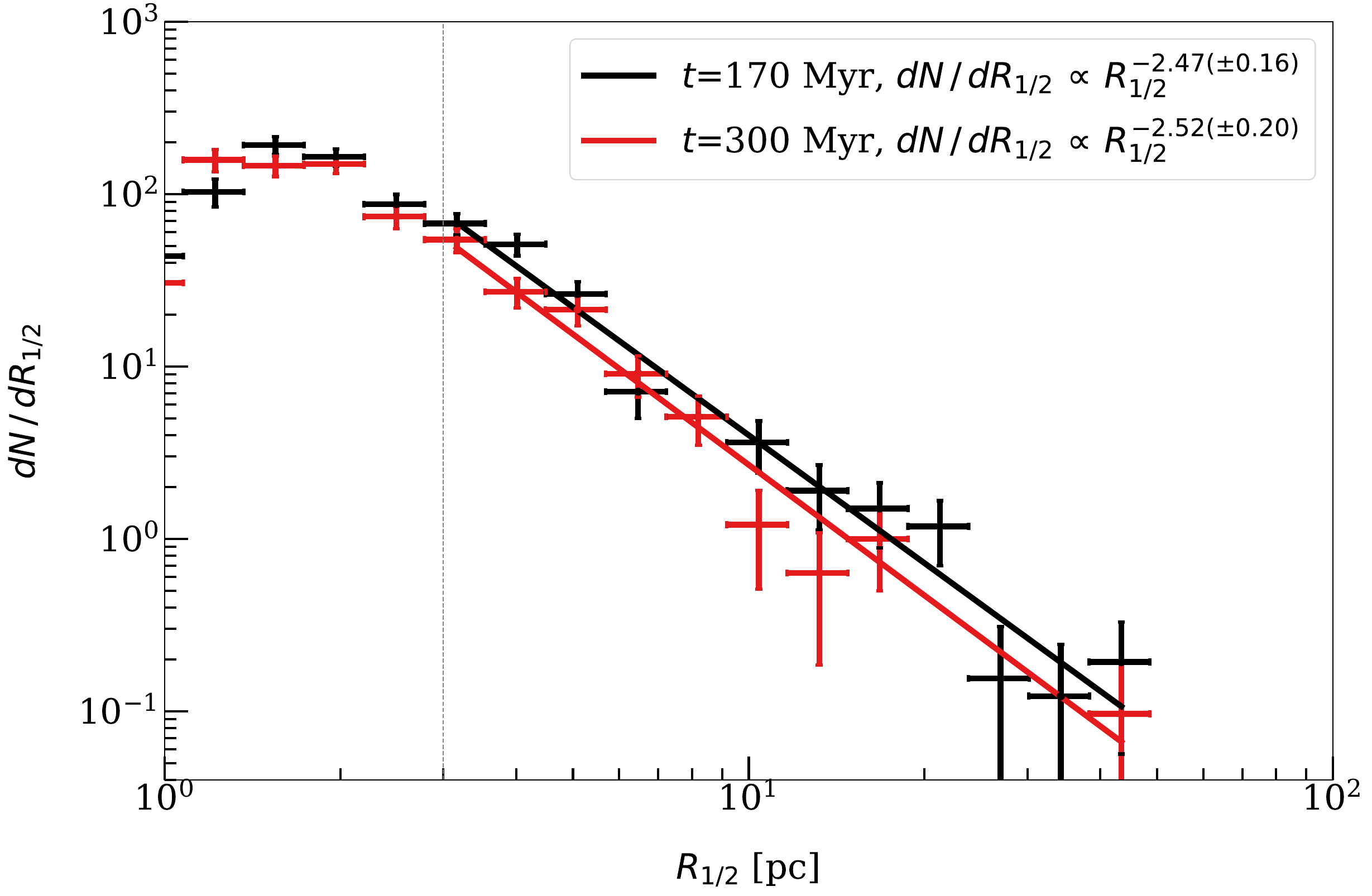}
\caption{Differential size distribution of the cold clouds detected at \mbox{170 Myr} (black) and at 300 Myr (red). The size of the clouds is represented by the projected half mass radius $R_{1/2}$. The differential size distribution of the clouds is a power law distribution with index around -2.5, for both 170 Myr and 300 Myr. The vertical dotted line indicates the observational limit of \citet{Heyer2001} at $R_{1/2}$ $\sim$ 3 pc. 
The error bars in the x-axis represent the widths of the respective size bins: $\delta x = dR_{1/2}$. The error bars in the y-axis represent the standard errors of the size distribution: $\delta y = \sqrt{dN/dR_{1/2}} / dR_{1/2}$. 
}
\label{fig:size_distibution}
\end{figure}
%%%%%%%%%%%%%%%%%%%%%%%%%%%%%%%%%%%%%%%%%%%%%%%%%%

In Fig. \ref{fig:size_distibution}, we plot the cloud size function of the cold clouds at 170 Myr (black) and 300 Myr (red). Similarly to the CMF, the cloud size function is defined as the number of clouds in a specific size range divided by that range, i.e. $dN/dR_{1/2}$, plotted against the size $R_{1/2}$. As a measure for the cloud size, we use the projected half mass radius $R_{1/2}$, which is calculated from the respective centre of each cloud. Each cloud is centred on its densest region (see Fig. \ref{fig:hateful8_300Myr}).
To determine the half-mass radius $R_{1/2}$ of each cloud, we first calculate the total mass of the cloud and the radial distances of each gas particle from the densest point of the cloud, on the $x-y$ plane. Subsequently, we sort the array of radial distances from minimum to maximum and the corresponding array of gas particle masses following the sorting of the radial distances array. This ensures that each element of both arrays corresponds to properties of the same gas particle. By summing the gas particle masses along the mass array until the half of the total mass of cloud is reached (or at least not exceeded), we identify the index at which this occurs. The radial distance associated with this index in the sorted array represents the projected half-mass radius on the x-y plane $R_{1/2}$. This process is repeated for all clouds and for the $y-z$ and $z-x$ planes, with minimal variation observed between the resulting half-mass radii in each projection. The sizes of the simulated cold clouds range from a few parsecs to $\sim$ 50 pc. The cloud size function follows a power-law distribution $dN/dR_{1/2} \propto R_{1/2}^\alpha$ with index $\alpha = -2.47(\pm 0.16)$ at \mbox{170 Myr} (black) and $\alpha = -2.52(\pm 0.20)$ at \mbox{300 Myr} (red). These values are shallower than that of the observations of \citet{Heyer2001} who find a slope of $\alpha_{\mathrm{obs}} = -3.2 (\pm 0.1)$. To directly compare with the observations with a detection threshold of 3 pc \citep{Heyer2001}, we have only used clouds with $R_{1/2} >$ 3 pc in our analysis, but we display the full range of our sample in Fig. \ref{fig:size_distibution}. The reason for the shallower size distribution slopes of the simulated clouds are not obvious and will have to be investigated in future studies. Here, the cold cloud structure identification, the tracer, the stellar feedback implementation, and the galactic environment could be of importance. With observations,  molecular cloud sizes are also difficult to measure  and not many observations exist. The observational tracer, the algorithm to determine the sizes, or the galactic environment can have an impact on the size distribution \citep[see e.g. discussions in][]{Heyer2015,2020A&A...642A.177D}.
From now on we will use only clouds with $R_{1/2} >$ 3 pc in our analysis, unless stated otherwise and we will display a vertical dotted line at $R_{1/2}$ $\sim$ 3 pc in the respective figures to indicate the observational limit of \citet{Heyer2001}.

\section{Cold cloud scaling relations}\label{sec:ColdCloudScalingRelations}
%%%%%%%%%%%%%%%%%%%%%%%%%%%%%%%%%%%%%%%%%%%%%%%%%%
\begin{figure} 
\includegraphics[width=1\columnwidth]{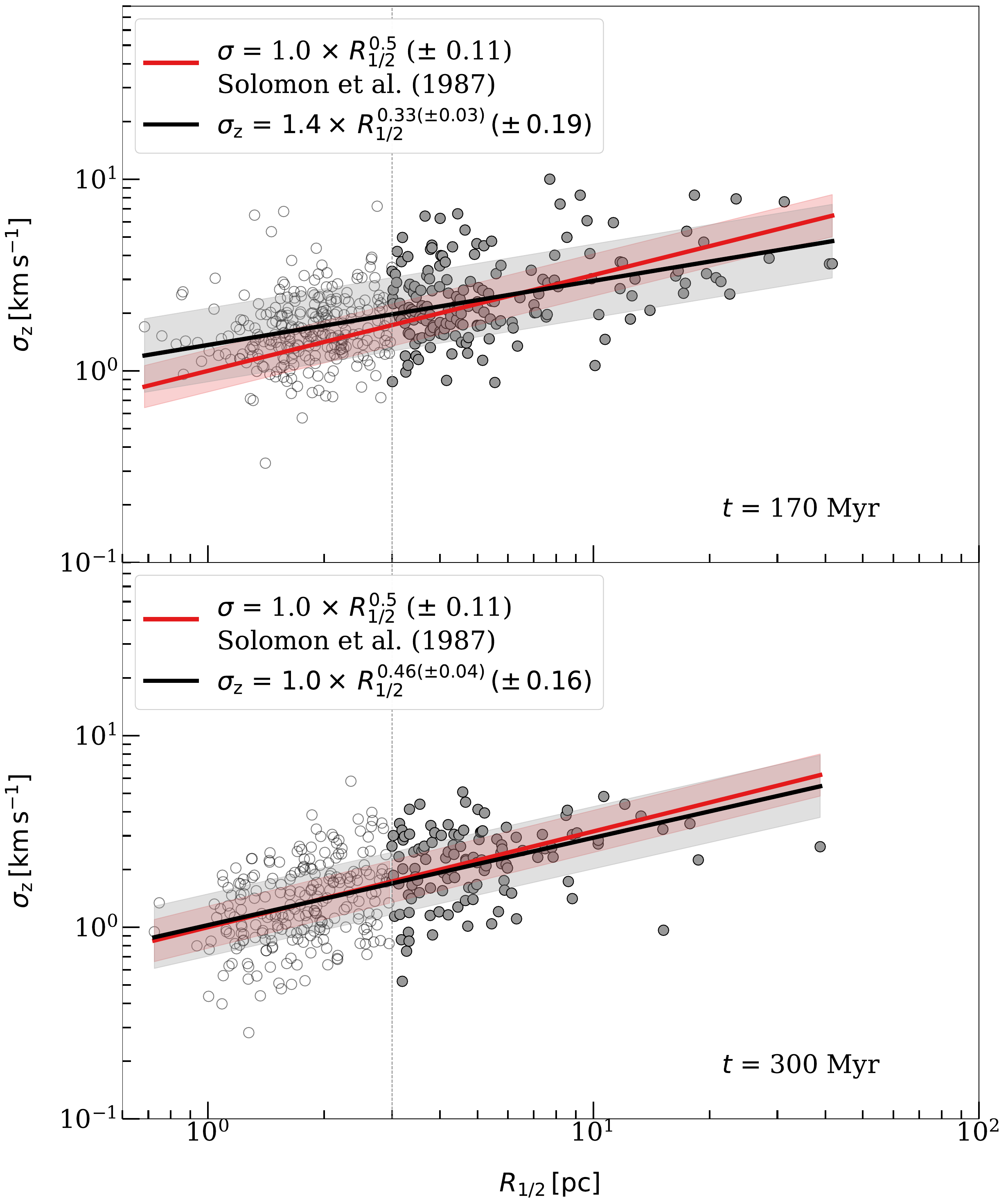}
\caption{Line-of-sight velocity dispersion in the $z$-direction $\sigma_\mathrm{z}$, as a function of the size of the identified cold clouds in terms of the projected half-mass-radius $R_{1/2}$ on the $x-y$ plane at 170 Myr (top panel) and 300 Myr (bottom panel). The vertical dashed line marks the 3 pc detection threshold of \citet{Heyer2001}. The empty circles represent clouds with $R_{1/2} \leq 3$ pc, whereas the filled circles represent clouds with $R_{1/2} > 3$ pc. The solid red line shows the best-fit to the \citet{Solomon1987} points and the shaded red area represents the scatter. The solid black line represents the best fit to the simulation data while the shaded black area denotes the standard deviation of the cloud distribution with sizes larger than 3 pc.
}
\label{fig:scaling_plot_1}
\end{figure}
%%%%%%%%%%%%%%%%%%%%%%%%%%%%%%%%%%%%%%%%%%%%%%%%%%

The observed cold cloud scaling relations introduced by \citet{Larson1981} can be understood as an indication that cold clouds are self-gravitating and, possibly, in some form of virial equilibrium \citep[see][]{Heyer2001, Heyer2009, Heyer2015}. To determine if a system -- like a cold cloud or clump -- is in virial equilibrium we use the virial parameter $\alpha_\mathrm{vir}$, which is a dimensionless parameter that quantifies the ratio of thermal and kinetic energies to gravitational energy in the system. It is mathematically expressed as \citep{Bertoldi1992}:
%*************************************************
\begin{equation}
 \alpha_\mathrm{vir} \equiv \frac{5 \sigma^2_\mathrm{v} R}{G M_{\mathrm{vir}}},
 \label{eq:virial_parameter}
\end{equation}
%*************************************************
where $G$ is the gravitational constant, $R$ is the radius, $M_\mathrm{vir}$ is the corresponding virial mass and $\sigma_v$ is the average one dimensional velocity dispersion of the system.
When the gravitational energy of a cloud becomes comparable to its kinetic energy, then $\alpha_\mathrm{vir} \sim 1$. However, when the surface pressure confines the cloud and self-gravity is negligible, then the virial parameter is significantly greater than one: $\alpha_\mathrm{vir} > 1$.
The virial mass $M_\mathrm{vir}$ estimate from Eq. \ref{eq:virial_parameter} and the mass of a cold cloud $M_\mathrm{cc}$ estimated from a measured effective (projected half-mass) radius $R_{1/2}$, an effective surface mass density $\Sigma_{1/2}$ (the mass within the effective radius divided by the area) are:
%*************************************************
\begin{equation}
M_{\mathrm{vir}} = 5 \frac{\sigma^2_\mathrm{v} R_{1/2}}{G} \; \mathrm{and} \; M_{\mathrm{cc}} = \pi \, \Sigma_{1/2} \,R_{1/2}^2 , 
\label{eq:cloud_and_virial_masses}
\end{equation}
%*************************************************
\noindent respectively. 
For clouds in virial equilibrium $\alpha_{\mathrm{vir}} = M_{\mathrm{vir}}/M_{\mathrm{cc}} =1$
%*************************************************
and 
\begin{equation}
\sigma_{v} = \sqrt{\frac{\pi G}{5}}\, R_{1/2}^{0.5}\,\Sigma_{1/2}^{0.5} .
\label{eq:vir}
\end{equation}
%*************************************************

%%%%%%%%%%%%%%%%%%%%%%%%%%%%%%%%%%%%%%%%%%%%%%%%%%
\begin{figure} 
\includegraphics[width=1\columnwidth]{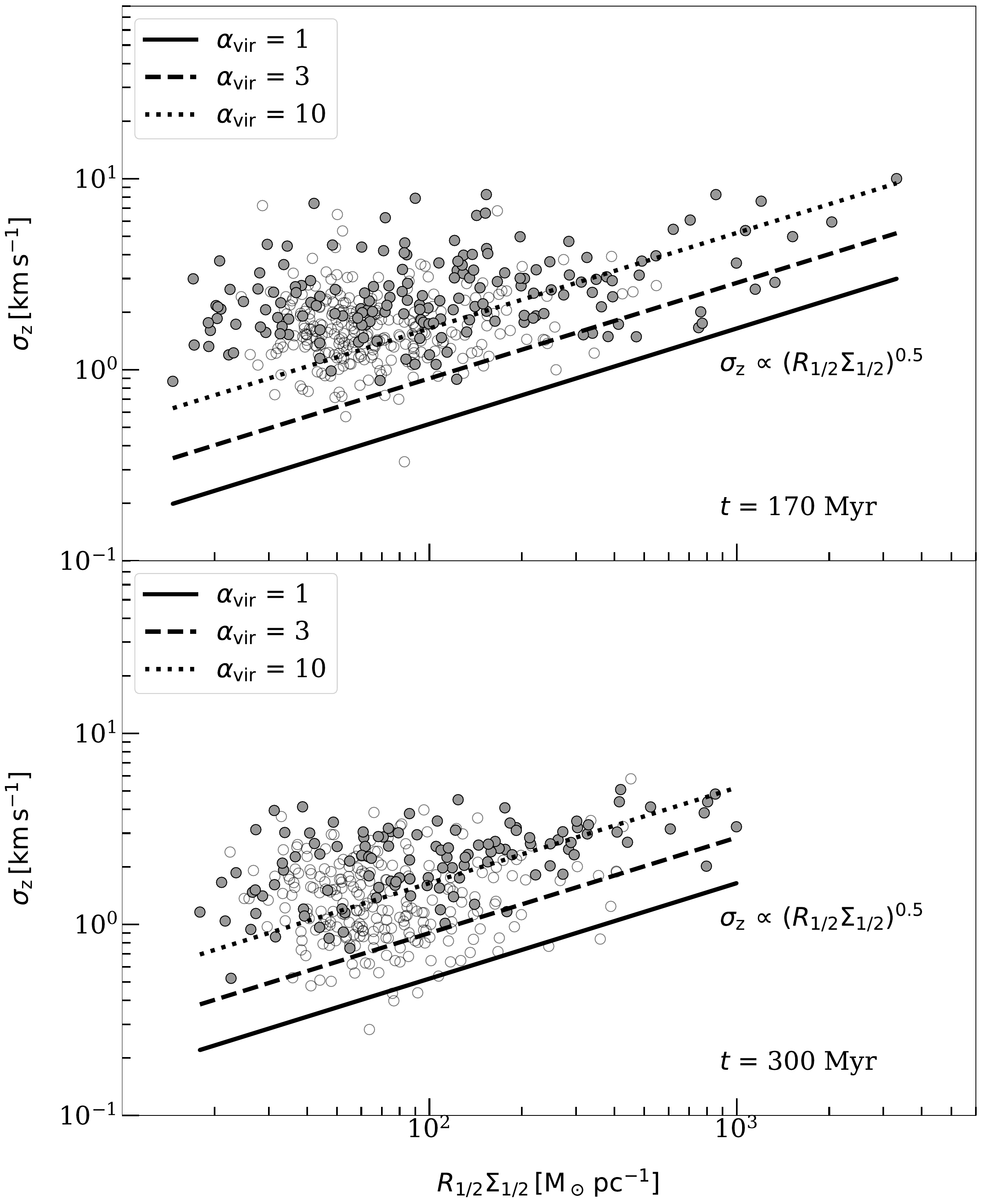}
\caption{Line-of-sight velocity dispersion in the $z$-direction $\sigma_\mathrm{z}$ vs. the product of the size of the identified cold clouds in terms of the face-on projected half-mass-radius $R_{1/2}$ on the $x-y$ plane and the effective surface density $\Sigma_{1/2}$ at 170 Myr (top panel) and 300 Myr (bottom panel). The empty circles represent clouds with $R_{1/2} \leq 3$ pc, whereas the filled circles represent clouds with $R_{1/2} > 3$ pc. The solid line represents Eq. \ref{eq:vir} for virial parameter $\alpha_\mathrm{vir}=1$. The dashed line is derived from Eq. \ref{eq:cloud_and_virial_masses} for $\alpha_\mathrm{vir}=3$ and the dotted line for $\alpha_\mathrm{vir}=10$. 
}
\label{fig:scaling_plot_2}
\end{figure}
%%%%%%%%%%%%%%%%%%%%%%%%%%%%%%%%%%%%%%%%%%%%%%%%%%

In Fig. \ref{fig:scaling_plot_1}, we show the first Larson relation for the clouds identified in the simulations with our fiducial density and temperature thresholds ($n_{\mathrm{gas}}>100\,\mathrm{cm^{-3}}$ and $T_{\mathrm{gas}}<300\,\mathrm{K}$). The one-dimensional line-of-sight velocity dispersion ($\sigma_\mathrm{z}$) for each cloud is computed from the mass-weighted gas particle velocities along the $z$-axis of the computational domain. The half mass radius $R_{1/2}$ is the projected radius enclosing half of the cloud mass from the densest region of the cloud. The effective mass surface density $\Sigma_{1/2}$ is the half-mass of each cloud divided by the half-mass projected area of the cloud around its densest point (Eq. \ref{eq:cloud_and_virial_masses}).
The relationship between the line-of-sight velocity dispersion and the size of the identified cold clouds shown in this figure was investigated at two characteristic times: right after the starburst at 170 Myr (top panel) and at 300 Myr (bottom panel) when the system has reached post-merger equilibrium. 
We show the best power-law fit to our simulated clouds and the standard deviation.
For comparison, we show the canonical relation derived in \citet{Solomon1987} and discussed in the review by \citet{Heyer2015}. It is represented by the solid red line and the shaded red area represents their scatter.  
The fitted slopes ($\alpha_\mathrm{170 \; Myr}$ = 0.33($\pm$ 0.03) and $\alpha_\mathrm{300 \; Myr}$ = 0.46($\pm$ 0.04) are somewhat shallower than the ``canonical'' slope of 0.5 reported in \citet{Solomon1987}. 
The slopes are well within the range reported from various observations (0.15 - 0.91, see Sec. \ref{sec:Introduction}). This indicates that the cold clouds in our simulation model have properties which are consistent with observed cold clouds.

Fig. \ref{fig:scaling_plot_2} shows the correlation between the one-dimensional line-of-sight velocity dispersion $\sigma_\mathrm{z}$ and the size of identified cold clouds in terms of their projected half-mass-radius $R_{1/2}$ and their effective surface density $\Sigma_{1/2}$ at 170 Myr (top panel) and 300 Myr (bottom panel). 
We show the virial parameters of 1, 3, and 10, respectively, assuming a virial constant of 5 (see Eq. \ref{eq:cloud_and_virial_masses} and \ref{eq:vir}). 
For the sake of completeness, we should note that when testing the fits on our whole sample of simulated cold clouds, the parameters of the fits (i.e. slope, normalisation of the power laws) did not change significantly for neither Fig. \ref{fig:scaling_plot_1} nor Fig. \ref{fig:scaling_plot_2}.
In addition, the results shown Fig. \ref{fig:scaling_plot_1} and Fig. \ref{fig:scaling_plot_2} are similar for line-of-sight projections along the x- and y-axis (i.e. $\sigma_\mathrm{x}$, $\sigma_\mathrm{y}$).

%%%%%%%%%%%%%%%%%%%%%%%%%%%%%%%%%%%%%%%%%%%%%%%%%%
\begin{figure} 
\includegraphics[width=1\columnwidth]{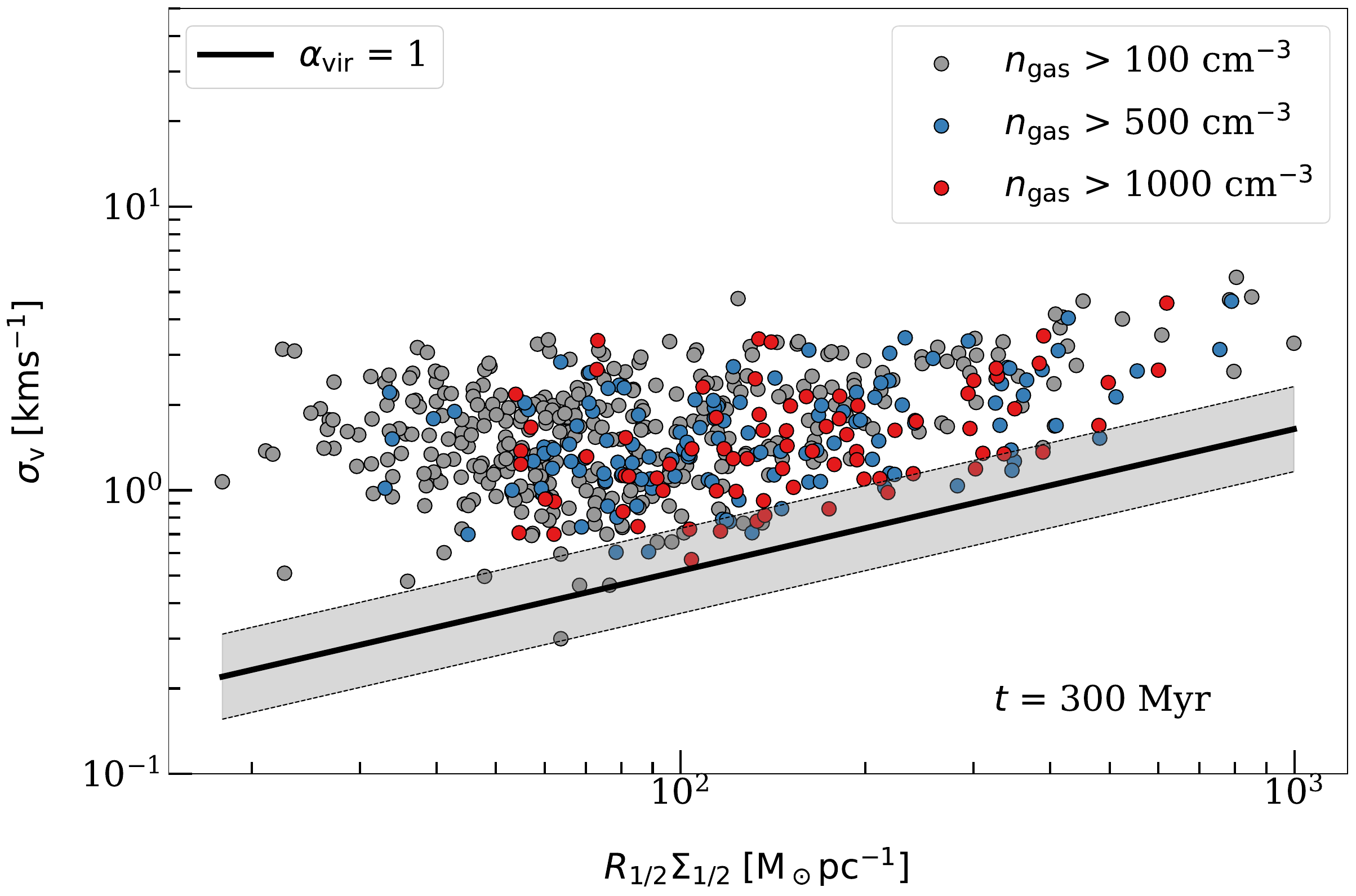}
\caption{Average one dimensional velocity dispersion $\sigma_\mathrm{v} = \sqrt{\sigma^2_\mathrm{x} + \sigma^2_\mathrm{y} + \sigma^2_\mathrm{z}} / \sqrt{3}$ as a function of the product of the identified cold clouds size in terms of the projected half-mass-radius $R_{1/2}$ and the effective surface density $\Sigma_{1/2}$ at 300 Myr for different gas density thresholds: $n_\mathrm{gas} > 100$ $\mathrm{cm}^{-3}$ (grey), $n_\mathrm{gas} > 500$ $\mathrm{cm}^{-3}$ (blue) and $n_\mathrm{gas} > 1000$ $\mathrm{cm}^{-3}$ (red). The solid line represents Eq. \ref{eq:vir} for virial parameter $\alpha_\mathrm{vir}=1$. A grey shaded area of 0.3 dex around the $\alpha_\mathrm{vir}=1$ line is plotted to visually indicate the number of clouds in approximate virial equilibrium. As the total gas number density threshold increases from $n_\mathrm{gas} > 100\,\mathrm{cm}^{-3}$ to $n_\mathrm{gas} > 1000\,\mathrm{cm}^{-3}$, the total number of identified clouds decreases, but the number of clouds in virial equilibrium (i.e. the clouds in the shaded area) increases.}
\label{fig:scaling_plot_2_evans}
\end{figure}
%%%%%%%%%%%%%%%%%%%%%%%%%%%%%%%%%%%%%%%%%%%%%%%%%%

In Fig. \ref{fig:scaling_plot_2_evans} we show the average one dimensional velocity dispersion $\sigma_\mathrm{v}$ of all identified clouds as a function of the product of cold cold clouds size, given by the projected half-mass-radius $R_{1/2}$, and the effective surface density $\Sigma_{1/2}$, for different gas number density thresholds: $n_\mathrm{gas} > 100$ $\mathrm{cm}^{-3}$ (grey), $n_\mathrm{gas} > 500$ $\mathrm{cm}^{-3}$ (blue), and $n_\mathrm{gas} > 1000$ $\mathrm{cm}^{-3}$ (red). 
The solid line in the plot represents Eq. \ref{eq:vir} for a virial parameter $\alpha_\mathrm{vir}=1$ and the grey-shaded indicates the range of $0.5 < \alpha_\mathrm{vir} < 2$ which might be considered close to virial equilibrium. Values of $\alpha_\mathrm{vir} < 0.5$ would clearly indicate collapse, while $\alpha_\mathrm{vir} > 2$ points towards expansion. As the gas density threshold increases from $n_\mathrm{gas} > 100$ $\mathrm{cm}^{-3}$ to $n_\mathrm{gas} > 1000$ $\mathrm{cm}^{-3}$, the total number of identified clouds decreases. With the higher gas density threshold, the identified cold clouds are the very dense cores of the previously identified clouds at a lower threshold, without their less dense outer regions. Some clouds are already so disrupted by stellar feedback or not located in suitable environmental conditions to form such dense cores, which leads to a decrease of the total number of identified clouds. However, the relative number of clouds in virial equilibrium, which are represented by the clouds in the shaded area, increases with the increase in density threshold. The dense regions of clouds, when outer less dense regions are ignored, are closer to virial equilibrium.
Using only the identified clouds with densities higher than 1000 $\mathrm{cm}^{-3}$ and in virial equilibrium, i.e. the cold clouds represented by red scatter points that fall into the shaded area in Fig. \ref{fig:scaling_plot_2_evans}, we calculate the SFR. To do so, first, we calculate the free-fall time of the selected clouds using the formula \citep{Evans2021}: 
%*************************************************
\begin{equation}
 t_\mathrm{ff} = \left(\frac{3 \pi}{32 G \rho}\right)^{0.5} ,
 \label{eq:free-fall time}
\end{equation}
%*************************************************
where $G$ is the gravitational constant and $\rho$ is the cloud's average volume density, weighted by the SPH-particles mass for each cloud.
The SFR$_\mathrm{cc}$ derived by the selected cold clouds can be calculated using the sum of the masses of the clouds divided by the free-fall times: 
%*************************************************
\begin{equation}
 \mathrm{SFR_{cc}} = \sum_{i=1}^{N} \frac{M_\mathrm{cc}}{t_\mathrm{ff}} ,
 \label{eq:SFR}
\end{equation}
%*************************************************
where $N$ is the number of the selected clouds.
By performing these calculations for the selected dense clouds ($n_\mathrm{gas} > 1000$ $\mathrm{cm}^{-3}$) in virial equilibrium we find that SFR$_\mathrm{cc}$ = 0.04 M$_\odot$yr$^{-1}$. The actual SFR in our simulation at 300 Myr is indeed 0.04 M$_\odot$yr$^{-1}$ (see also the bottom panel of Fig. \ref{fig:FoF_cold_mass_evolution} for a quick estimate of the actual SFR at $t$ = 300 Myr). The fact that the SFR$_\mathrm{cc}$ calculated from the free-fall time of the selected clouds is equal to the actual SFR of the simulation indicates that the star formation in our simulation comes directly and exclusively from high-density clouds close to virial equilibrium.
Alternatively, we identified all the star particles at $t$ = 300 Myr that were created during the last 1 Myr and identified their progenitor gas particles at the previous snapshot of our simulation, i.e. at $t$ = 299 Myr. In the same snapshot, we identified the cold dense clouds with $n_\mathrm{gas} > 1000$ $\mathrm{cm}^{-3}$. These are the dense regions of the cold clouds, as we explained above. This analysis showed that these dense cores include 57 per cent of the gas particles that will turn into stars at $t$ = 300 Myr. By performing the same analysis at $t$ = 299 Myr for the cold gas with the lower density threshold of 100 $\mathrm{cm}^{-3}$, we find that around 94 per cent of the gas particles that will crate stars at the next 1 Myr belong to the identified clouds. It is evident that in our simulation star formation takes place in the cold clouds and more than half of the stars are located in the dense regions of these clouds. To gain a statistical overview, we performed the same analysis for five snapshots around the time of the merger (170 Myr) and for five snapshots around the more quiescent stage of the system (300 Myr). Our analysis showed that on average around 60 per cent of the gas particles that create stars within 1 Myr are located in the denser regions of the identified cold clouds.

%%%%%%%%%%%%%%%%%%%%%%%%%%%%%%%%%%%%%%%%%%%%%%%%%%
\begin{figure} 
\includegraphics[width=1\columnwidth]{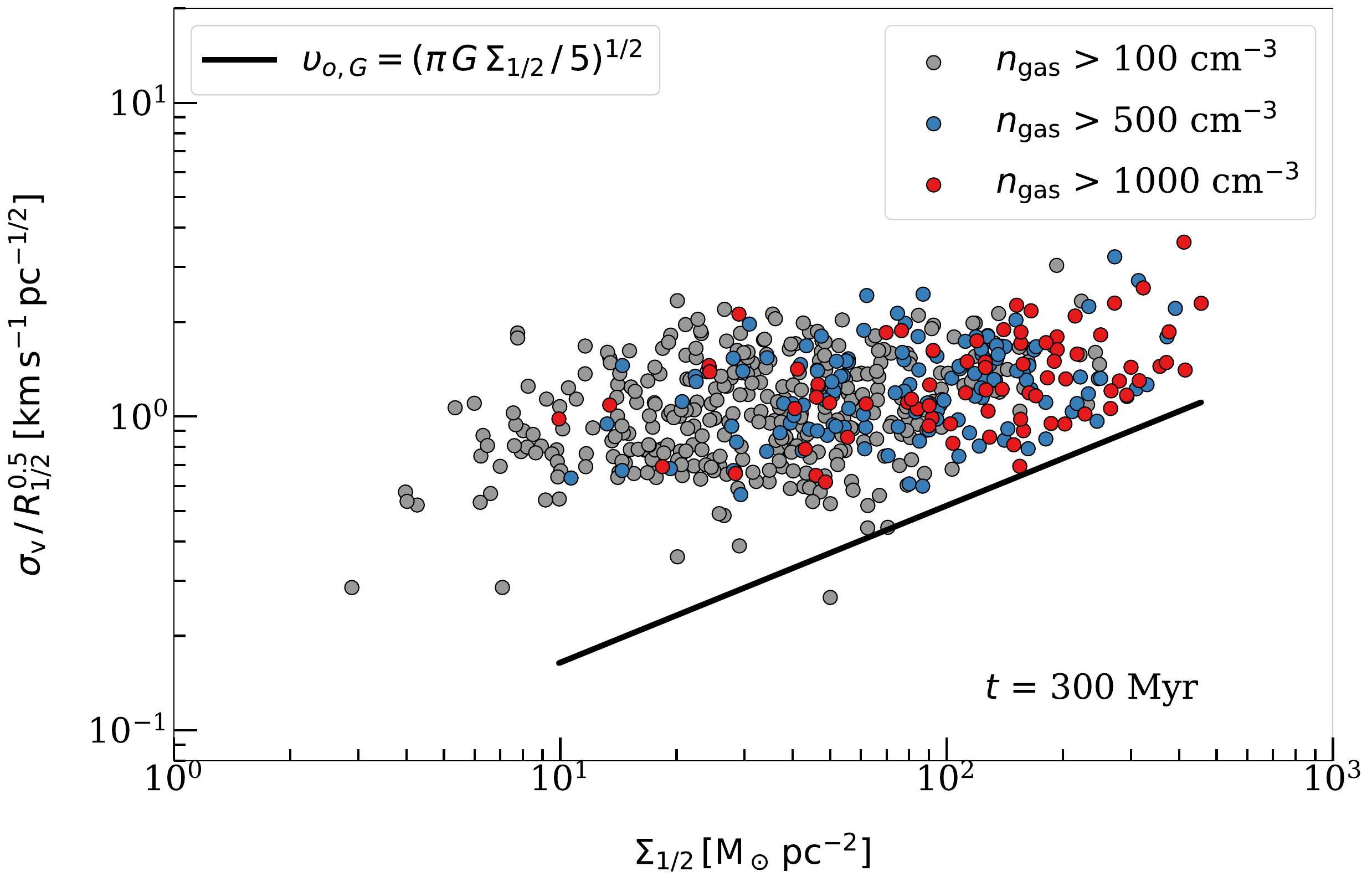}
\caption{The scaling coefficient $\upsilon_\mathrm{o}=\sigma_\mathrm{v} / R^{0.5}_{1/2}$ as a function of the effective surface density $\Sigma_{1/2}$ at 300 Myr for different gas density thresholds: $n_\mathrm{gas} > 100$ $\mathrm{cm}^{-3}$ (grey), $n_\mathrm{gas} > 500$ $\mathrm{cm}^{-3}$ (blue) and $n_\mathrm{gas} > 1000$ $\mathrm{cm}^{-3}$ (red). The average line-of-sight velocity dispersion is $\sigma_\mathrm{v} = \sqrt{\sigma^2_\mathrm{x} + \sigma^2_\mathrm{y} + \sigma^2_\mathrm{z}} / \sqrt{3}$ and $R_{1/2}$ is the projected half-mass-radius of the identified cold clouds. The solid line shows the loci of gravitationally bound clouds.
}
\label{fig:scaling_plot_2_heyer2}
\end{figure}
%%%%%%%%%%%%%%%%%%%%%%%%%%%%%%%%%%%%%%%%%%%%%%%%%%
While the original third Larson relation indicates that all clouds have similar surface densities, $\Sigma_0$, this is not the case in many observed systems and in our simulations (see Sec. \ref{sec:Introduction}). From Eq. \ref{eq:vir}, a constant surface density would imply that the scaling coefficient $\upsilon_\mathrm{o,G}=\sigma_\mathrm{v} / R^{0.5}_{1/2} = (\pi G \Sigma_{1/2} / 5)^{1/2}$, which relates the velocity dispersion and cloud size in virial equilibrium, is constant with a value of $(\pi G \Sigma_{0}/ 5)^{1/2}$. In Fig. \ref{fig:scaling_plot_2_heyer2} we show the scaling coefficient  $\upsilon_\mathrm{o}=\sigma_\mathrm{v} / R^{0.5}_{1/2}$ for the simulated clouds at $t$ = 300 Myr at the same density thresholds as Fig. \ref{fig:scaling_plot_2_evans}. The simulated clouds cover more than one order of magnitude in surface density, similar to observational estimates \citep[e.g.][]{Heyer2009,2013ApJ...779...46H,Duarte-Cabral2021} and they also do not seem to follow the equilibrium scaling in general. In particular low surface density clouds deviate significantly from the equilibrium expectation indicated in Fig. \ref{fig:scaling_plot_2_heyer2}. 

\section{The cold cloud life cycle}\label{sec:TheColdCloudlifecycle}
%%%%%%%%%%%%%%%%%%%%%%%%%%%%%%%%%%%%%%%%%%%%%%%%%%
\begin{figure} 
\includegraphics[width=1\columnwidth]{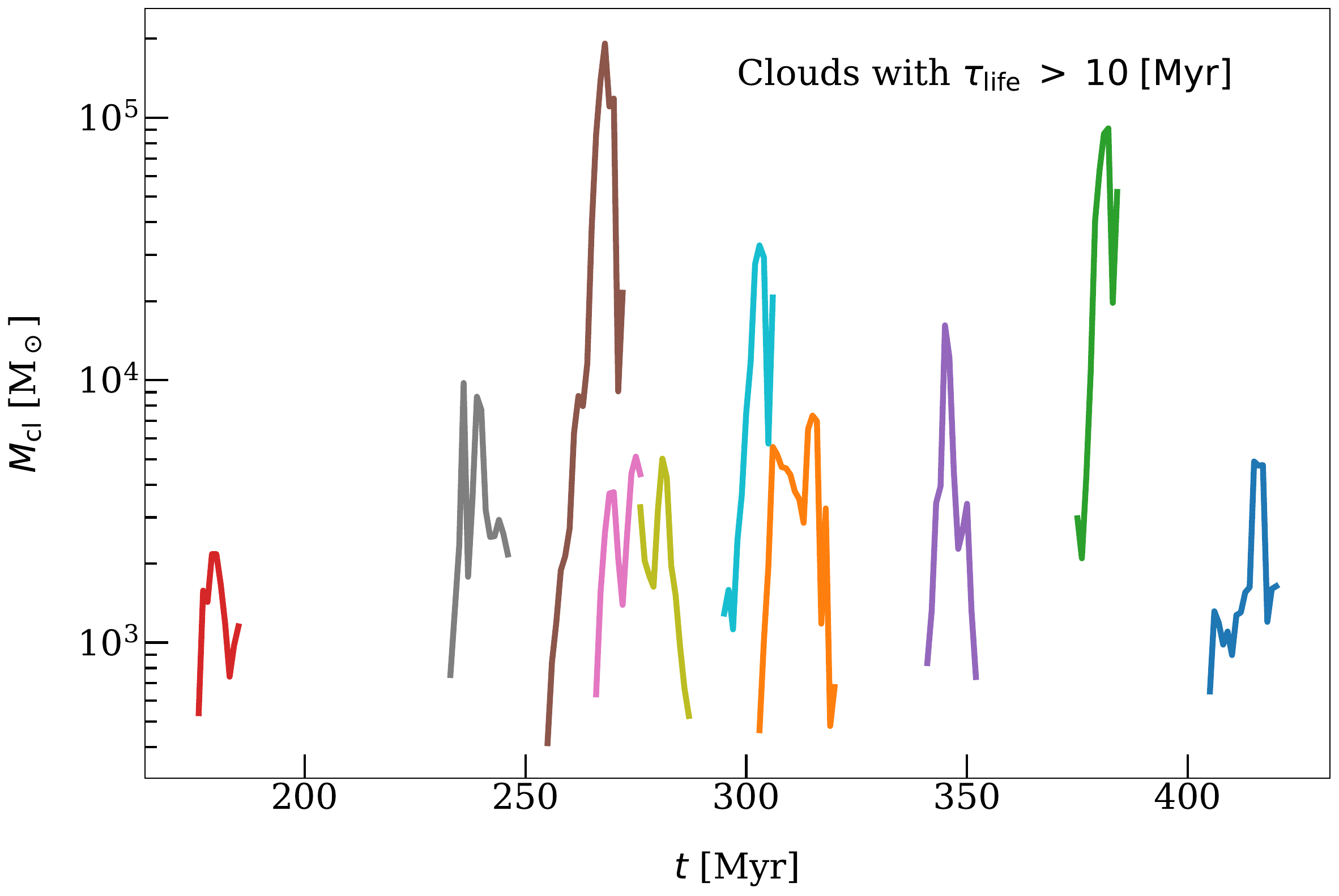}
\caption{The mass evolution for 10 cold clouds with lifetimes larger than 10 Myr, randomly selected from our sample of traced clouds. In their growth phase the clouds gain enough mass for gravitational collapse and the beginning of star formation. In their destruction phase the energy and momentum input from newly formed stars lead to the dispersal of the clouds. Both the growth and destruction phases of the clouds are visible on the selected mass traces.
}
\label{fig:tracer_mass_evolution}
\end{figure}
%%%%%%%%%%%%%%%%%%%%%%%%%%%%%%%%%%%%%%%%%%%%%%%%%%
We have used the method described in Sec. \ref{sec:cloud-identification} to follow the tracks of 31222 individual clouds starting shortly before the peak starburst (150 Myr) until the end of our simulation (457 Myr). The cold cloud life cycle ends as soon as the cloud cannot be detected anymore. 61 per cent of the cold clouds (i.e 19123 traced cold clouds) have lifetimes longer than 1 Myr. 91 per cent of the cold clouds that are detected in two consecutive snapshots (1 Myr) have lifetimes shorter than 10 Myr and no cloud lives longer than $\sim 30\,\mathrm{Myr}$. As an example, we show the mass evolution of 10 randomly chosen clouds with lifetimes longer than 10 Myr is shown in Fig. \ref{fig:tracer_mass_evolution}. The cloud mass at first identification can be relatively high as the gas has to cross the density threshold to be detected. Clouds typically grow in mass after they are first detected (e.g. the first cloud on the left in Fig. \ref{fig:tracer_mass_evolution}). Thereafter, the clouds lose mass due to star formation and stellar feedback. Individual cold clouds, however, can have more complex histories. Once the clouds are not detected any more their life-cycle is complete.

%%%%%%%%%%%%%%%%%%%%%%%%%%%%%%%%%%%%%%%%%%%%%%%%%%
\begin{figure*} 
\includegraphics[width=\textwidth]{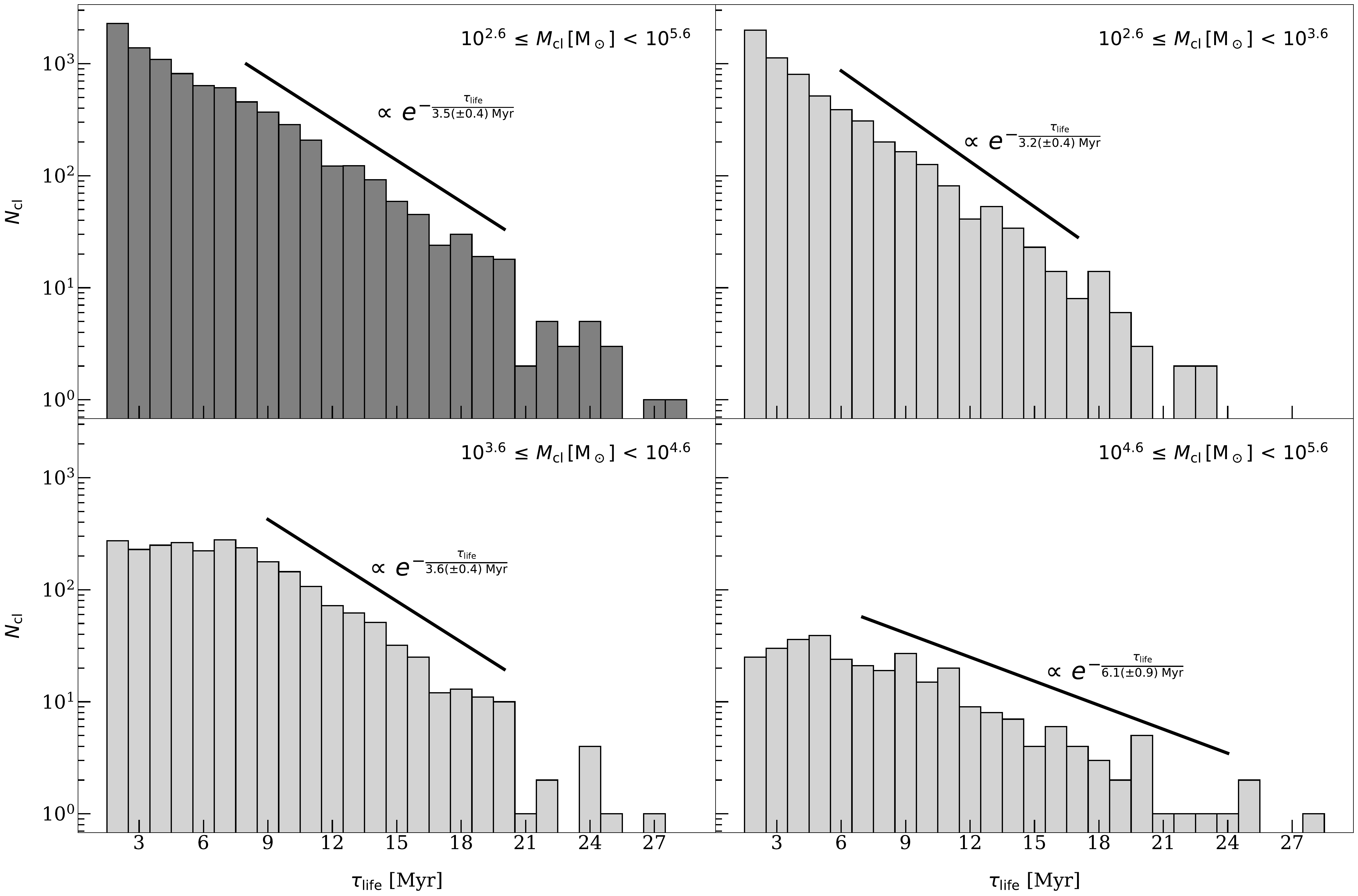}
\caption{{\it Top left}: Histogram of the lifetimes of the total number of cold clouds that survived for at least 3 Myr traced in our simulation, i.e. in the mass range $10^{2.6} \, \leq \, M_\mathrm{cl} \, [\mathrm{M}_\odot] \, < \, 10^{5.6}$. The e-folding timescale is $3.5 (\pm 0.4)$ Myr. {\it Top right}: Histogram of the cold clouds traced in the mass range $10^{2.6} \, \leq \, M_\mathrm{cl} \, [\mathrm{M}_\odot] \, < \, 10^{3.6}$ with e-folding timescale $3.2 (\pm 0.4)$ Myr. {\it Bottom left}: Histogram of the cold clouds traced in the mass range $10^{3.6} \, \leq \, M_\mathrm{cl} \, [\mathrm{M}_\odot] \, < \, 10^{4.6}$ with e-folding timescale $3.6 (\pm 0.4)$ Myr. {\it Bottom right}: Histogram of the cold clouds traced in the mass range $10^{4.6} \, \leq \, M_\mathrm{cl} \, [\mathrm{M}_\odot] \, < \, 10^{5.6}$ with e-folding timescale $6.1 (\pm 0.9)\, \mathrm{Myr}$.}
\label{fig:tracer_lifespan_histogram}
\end{figure*} 
%%%%%%%%%%%%%%%%%%%%%%%%%%%%%%%%%%%%%%%%%%%%%%%%%%

In Fig. \ref{fig:tracer_lifespan_histogram}, we show the lifetime distribution for all traced cold clouds that survive for at least 3 Myr (top left panel) as well as the distribution in three different mass bins covering the mass range of identified clouds, i.e. in mass bins of $10^{2.6} \,\mathrm{M}_\odot \lesssim M_\mathrm{cl} < 10^{3.6}\,\mathrm{M}_\odot$ (top right panel), $10^{3.6} \,\mathrm{M}_\odot \lesssim M_\mathrm{cl} < 10^{4.6}\,\mathrm{M}_\odot$ (bottom left panel), and $10^{4.6} \mathrm{M}_\odot \lesssim M_\mathrm{cl} < 10^{5.6}\,\mathrm{M}_\odot$ (bottom right panel). In general, the cold clouds follow an exponential lifetime distribution with an e-folding timescale of $\sim 3.5 (\pm 0.4)$ Myr. In the mass range of $10^{2.6} \,\mathrm{M}_\odot \lesssim M_\mathrm{cl} < 10^{3.6}\,\mathrm{M}_\odot$ the e-folding timescale is $\sim 3.2 (\pm 0.4)\,\mathrm{Myr}$, in the mass range $10^{3.6} \,\mathrm{M}_\odot \lesssim M_\mathrm{cl} < 10^{4.6}\,\mathrm{M}_\odot$ the e-folding timescale is $\sim 3.6 (\pm 0.4)$ Myr and on the high mass end of $10^{4.6} \mathrm{M}_\odot \lesssim M_\mathrm{cl} < 10^{5.6}\,\mathrm{M}_\odot$ the e-folding timescale is $\sim 6.1 (\pm 0.9)$ Myr. There is a moderate trend for lower- and medium-mass clouds to have a shorter e-folding timescale whereas more massive clouds have slightly longer typical lifetimes.

%%%%%%%%%%%%%%%%%%%%%%%%%%%%%%%%%%%%%%%%%%%%%%%%%%
\begin{figure} 
\includegraphics[width=1\columnwidth]{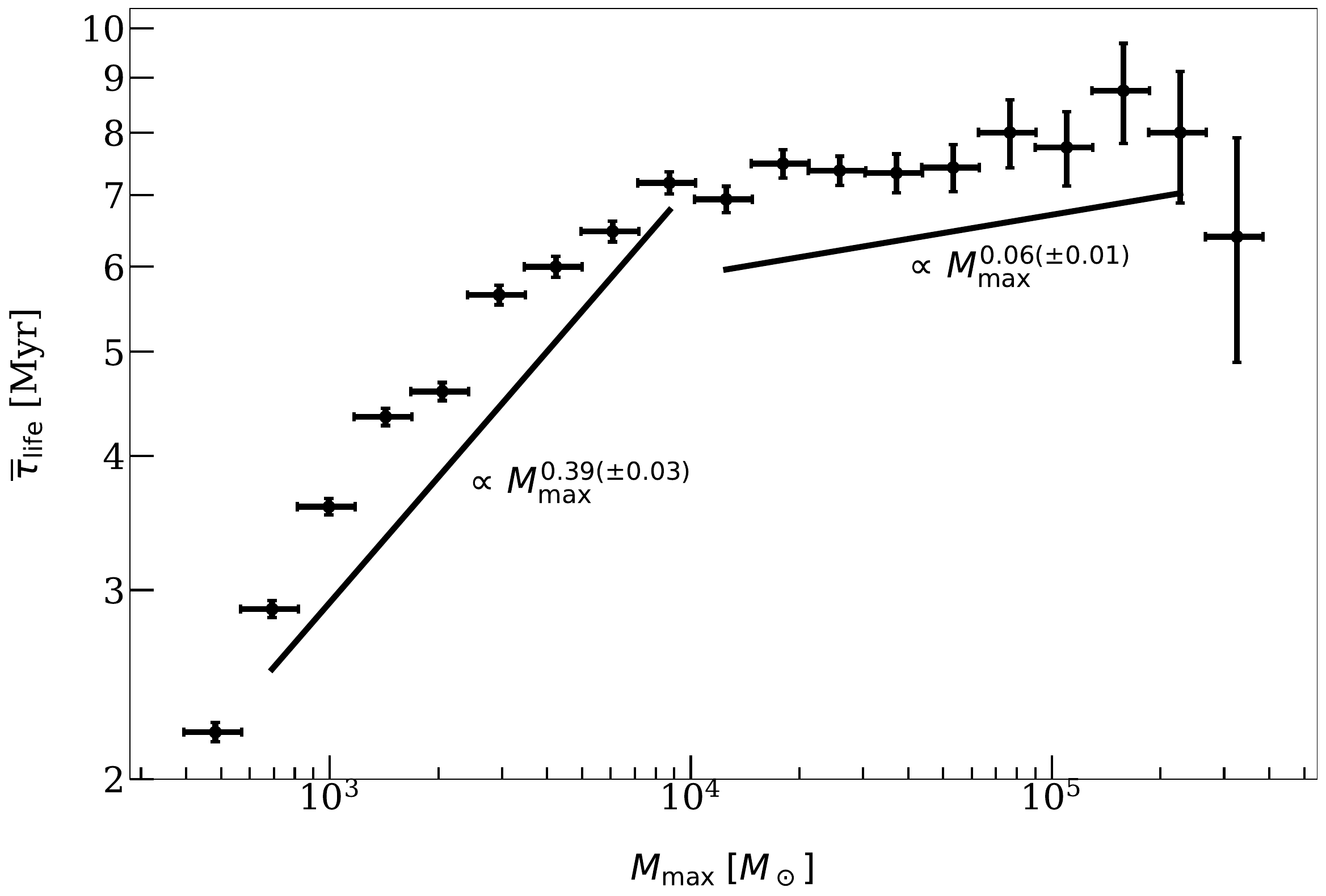}
\caption{Peak mass of the traced clouds during their lifetime $M_\mathrm{max}$ binned in 20 logarithmic bins vs. the average lifetime of the clouds $\overline{\tau}_\mathrm{life}$ in each bin. Each scatter point represents each bin. The error bars in the $x$-axis represent the size of each bin, while in the $y$-axis they stand for the standard error of the average value of the lifetimes in each bin. The solid lines indicate the indexes of the power-law distributions that represent our sample. For clouds with peak masses $M_\mathrm{max} \, < \, 10^4 \, \mathrm{M_\odot}$ we have $\overline{\tau}_\mathrm{life} \propto M_\mathrm{max}^{0.39(\pm 0.03)}$. For clouds with peak sizes $M_\mathrm{max} \, > \, 10^4 \, \mathrm{M_\odot}$ we have $\overline{\tau}_\mathrm{life} \propto M_\mathrm{max}^{0.06(\pm 0.01)}$. The trend is that more massive clouds have longer lifetimes and for clouds with peak masses larger than $\mathrm{10^4 \, M_\odot}$ the lifetimes tend to stabilise around 8 Myr.
}
\label{fig:tracer_lifetime_peakmass}
\end{figure}
%%%%%%%%%%%%%%%%%%%%%%%%%%%%%%%%%%%%%%%%%%%%%%%%%%

To quantify this and check if the trend holds, we make a finer binning of the cold cloud masses in our sample and we calculate the average lifetime of the clouds in each mass bin $\overline{\tau}_\mathrm{life}$. As a representative value of the masses, we choose the peak mass that each cloud reaches during its lifetime $M_\mathrm{max}$\footnote{The peak mass of each cloud results from an interplay of the cloud's environment and the star formation process within the cloud. This mass is the only representative mass for which we can make meaningful assessments. The birth mass, for example, would be an ill-defined variable, as it is nearly the same for all clouds (see e.g. Fig. \ref{fig:tracer_mass_evolution}), due to the numerical realisation of clouds in our simulation and the way we identify clouds with the FoF algorithm in our analysis. An initially more massive cloud at a given time step $t = t_0$ would most likely be identified by FoF at an earlier time step $t_0 - 1$ with a lesser amount of connected gas particles.}. The results of our analysis are shown in Fig. \ref{fig:tracer_lifetime_peakmass} and for the binning, we use 20 logarithmic bins. 
Approximating the slope of the power-law distributions that fit the low mass range in our sample (i.e. clouds with peak masses $M_\mathrm{max} \, < \, 10^4 \, \mathrm{M_\odot}$), we find that there is a power-law dependency between the peak mass $M_\mathrm{max}$ and the average lifetime $\overline{\tau}_\mathrm{life}$ such that $\overline{\tau}_\mathrm{life} \propto M_\mathrm{max}^{0.39(\pm0.03)}$. This dependency indicates that the cold clouds at the lower mass end of our sample have longer lifetimes the more massive they are. 
For the higher mass range in our sample (i.e. clouds with peak sizes $M_\mathrm{max} \, > \, 10^4 \, \mathrm{M_\odot}$), we have that $\overline{\tau}_\mathrm{life} \propto M_\mathrm{max}^{0.06(\pm0.01)}$\footnote{Splitting the data set into two mass ranges is motivated by the analysis in Fig. \ref{fig:tracer_lifespan_histogram}. The behaviour of the lifetimes of the clouds changes qualitatively in the upper mass bin. Therefore we analyse this mass bin separately from the total sample.}. This indicates that the more massive cold clouds in our sample only have slight differences between their average lifetimes which tend to stabilise around 8 Myr. 
This might be explained by a maximum lifetime for the most massive clouds which is regulated by the onset of SNe. We analyse this possibility further down. We also find no evidence for a change of the cloud lifetime distribution depending on the state of the merger (peak starburst vs. later evolution, see Fig. \ref{tracer_lifespan_comparison} in the Appendix). This indicates that the cloud lifetimes are not affected by the environment. 

%%%%%%%%%%%%%%%%%%%%%%%%%%%%%%%%%%%%%%%%%%%%%%%%%%
\begin{figure} 
\includegraphics[width=1\columnwidth]{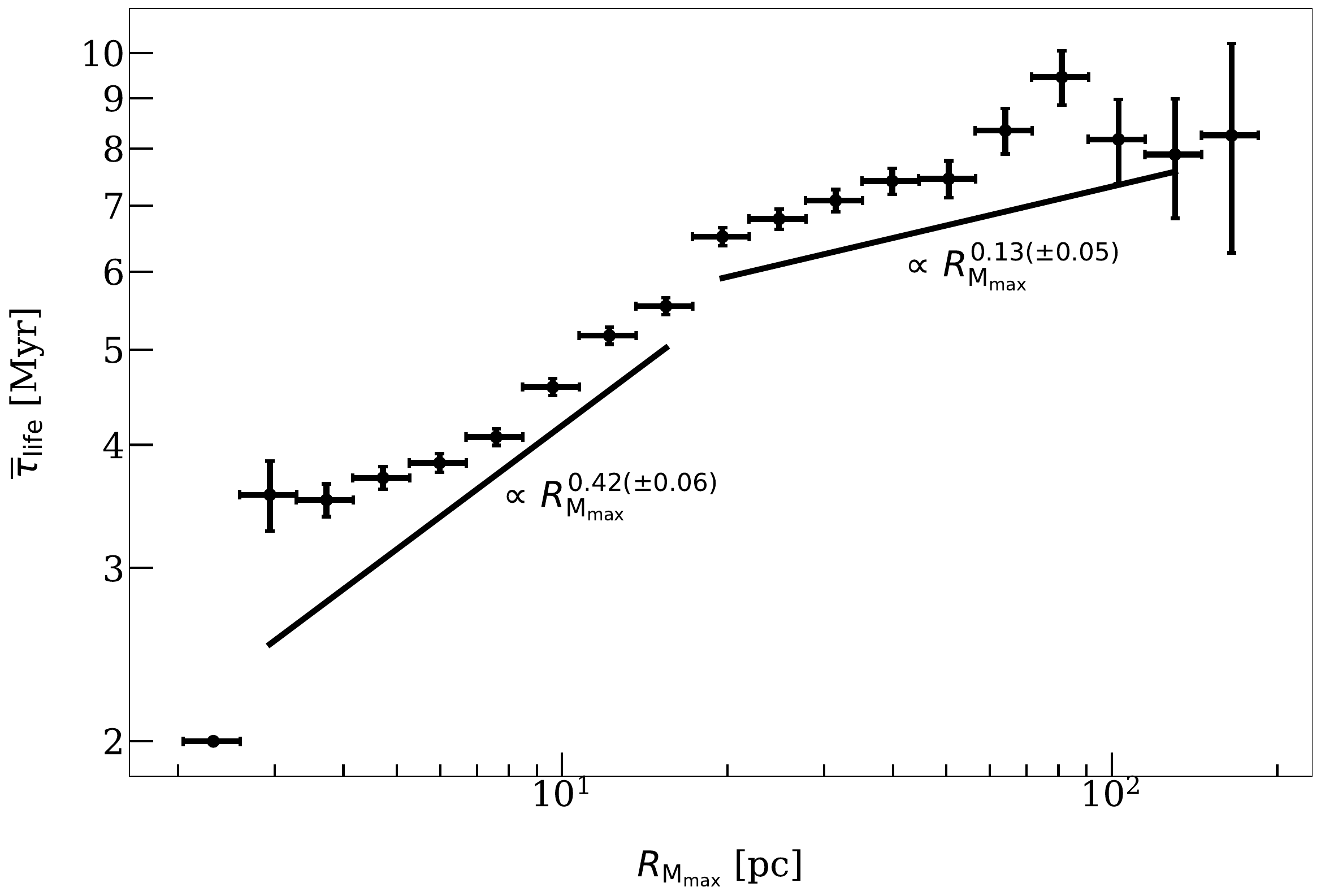}
\caption{Size at peak mass $R_\mathrm{M_{max}}$ of the traced clouds during their lifetime binned in 20 logarithmic bins vs. the average lifetime of the clouds $\overline{\tau}_\mathrm{life}$ in each bin. Each scatter point represents each bin. The error bars in the $x$-axis indicate the size of the bin and in the $y$-axis they represent the standard error of the average value of the lifetimes in each bin. The solid lines indicate the indexes of the power-law distributions that represent our sample. For clouds with sizes at their peak mass $R_\mathrm{M_{max}} < 20$ pc we have $\overline{\tau}_\mathrm{life} \propto R_\mathrm{M_{max}}^{0.42(\pm 0.06)}$. For clouds with sizes at their peak mass $R_\mathrm{M_{max}} > 20\, \mathrm{pc}$ we have $\overline{\tau}_\mathrm{life} \propto R_\mathrm{M_{max}}^{0.13(\pm 0.05)}$. The trend is that smaller clouds have shorter lifetimes whereas for clouds with sizes larger than 20 pc the lifetimes tend to stabilise around 8 Myr.}
\label{fig:tracer_lifetime_size_at_peak_mass}
\end{figure}
 %%%%%%%%%%%%%%%%%%%%%%%%%%%%%%%%%%%%%%%%%%%%%%%%%%
 
Similar as in Fig. \ref{fig:tracer_lifetime_peakmass}, we perform a logarithmic binning of the sizes of the traced cold clouds. As a measure of size, we use the maximum distance that a SPH gas particle belonging to the cloud has from the centre of mass of the cloud. As a representative size value for the binning we chose the size each cloud has when it reaches its peak mass $R_\mathrm{M_{max}}$ and we calculate the average lifetime of the clouds in each size bin $\overline{\tau}_\mathrm{life}$ (we work with 20 logarithmic bins as in Fig. \ref{fig:tracer_lifetime_peakmass}). The results are shown in Fig. \ref{fig:tracer_lifetime_size_at_peak_mass}. Each scatter point represents each size bin. The error in the $x$-axis is represented by the horizontal error bars which are drawn based on the width of each logarithmic bin. For the error in the $y$-axis we use the standard error of the average value of the lifetimes $\overline{\tau}_\mathrm{life}$ in each bin which is visualised by the vertical error bars. As in Fig. \ref{fig:tracer_lifetime_peakmass}, we approximate the slope of the power-law distributions that fit the smaller size range in our sample (i.e. clouds with sizes at peak mass $R_\mathrm{M_{max}} < 20\,\mathrm{pc}$) and we find a power-law dependency between the size at peak mass $R_\mathrm{M_{max}}$ and the average lifetime $\overline{\tau}_\mathrm{life}$, i.e. 
$\overline{\tau}_\mathrm{life} \propto R_\mathrm{M_{max}}^{0.42(\pm0.06)}$. This dependency indicates that in the small-size-end of our sample the larger the clouds the longer they live.
For the larger cold clouds in our sample (that is clouds with sizes at peak mass $R_\mathrm{M_{max}} > 20$ pc) we find a shallower $\overline{\tau}_\mathrm{life} \propto R_\mathrm{M_{max}}^{0.13(\pm0.05)}$.

%%%%%%%%%%%%%%%%%%%%%%%%%%%%%%%%%%%%%%%%%%%%%%%%%%
\begin{figure} 
\includegraphics[width=1\columnwidth]{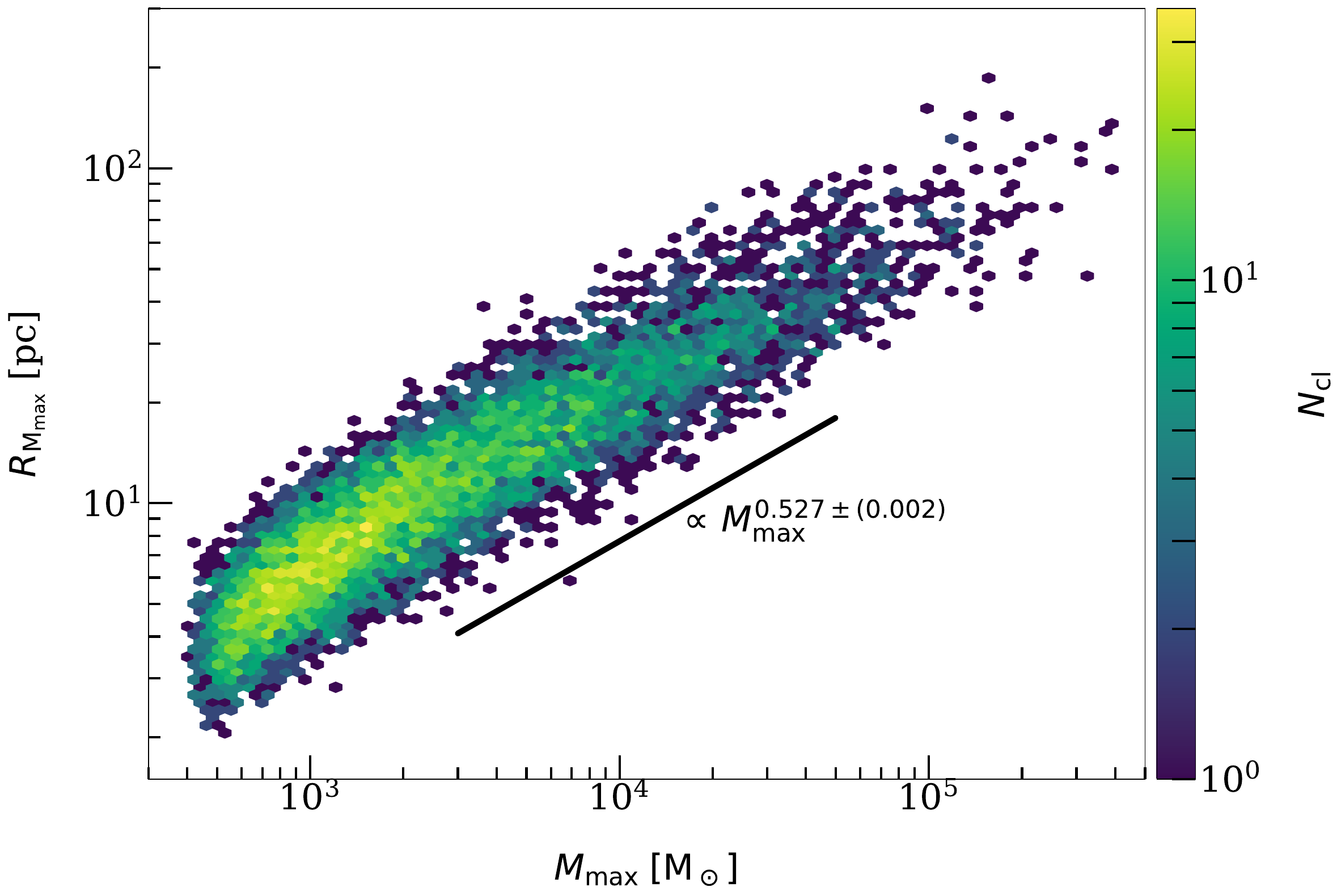}
\caption{Peak mass $M_\mathrm{max}$ of each cold cloud during its lifetime vs. its size at that given point of its life $R_\mathrm{M_{max}}$. The colour indicates the number of clouds in the respective region. The best fitting mass size relation, $R_\mathrm{M_{max}} \propto M_\mathrm{max}^{0.527(\pm0.002)}$, is indicated. 
}
\label{fig:tracer_size_at_peak_mass}
\end{figure}
%%%%%%%%%%%%%%%%%%%%%%%%%%%%%%%%%%%%%%%%%%%%%%%%%% 

In Fig. \ref{fig:tracer_size_at_peak_mass} we plot the peak mass, $M_\mathrm{max}$, against the corresponding size, $R_\mathrm{M_{max}}$, for each traced cold cloud. The color indicates the number of clouds in each pixel. The cloud peak size scales very well with the square root of its mass, following the relation $R_\mathrm{M_{max}} \propto M_\mathrm{max}^{0.527(\pm0.002)}$. There is no obvious change in the mass-size relation at the mass and size scales ($\sim 10^4 M_\odot, \sim 20 pc$) at which the lifetime slopes are changing (see Figs. \ref{fig:tracer_lifetime_peakmass} and \ref{fig:tracer_lifetime_size_at_peak_mass}).

%%%%%%%%%%%%%%%%%%%%%%%%%%%%%%%%%%%%%%%%%%%%%%%%%%
\begin{figure} 
\includegraphics[width=1\columnwidth]{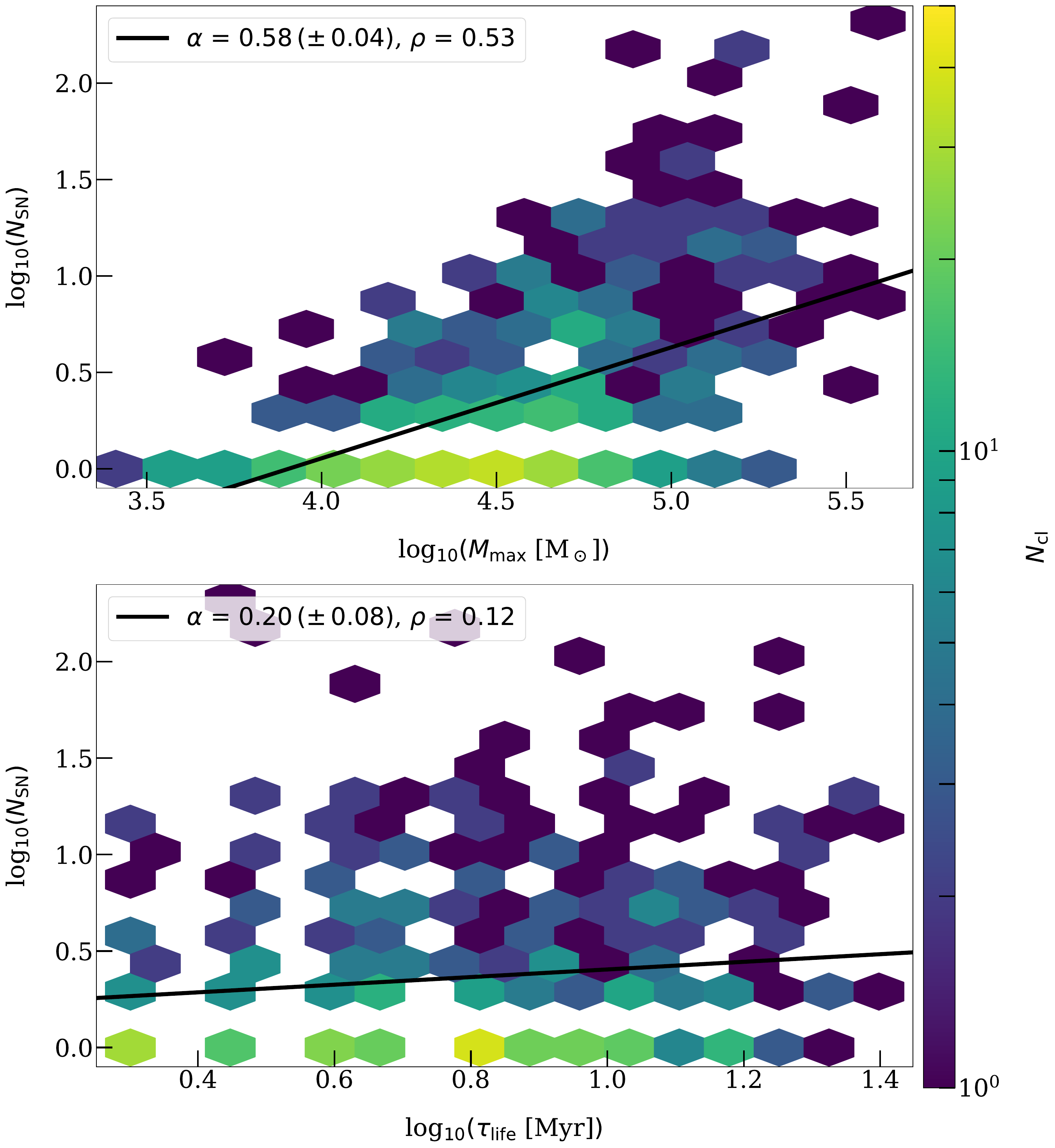}
\caption{Number of SNe, $N_\mathrm{SN}$, which explode within a cold cloud as a function of the cloud peak mass, $M_\mathrm{max}$, and lifetime, $\tau_\mathrm{life}$. We indicate the slope, $\alpha$, of power-law fits as well as the Pearson correlation coefficients, $\rho$.}
\label{fig:SN}
\end{figure}
 %%%%%%%%%%%%%%%%%%%%%%%%%%%%%%%%%%%%%%%%%%%%%%%%%%

We inspect how many SNe explode in each cloud and correlate this number with the clouds peak mass and lifetime in Fig. \ref{fig:SN}. We find a moderate correlation ($\rho = 0.53$) for the $M_\mathrm{max}-N_\mathrm{SN}$ relation with a sub-linear power-law slope $\alpha = 0.58 (\pm 0.04)$. The $\tau_\mathrm{life}-N_\mathrm{SN}$ relation is not correlated ($\rho = 0.12$) and follows a flat distribution with $\alpha = 0.20 (\pm 0.08)$. The majority of clouds withstand not more than 1-2 SN explosions before they disperse (see accumulation of $N_\mathrm{clouds}$ at the bottom of both panels in Fig. \ref{fig:SN}). The moderate correlation between $N_\mathrm{SN}$ and $M_\mathrm{max}$ with the sub-linear power-law slope indicates that more massive clouds can host more SNe. This is understandable with the fact that the SNe tend to be very closely clustered in space and time. The more massive clouds host a bigger stellar population. Those stars, however, form at approximately the same time and multiple SNe can therefore explode in very short succession while the cloud is being dispersed. This finding is supported by the fact that there is no correlation between $\tau_\mathrm{life}$ and $N_\mathrm{SN}$. The average time between the first and the last SNe in a cloud for clouds with at least 2 SNe occurring in them is $dt_\mathrm{SN}^\mathrm{avg} = 2.26\,\mathrm{Myr}$, and the median time difference is $dt_\mathrm{SN}^\mathrm{med} = 1\,\mathrm{Myr}$. In conclusion, the lifetime of the clouds is regulated by the onset of SNe. Multiple SNe, closely clustered in space and time, can occur within a cloud. More massive clouds have a slight tendency to host more SNe, because they also host a bigger stellar population.

\section{Discussion}\label{sec:Discussion}
As we mentioned in Sec. \ref{sec:Introduction} the three Larson relations are not independent but algebraically bound, i.e. if any two are valid the third one is valid, too. In Sec. \ref{sec:ColdCloudScalingRelations}, using this bond and equating the mass of the clouds to the virial mass we derived the cohesive Eq. \ref{eq:vir} which we use in Fig. \ref{fig:scaling_plot_2} to check the virialisation of the identified cold clouds. 
Observed molecular clouds in the Milky Way disk typically fall in between a virial parameter $\alpha_\mathrm{vir}$ of 1 and 3 \citep{Heyer2009, Heyer2015} and the trend is similar for molecular clouds in nearby galaxies \citep{Bolatto2008, Leroy2015}. However, in the Milky Way centre the values of $\alpha_\mathrm{vir}$ are between 3 and 10 \citep{Oka2001}.
Our simulated cold clouds have a virial parameter of $\alpha_\mathrm{vir} \sim$ 10 (see Fig. \ref{fig:scaling_plot_2}) and they follow the first Larson relation, but not Eq. \ref{eq:vir}. This means that they still follow a size-velocity dispersion relationship, but one that is independent of the surface density introduced in Eq. \ref{eq:vir}.
By examining also Figs. \ref{fig:scaling_plot_2_evans} and \ref{fig:scaling_plot_2_heyer2}, which provide a succinct illustration of the Larson scaling relations, we can infer that the complete sample of our simulated clouds does not satisfy the condition of virial equilibrium. However, when we increase the gas density threshold to 1000 $\mathrm{cm}^{-3}$ (red scatter points in Figs. \ref{fig:scaling_plot_2_evans} and \ref{fig:scaling_plot_2_heyer2}), we notice that only a fraction of dense substructures (13 per cent, see Fig. \ref{fig:scaling_plot_2_evans}) are in approximate virial equilibrium, indicating that denser regions within our identified cold clouds are gravitationally bound. This finding suggests that the clouds contain dense virialised cores surrounded by diffuse, lower density cold gas that forms a sort of cocoon around the cores. Our result is in agreement with the observational results of \citet{Evans2021}, who confirmed -- among others -- that the value of the virial parameter varies in general with the surface density and the mass of the bound structures. 
Calculating the free-fall time of the dense cores that are in approximate virial equilibrium (see Eq. \ref{eq:free-fall time}) and from it the SFR (see Eq. \ref{eq:SFR}), we find that this SFR is equal to the actual SFR of our simulation. This suggests that the dense cores within the cold clouds are the sites of star formation in our system.

Previous studies in the literature suggest that clouds similar to ours, i.e. clouds which are not gravitationally bound but follow the linewidth-size relation exist if they are bound by external pressure \citep{Elmegreen1989, Bertoldi1992, McKee2003, Field2011, McKee2010, Hennebelle2012, Heyer2015}. 
In more detail, \citet{Bertoldi1992} showed that in giant molecular clouds (GMCs), smaller substructures with high virial parameter values ($\alpha_\mathrm{vir} \gg 1$) are bound by the external pressure of the surrounding interstellar medium. Some of these clumps contain cores which are in gravitational virial equilibrium, but these cores contain only a small fraction of the mass \citep[][]{Myers1985}. 
Another idea explored in the literature is that such clouds with high virial parameters can still follow the size-velocity dispersion relation if they have short lifetimes compared to their dynamical timescales \citep{Field2011, Heyer2015}. Most of our simulated clouds live for a couple of Myr with an average e-folding timescale of 3.5 Myr (see Fig. \ref{fig:tracer_lifespan_histogram}), so this theory could describe the regime of our clouds and the reason of the discrepancy between non-virialisation and adherence to the first Larson relation.

The variance between $\upsilon_\mathrm{o}=\sigma_\mathrm{v} / R^{0.5}_{1/2}$ with $\Sigma^{0.5}_{1/2}$ that we encounter in Fig. \ref{fig:scaling_plot_2_heyer2} agrees with the results of \citet{Heyer2009} and suggests a deviation from strict adherence to the Larson relations. \citet{Heyer2009} attributes the displacement $\upsilon_\mathrm{o}=\sigma_\mathrm{v} / R^{0.5}_{1/2}$ with $\Sigma^{0.5}_{1/2}$ to the systematic underestimation of the surface densities of the clouds due to observational constraints (more specifically they refer to their values as ``lower limits"). \citet{Dobbs2011} propose that the displacement is because the clouds have too large $\sigma_\mathrm{v}$ to satisfy virial equilibrium and they resort to arguments of external pressure confinement of the gravitationally unbound clouds.
\citet{Ballesteros-Paredes2011} argue that the displacement of $\upsilon_\mathrm{o}=\sigma_\mathrm{v} / R^{0.5}_{1/2}$ with $\Sigma^{0.5}_{1/2}$ is not due to turbulence, but is rather caused by internal motions of the gas in the clouds undergoing gravitational collapse as a whole, while experiencing individual gravitational collapse in their cores. 

Following the identification of the cold clouds in our simulation and the detailed study of their physical properties, we proceeded to track their life cycles to construct a comprehensive picture of their lifespan. Cold clouds are destroyed by stellar feedback thus having lifetimes in the range of 10-30 Myr and within $\sim$ 5 Myr after the emergence of very massive stars \citep[see e.g.][]{Kruijssen2019,Chevance2020, Chevance2022}. 
The simulations of Milky Way-like systems by \citet{Jeffreson2023} put the average lifetime of molecular clouds at $\sim$ 25 Myr independent from the typical chemical lifetime of H$_2$, which they found to be only 4 Myr and independent from the lifetime of the host cloud. 
Our simulated clouds live up to 28 Myr with most clouds in our sample being short-lived and surviving for only a couple Myr and $\sim$ 8 Myr after the emergence of very massive stars. This difference could be attributed to the more massive galaxies that \citet{Jeffreson2023} simulate compared to our simulation of a dwarf merger system.

In \citet{Jeffreson2021} it is supported that the distribution of the number of clouds has exponential form provided that the rate of cloud formation and the lifespan of molecular clouds are time-invariant (e.g. if the total SFR remains relatively stable over time and the population of clouds does not change significantly over time, then this hypothesis is valid).
Indeed, the distribution of the number of clouds in our sample has exponential form, with an average e-folding timescale $\overline{\tau}_\mathrm{life}$ = 3.5 Myr (see Fig. \ref{fig:tracer_lifespan_histogram}).
Calculating the e-folding timescales for different mass ranges, we find that on the larger mass end ($10^{4.6} \, \leq \, M_\mathrm{cl} \, [\mathrm{M}_\odot] \, < \, 10^{5.6}$) the average e-folding timescale rises to 6.1 Myr (bottom right panel of Fig. \ref{fig:tracer_lifespan_histogram}). Exploring this correlation between the mass of the cold clouds and their lifetimes further, we find a trend: more massive clouds have longer lifetimes (see Fig. \ref{fig:tracer_lifetime_peakmass}). In more detail, the lifetime $\overline{\tau}_\mathrm{life}$ of clouds with peak masses ($M_\mathrm{max}$) below $10^4 \, \mathrm{M_\odot}$ scales with $M_\mathrm{max}$ as: $\overline{\tau}_\mathrm{life} \propto M_\mathrm{max}^{0.39}$, whereas the lifetime of clouds with $M_\mathrm{max} \, > \, 10^4 \, \mathrm{M_\odot}$ tends to stabilise around 8 Myr. This happens because in our simulation the delay times of SNe are at around 8 Myr, therefore this is the timescale that even the most massive clouds will be disrupted not being able to resist to the explosive feedback of the SNe. 

We perform a similar analysis for the cloud size $R_\mathrm{M_{max}}$, which is the maximum distance of a SPH gas particle of the cloud from the centre of mass of the cloud 
at the time that the cloud has reached its peak mass. We find that the average lifetime for clouds with sizes below 20 pc correlates with the size as: $\overline{\tau}_\mathrm{life} \propto R_\mathrm{M_{max}}^{0.42}$, whereas for clouds with sizes larger than 20 pc the correlation is $\overline{\tau}_\mathrm{life} \propto R_\mathrm{M_{max}}^{0.13}$ and their lifetime stabilises around 8 Myr.
\citet{Jeffreson2021}, using simulations of Milky Way-like isolated disc galaxies, find an anti-correlation between the lifetime and the size $\ell$ of their traced molecular clouds below the disc scale-height: $\tau_\mathrm{life} \propto \ell^{-0.3}$. The disagreement is possible to lie in the differences in the tracing methods we use: our method is three-dimensional and we cannot account for cloud mergers with our tracing algorithm, whereas the \citet{Jeffreson2021} method is two-dimensional and accounting for cloud mergers. Both studies use similar star formation efficiencies $\epsilon_\mathrm{ff}$: \citet{Jeffreson2021} use $\epsilon_\mathrm{ff}$ of 1 per cent and in our simulation $\epsilon_\mathrm{ff}$ = 2 per cent is used. Furthermore, there is a difference in the simulated systems: \citet{Jeffreson2021} simulate more massive and isolated disc galaxies, whereas we simulate a dwarf starburst which creates cold clouds very efficiently after the major merger.

\section{Conclusions}\label{sec:Conclusions}

In this work, we analysed the cold and dense interstellar medium of a low metallicity starburst generated through a high-resolution hydrodynamical simulation of a gas-rich dwarf galaxy merger. The simulations resolve the multi-phase interstellar medium at 4 M$_\odot$ gas phase mass resolution and $\sim$ 0.1 pc spatial resolution. The properties of the ISM are followed with a non-equilibrium chemical heating/cooling network at temperatures below 10$^4$ K. Massive stars are sampled individually and interact with the ISM through a model implementation for the formation of HII regions and mostly resolved supernova explosions. Based on the definitions and methods presented in the paper, we identified and traced the formation and disruption of cold clouds. We summarise our main findings below and emphasise the remaining uncertainties in a consistent comparison with observations: 
\begin{itemize}
 \item The cold ISM of the simulation evolves from a sub-dominant component in the warm gas-dominated isolated galaxy ISM to a highly filamentary dominant and clumpy ISM component in the merger-triggered starburst. We identified hundreds of cold clouds with masses of $10^{2.6}$ – $10^{5.6} M_\odot$ following a power law mass function, $dN/dM \propto M^\alpha$ with $\alpha = -1.78 (\pm 0.08) $ and realistic size distributions. We find no statistical evidence for a change in mass function slope with the post-merger SFR of the system. The mass of the most massive cloud formed at any given time, however, correlates with the SFR of the system.
 \item The identified cold clouds are highly structured. The slope of the CMF remains roughly constant independent of the assumed density threshold. This indicates a self-similar density substructure.
 \item The identified clouds have virial parameters $\alpha_\mathrm{vir} > 1$ and approximately follow the Larson relations. The effective surface densities of the clouds is not constant. It varies from a few to a few hundred solar masses per square parsec. 
 \item All cold clouds are short-lived. Their lifetime distributions are approximately exponential with e-folding timescales in the range from 3.2 Myr for lower-mass clouds up to 6.1 Myr for higher-mass clouds. The cloud lifetimes do not seem to be affected by the environment, i.e. the actual state of the ongoing merger. 
 \item The lifetime of the cold clouds and their peak mass, $M_\mathrm{max}$, follows a power-law distribution $\tau_\mathrm{life} \propto M_\mathrm{max}^\alpha$. Cold clouds with peak masses $M_\mathrm{max}$ up to 10$^4$ M$_\odot$ have a power-law index $\alpha=0.39$, whereas clouds with peak masses higher than 10$^4$ M$_\odot$ have $\alpha=0.06$. The more massive the clouds the longer they live up to \mbox{10$^4$ M$_\odot$} and then their lifetimes saturate at $\sim$ 8 Myr. 
 \item The lifetime of the cold clouds and their size at peak mass, $R_\mathrm{M_{max}}$, follows a power-law distribution $\tau_\mathrm{life} \propto R_\mathrm{M_{max}}^\alpha$. Cold clouds with sizes at peak mass $R_\mathrm{M_{max}}$ up to \mbox{20 pc} have a power-law index $\alpha=0.48$, whereas clouds with sizes at peak mass larger than \mbox{20 pc} have $\alpha=0.13$. The larger the clouds the longer they live up to \mbox{20 pc} and then their lifetimes tend to stabilise around \mbox{8 Myr}.
 \item The size of the cold clouds correlates approximately with the square root of their peak mass: $R_\mathrm{M_{max}} \propto M_\mathrm{max}^{1/2}$.
\end{itemize}
Our analysis indicates that the assumed interstellar medium model forms realistic cold clouds with short lifetimes embedded in a multi-phase interstellar medium. In combination with the realistic star cluster properties forming in these clouds the same simulation \citep{Lahen2020a} it appears that the basic cycle of cold cloud formation and clustered star formation in the galactic ISM is well captured by the \textsc{GRIFFIN} model. The model supports the rapid formation and disruption of star-forming clouds by stellar radiation and supernovae on timescale less than 10 Myr in the low metallicity interstellar medium. 

\section*{Acknowledgements}
We would like to thank Chia-Yu Hu for his help in the realisation of the present work.
T. Naab acknowledges support from the Deutsche Forschungsgemeinschaft (DFG, German Research Foundation) under Germany’s Excellence Strategy - EXC-2094 - 390783311 from the DFG Cluster of Excellence ``ORIGINS''.
J.M. Hislop and P.H. Johansson acknowledge the support by the European Research Council via ERC Consolidator Grant KETJU (no. 818930) and the Academy of Finland through grant 339127. Part of the simulations were carried out at CSC—IT Center for Science Ltd. in Finland. We acknowledge the Gauss Centre for Supercomputing e.V. (www.gauss-centre.eu) for providing computing time on the GCS Supercomputer SUPERMUC-NG at Leibniz Supercomputing Centre (www.lrz.de) under project numbers pn49qi and pn72bu. The data analysis was performed on the FREYA cluster hosted by The Max Planck Computing and Data Facility (MPCDF) in Garching, Germany.
T.-E. Rathjen acknowledges support by the project ``NRW-Cluster for data-intensive radio astronomy: Big Bang to Big Data (B3D)'' funded through the programme ``Profilbildung 2020'', an initiative of the Ministry of Culture and Science of the State of North Rhine-Westphalia. The sole responsibility for the content of this publication lies with the authors.

%The Acknowledgements section is not numbered. Here you can thank helpful
%colleagues, acknowledge funding agencies, telescopes and facilities used etc.
\section*{Data availabiliy}
The data underlying this article will be shared on reasonable request to the corresponding author.


\bibliographystyle{mnras}
\bibliography{bibliography} 

\begin{thebibliography}{}
\makeatletter
\relax
\def\mn@urlcharsother{\let\do\@makeother \do\$\do\&\do\#\do\^\do\_\do\%\do\~}
\def\mn@doi{\begingroup\mn@urlcharsother \@ifnextchar [ {\mn@doi@} {\mn@doi@[]}}
\def\mn@doi@[#1]#2{\def\@tempa{#1}\ifx\@tempa\@empty \href {http://dx.doi.org/#2} {doi:#2}\else \href {http://dx.doi.org/#2} {#1}\fi \endgroup}
\def\mn@eprint#1#2{\mn@eprint@#1:#2::\@nil}
\def\mn@eprint@arXiv#1{\href {http://arxiv.org/abs/#1} {{\tt arXiv:#1}}}
\def\mn@eprint@dblp#1{\href {http://dblp.uni-trier.de/rec/bibtex/#1.xml} {dblp:#1}}
\def\mn@eprint@#1:#2:#3:#4\@nil{\def\@tempa {#1}\def\@tempb {#2}\def\@tempc {#3}\ifx \@tempc \@empty \let \@tempc \@tempb \let \@tempb \@tempa \fi \ifx \@tempb \@empty \def\@tempb {arXiv}\fi \@ifundefined {mn@eprint@\@tempb}{\@tempb:\@tempc}{\expandafter \expandafter \csname mn@eprint@\@tempb\endcsname \expandafter{\@tempc}}}

\bibitem[\protect\citeauthoryear{{Ballesteros-Paredes} \& {Mac Low}}{{Ballesteros-Paredes} \& {Mac Low}}{2002}]{Ballesteros-Paredes2002}
{Ballesteros-Paredes} J.,  {Mac Low} M.-M.,  2002, \mn@doi [\apj] {10.1086/339624}, \href {https://ui.adsabs.harvard.edu/abs/2002ApJ...570..734B} {570, 734}

\bibitem[\protect\citeauthoryear{{Ballesteros-Paredes}, {Hartmann}, {V{\'a}zquez-Semadeni}, {Heitsch}  \& {Zamora-Avil{\'e}s}}{{Ballesteros-Paredes} et~al.}{2011}]{Ballesteros-Paredes2011}
{Ballesteros-Paredes} J.,  {Hartmann} L.~W.,  {V{\'a}zquez-Semadeni} E.,  {Heitsch} F.,   {Zamora-Avil{\'e}s} M.~A.,  2011, \mn@doi [\mnras] {10.1111/j.1365-2966.2010.17657.x}, \href {https://ui.adsabs.harvard.edu/abs/2011MNRAS.411...65B} {411, 65}

\bibitem[\protect\citeauthoryear{{Benincasa}, {Tasker}, {Pudritz}  \& {Wadsley}}{{Benincasa} et~al.}{2013}]{2013ApJ...776...23B}
{Benincasa} S.~M.,  {Tasker} E.~J.,  {Pudritz} R.~E.,   {Wadsley} J.,  2013, \mn@doi [\apj] {10.1088/0004-637X/776/1/23}, \href {https://ui.adsabs.harvard.edu/abs/2013ApJ...776...23B} {776, 23}

\bibitem[\protect\citeauthoryear{{Benincasa} et~al.,}{{Benincasa} et~al.}{2020}]{2020MNRAS.497.3993B}
{Benincasa} S.~M.,  et~al., 2020, \mn@doi [\mnras] {10.1093/mnras/staa2116}, \href {https://ui.adsabs.harvard.edu/abs/2020MNRAS.497.3993B} {497, 3993}

\bibitem[\protect\citeauthoryear{{Bertoldi} \& {McKee}}{{Bertoldi} \& {McKee}}{1992}]{Bertoldi1992}
{Bertoldi} F.,  {McKee} C.~F.,  1992, \mn@doi [\apj] {10.1086/171638}, \href {https://ui.adsabs.harvard.edu/abs/1992ApJ...395..140B} {395, 140}

\bibitem[\protect\citeauthoryear{{Bigiel}, {Leroy}, {Walter}, {Brinks}, {de Blok}, {Madore}  \& {Thornley}}{{Bigiel} et~al.}{2008a}]{2008AJ....136.2846B}
{Bigiel} F.,  {Leroy} A.,  {Walter} F.,  {Brinks} E.,  {de Blok} W.~J.~G.,  {Madore} B.,   {Thornley} M.~D.,  2008a, \mn@doi [\aj] {10.1088/0004-6256/136/6/2846}, \href {https://ui.adsabs.harvard.edu/abs/2008AJ....136.2846B} {136, 2846}

\bibitem[\protect\citeauthoryear{{Bigiel}, {Leroy}, {Walter}, {Brinks}, {de Blok}, {Madore}  \& {Thornley}}{{Bigiel} et~al.}{2008b}]{Bigiel2008}
{Bigiel} F.,  {Leroy} A.,  {Walter} F.,  {Brinks} E.,  {de Blok} W.~J.~G.,  {Madore} B.,   {Thornley} M.~D.,  2008b, \mn@doi [\aj] {10.1088/0004-6256/136/6/2846}, \href {https://ui.adsabs.harvard.edu/abs/2008AJ....136.2846B} {136, 2846}

\bibitem[\protect\citeauthoryear{{Bigiel} et~al.,}{{Bigiel} et~al.}{2011}]{Bigiel2011}
{Bigiel} F.,  et~al., 2011, \mn@doi [\apjl] {10.1088/2041-8205/730/2/L13}, \href {https://ui.adsabs.harvard.edu/abs/2011ApJ...730L..13B} {730, L13}

\bibitem[\protect\citeauthoryear{{Bisbas} et~al.,}{{Bisbas} et~al.}{2022}]{Bisbas2022}
{Bisbas} T.~G.,  et~al., 2022, \mn@doi [\apj] {10.3847/1538-4357/ac7960}, \href {https://ui.adsabs.harvard.edu/abs/2022ApJ...934..115B} {934, 115}

\bibitem[\protect\citeauthoryear{{Blitz}, {Fukui}, {Kawamura}, {Leroy}, {Mizuno}  \& {Rosolowsky}}{{Blitz} et~al.}{2007}]{2007prpl.conf...81B}
{Blitz} L.,  {Fukui} Y.,  {Kawamura} A.,  {Leroy} A.,  {Mizuno} N.,   {Rosolowsky} E.,  2007, in {Reipurth} B.,  {Jewitt} D.,   {Keil} K.,  eds, Protostars and Planets V. p.~81 (\mn@eprint {arXiv} {astro-ph/0602600}), \mn@doi{10.48550/arXiv.astro-ph/0602600}

\bibitem[\protect\citeauthoryear{{Bolatto}, {Leroy}, {Rosolowsky}, {Walter}  \& {Blitz}}{{Bolatto} et~al.}{2008a}]{2008ApJ...686..948B}
{Bolatto} A.~D.,  {Leroy} A.~K.,  {Rosolowsky} E.,  {Walter} F.,   {Blitz} L.,  2008a, \mn@doi [\apj] {10.1086/591513}, \href {https://ui.adsabs.harvard.edu/abs/2008ApJ...686..948B} {686, 948}

\bibitem[\protect\citeauthoryear{{Bolatto}, {Leroy}, {Rosolowsky}, {Walter}  \& {Blitz}}{{Bolatto} et~al.}{2008b}]{Bolatto2008}
{Bolatto} A.~D.,  {Leroy} A.~K.,  {Rosolowsky} E.,  {Walter} F.,   {Blitz} L.,  2008b, \mn@doi [\apj] {10.1086/591513}, \href {https://ui.adsabs.harvard.edu/abs/2008ApJ...686..948B} {686, 948}

\bibitem[\protect\citeauthoryear{{Bond}, {Cole}, {Efstathiou}  \& {Kaiser}}{{Bond} et~al.}{1991}]{1991ApJ...379..440B}
{Bond} J.~R.,  {Cole} S.,  {Efstathiou} G.,   {Kaiser} N.,  1991, \mn@doi [\apj] {10.1086/170520}, \href {https://ui.adsabs.harvard.edu/abs/1991ApJ...379..440B} {379, 440}

\bibitem[\protect\citeauthoryear{{Chevance} et~al.,}{{Chevance} et~al.}{2020a}]{2020SSRv..216...50C}
{Chevance} M.,  et~al., 2020a, \mn@doi [\ssr] {10.1007/s11214-020-00674-x}, \href {https://ui.adsabs.harvard.edu/abs/2020SSRv..216...50C} {216, 50}

\bibitem[\protect\citeauthoryear{{Chevance}, {Kruijssen}, {Hygate}, {Schruba}, {Longmore}  et~al.}{{Chevance} et~al.}{2020b}]{Chevance2020}
{Chevance} M.,  {Kruijssen} J.~M.~D.,  {Hygate} A. P.~S.,  {Schruba} A.,  {Longmore} S.~N.,   et~al., 2020b, \mn@doi [\mnras] {10.1093/mnras/stz3525}, \href {https://ui.adsabs.harvard.edu/abs/2020MNRAS.493.2872C} {493, 2872}

\bibitem[\protect\citeauthoryear{{Chevance}, {Kruijssen}, {Krumholz}, {Groves}, {Keller}  et~al.}{{Chevance} et~al.}{2022}]{Chevance2022}
{Chevance} M.,  {Kruijssen} J.~M.~D.,  {Krumholz} M.~R.,  {Groves} B.,  {Keller} B.~W.,   et~al., 2022, \mn@doi [\mnras] {10.1093/mnras/stab2938}, \href {https://ui.adsabs.harvard.edu/abs/2022MNRAS.509..272C} {509, 272}

\bibitem[\protect\citeauthoryear{{Chieffi} \& {Limongi}}{{Chieffi} \& {Limongi}}{2004}]{Chieffi2004}
{Chieffi} A.,  {Limongi} M.,  2004, \mn@doi [\apj] {10.1086/392523}, \href {https://ui.adsabs.harvard.edu/abs/2004ApJ...608..405C} {608, 405}

\bibitem[\protect\citeauthoryear{{Clark}, {Glover}  \& {Klessen}}{{Clark} et~al.}{2012}]{Clark2012}
{Clark} P.~C.,  {Glover} S. C.~O.,   {Klessen} R.~S.,  2012, \mn@doi [\mnras] {10.1111/j.1365-2966.2011.20087.x}, \href {https://ui.adsabs.harvard.edu/abs/2012MNRAS.420..745C} {420, 745}

\bibitem[\protect\citeauthoryear{{Dib} et~al.,}{{Dib} et~al.}{2020}]{2020A&A...642A.177D}
{Dib} S.,  et~al., 2020, \mn@doi [\aap] {10.1051/0004-6361/202038849}, \href {https://ui.adsabs.harvard.edu/abs/2020A&A...642A.177D} {642, A177}

\bibitem[\protect\citeauthoryear{{Dobbs}}{{Dobbs}}{2008}]{2008MNRAS.391..844D}
{Dobbs} C.~L.,  2008, \mn@doi [\mnras] {10.1111/j.1365-2966.2008.13939.x}, \href {https://ui.adsabs.harvard.edu/abs/2008MNRAS.391..844D} {391, 844}

\bibitem[\protect\citeauthoryear{{Dobbs} \& {Pringle}}{{Dobbs} \& {Pringle}}{2013}]{2013MNRAS.432..653D}
{Dobbs} C.~L.,  {Pringle} J.~E.,  2013, \mn@doi [\mnras] {10.1093/mnras/stt508}, \href {https://ui.adsabs.harvard.edu/abs/2013MNRAS.432..653D} {432, 653}

\bibitem[\protect\citeauthoryear{{Dobbs}, {Burkert}  \& {Pringle}}{{Dobbs} et~al.}{2011a}]{2011MNRAS.413.2935D}
{Dobbs} C.~L.,  {Burkert} A.,   {Pringle} J.~E.,  2011a, \mn@doi [\mnras] {10.1111/j.1365-2966.2011.18371.x}, \href {https://ui.adsabs.harvard.edu/abs/2011MNRAS.413.2935D} {413, 2935}

\bibitem[\protect\citeauthoryear{{Dobbs}, {Burkert}  \& {Pringle}}{{Dobbs} et~al.}{2011b}]{Dobbs2011}
{Dobbs} C.~L.,  {Burkert} A.,   {Pringle} J.~E.,  2011b, \mn@doi [\mnras] {10.1111/j.1365-2966.2011.18371.x}, \href {https://ui.adsabs.harvard.edu/abs/2011MNRAS.413.2935D} {413, 2935}

\bibitem[\protect\citeauthoryear{{Dobbs}, {Burkert}  \& {Pringle}}{{Dobbs} et~al.}{2011c}]{2011MNRAS.417.1318D}
{Dobbs} C.~L.,  {Burkert} A.,   {Pringle} J.~E.,  2011c, \mn@doi [\mnras] {10.1111/j.1365-2966.2011.19346.x}, \href {https://ui.adsabs.harvard.edu/abs/2011MNRAS.417.1318D} {417, 1318}

\bibitem[\protect\citeauthoryear{{Duarte-Cabral} et~al.,}{{Duarte-Cabral} et~al.}{2021}]{Duarte-Cabral2021}
{Duarte-Cabral} A.,  et~al., 2021, {The SEDIGISM survey: molecular clouds in the inner Galaxy}, Monthly Notices of the Royal Astronomical Society, Volume 500, Issue 3, pp.3027-3049 (\mn@eprint {arXiv} {2012.01502}), \mn@doi{10.1093/mnras/staa2480}

\bibitem[\protect\citeauthoryear{{Elmegreen}}{{Elmegreen}}{1989}]{Elmegreen1989}
{Elmegreen} B.~G.,  1989, \mn@doi [\apj] {10.1086/167192}, \href {https://ui.adsabs.harvard.edu/abs/1989ApJ...338..178E} {338, 178}

\bibitem[\protect\citeauthoryear{{Elmegreen} \& {Efremov}}{{Elmegreen} \& {Efremov}}{1997}]{Elmegreen1997}
{Elmegreen} B.~G.,  {Efremov} Y.~N.,  1997, \mn@doi [\apj] {10.1086/303966}, \href {https://ui.adsabs.harvard.edu/abs/1997ApJ...480..235E} {480, 235}

\bibitem[\protect\citeauthoryear{{Elmegreen} \& {Falgarone}}{{Elmegreen} \& {Falgarone}}{1996}]{Elmegreen1996}
{Elmegreen} B.~G.,  {Falgarone} E.,  1996, \mn@doi [\apj] {10.1086/178009}, \href {https://ui.adsabs.harvard.edu/abs/1996ApJ...471..816E} {471, 816}

\bibitem[\protect\citeauthoryear{{Elmegreen} \& {Scalo}}{{Elmegreen} \& {Scalo}}{2004}]{2004ARA&A..42..211E}
{Elmegreen} B.~G.,  {Scalo} J.,  2004, \mn@doi [\araa] {10.1146/annurev.astro.41.011802.094859}, \href {https://ui.adsabs.harvard.edu/abs/2004ARA&A..42..211E} {42, 211}

\bibitem[\protect\citeauthoryear{{Evans}, {Heyer}, {Miville-Desch{\^e}nes}, {Nguyen-Luong}  \& {Merello}}{{Evans} et~al.}{2021}]{Evans2021}
{Evans} Neal~J. I.,  {Heyer} M.,  {Miville-Desch{\^e}nes} M.-A.,  {Nguyen-Luong} Q.,   {Merello} M.,  2021, \mn@doi [\apj] {10.3847/1538-4357/ac1425}, \href {https://ui.adsabs.harvard.edu/abs/2021ApJ...920..126E} {920, 126}

\bibitem[\protect\citeauthoryear{{Field}, {Blackman}  \& {Keto}}{{Field} et~al.}{2011}]{Field2011}
{Field} G.~B.,  {Blackman} E.~G.,   {Keto} E.~R.,  2011, \mn@doi [\mnras] {10.1111/j.1365-2966.2011.19091.x}, \href {https://ui.adsabs.harvard.edu/abs/2011MNRAS.416..710F} {416, 710}

\bibitem[\protect\citeauthoryear{{Georgy} et~al.,}{{Georgy} et~al.}{2013}]{Georgy2013}
{Georgy} C.,  et~al., 2013, \mn@doi [\aap] {10.1051/0004-6361/201322178}, \href {https://ui.adsabs.harvard.edu/abs/2013A&A...558A.103G} {558, A103}

\bibitem[\protect\citeauthoryear{{Glover} \& {Clark}}{{Glover} \& {Clark}}{2012}]{Glover2012}
{Glover} S. C.~O.,  {Clark} P.~C.,  2012, \mn@doi [\mnras] {10.1111/j.1365-2966.2011.19648.x}, \href {https://ui.adsabs.harvard.edu/abs/2012MNRAS.421....9G} {421, 9}

\bibitem[\protect\citeauthoryear{{Glover} \& {Mac Low}}{{Glover} \& {Mac Low}}{2007}]{Glover2007}
{Glover} S. C.~O.,  {Mac Low} M.-M.,  2007, \mn@doi [\apjs] {10.1086/512238}, \href {https://ui.adsabs.harvard.edu/abs/2007ApJS..169..239G} {169, 239}

\bibitem[\protect\citeauthoryear{{Grisdale}, {Agertz}, {Renaud}  \& {Romeo}}{{Grisdale} et~al.}{2018}]{2018MNRAS.479.3167G}
{Grisdale} K.,  {Agertz} O.,  {Renaud} F.,   {Romeo} A.~B.,  2018, \mn@doi [\mnras] {10.1093/mnras/sty1595}, \href {https://ui.adsabs.harvard.edu/abs/2018MNRAS.479.3167G} {479, 3167}

\bibitem[\protect\citeauthoryear{{Haardt} \& {Madau}}{{Haardt} \& {Madau}}{1996}]{Haardt1996}
{Haardt} F.,  {Madau} P.,  1996, \mn@doi [\apj] {10.1086/177035}, \href {https://ui.adsabs.harvard.edu/abs/1996ApJ...461...20H} {461, 20}

\bibitem[\protect\citeauthoryear{{Hennebelle} \& {Falgarone}}{{Hennebelle} \& {Falgarone}}{2012}]{Hennebelle2012}
{Hennebelle} P.,  {Falgarone} E.,  2012, \mn@doi [\aapr] {10.1007/s00159-012-0055-y}, \href {https://ui.adsabs.harvard.edu/abs/2012A\&ARv..20...55H} {20, 55}

\bibitem[\protect\citeauthoryear{{Hernquist}}{{Hernquist}}{1990}]{Hernquist1990}
{Hernquist} L.,  1990, \mn@doi [\apj] {10.1086/168845}, \href {https://ui.adsabs.harvard.edu/abs/1990ApJ...356..359H} {356, 359}

\bibitem[\protect\citeauthoryear{{Herrera}, {Boulanger}, {Nesvadba}  \& {Falgarone}}{{Herrera} et~al.}{2012}]{Herrera2012}
{Herrera} C.~N.,  {Boulanger} F.,  {Nesvadba} N.~P.~H.,   {Falgarone} E.,  2012, \mn@doi [\aap] {10.1051/0004-6361/201118317}, \href {https://ui.adsabs.harvard.edu/abs/2012A\&A...538L...9H} {538, L9}

\bibitem[\protect\citeauthoryear{{Heyer} \& {Dame}}{{Heyer} \& {Dame}}{2015}]{Heyer2015}
{Heyer} M.,  {Dame} T.~M.,  2015, \mn@doi [\araa] {10.1146/annurev-astro-082214-122324}, \href {https://ui.adsabs.harvard.edu/abs/2015ARA\&A..53..583H} {53, 583}

\bibitem[\protect\citeauthoryear{Heyer, Carpenter  \& Snell}{Heyer et~al.}{2001}]{Heyer2001}
Heyer M.~H.,  Carpenter J.~M.,   Snell R.~L.,  2001, \mn@doi [The Astrophysical Journal] {10.1086/320218}, 551, 852

\bibitem[\protect\citeauthoryear{{Heyer}, {Krawczyk}, {Duval}  \& {Jackson}}{{Heyer} et~al.}{2009}]{Heyer2009}
{Heyer} M.,  {Krawczyk} C.,  {Duval} J.,   {Jackson} J.~M.,  2009, \mn@doi [\apj] {10.1088/0004-637X/699/2/1092}, \href {https://ui.adsabs.harvard.edu/abs/2009ApJ...699.1092H} {699, 1092}

\bibitem[\protect\citeauthoryear{{Hislop}, {Naab}, {Steinwandel}, {Lah{\'e}n}, {Irodotou}, {Johansson}  \& {Walch}}{{Hislop} et~al.}{2022}]{Hislop2022}
{Hislop} J.~M.,  {Naab} T.,  {Steinwandel} U.~P.,  {Lah{\'e}n} N.,  {Irodotou} D.,  {Johansson} P.~H.,   {Walch} S.,  2022, \mn@doi [\mnras] {10.1093/mnras/stab3347}, \href {https://ui.adsabs.harvard.edu/abs/2022MNRAS.509.5938H} {509, 5938}

\bibitem[\protect\citeauthoryear{{Holtzman}, {Faber}, {Shaya}, {Lauer}, {Groth}  et~al.}{{Holtzman} et~al.}{1992}]{Holtzman1992}
{Holtzman} J.~A.,  {Faber} S.~M.,  {Shaya} E.~J.,  {Lauer} T.~R.,  {Groth} J.,   et~al., 1992, \mn@doi [\aj] {10.1086/116094}, \href {https://ui.adsabs.harvard.edu/abs/1992AJ....103..691H} {103, 691}

\bibitem[\protect\citeauthoryear{{Hopkins}, {Quataert}  \& {Murray}}{{Hopkins} et~al.}{2012}]{2012MNRAS.421.3488H}
{Hopkins} P.~F.,  {Quataert} E.,   {Murray} N.,  2012, \mn@doi [\mnras] {10.1111/j.1365-2966.2012.20578.x}, \href {https://ui.adsabs.harvard.edu/abs/2012MNRAS.421.3488H} {421, 3488}

\bibitem[\protect\citeauthoryear{{Hu}, {Naab}, {Walch}, {Moster}  \& {Oser}}{{Hu} et~al.}{2014}]{Hu2014}
{Hu} C.-Y.,  {Naab} T.,  {Walch} S.,  {Moster} B.~P.,   {Oser} L.,  2014, \mn@doi [\mnras] {10.1093/mnras/stu1187}, \href {https://ui.adsabs.harvard.edu/abs/2014MNRAS.443.1173H} {443, 1173}

\bibitem[\protect\citeauthoryear{{Hu}, {Naab}, {Walch}, {Glover}  \& {Clark}}{{Hu} et~al.}{2016}]{Hu2016}
{Hu} C.-Y.,  {Naab} T.,  {Walch} S.,  {Glover} S. C.~O.,   {Clark} P.~C.,  2016, \mn@doi [\mnras] {10.1093/mnras/stw544}, \href {https://ui.adsabs.harvard.edu/abs/2016MNRAS.458.3528H} {458, 3528}

\bibitem[\protect\citeauthoryear{{Hu}, {Naab}, {Glover}, {Walch}  \& {Clark}}{{Hu} et~al.}{2017}]{Hu2017}
{Hu} C.-Y.,  {Naab} T.,  {Glover} S. C.~O.,  {Walch} S.,   {Clark} P.~C.,  2017, \mn@doi [\mnras] {10.1093/mnras/stx1773}, \href {https://ui.adsabs.harvard.edu/abs/2017MNRAS.471.2151H} {471, 2151}

\bibitem[\protect\citeauthoryear{{Hughes} et~al.,}{{Hughes} et~al.}{2013}]{2013ApJ...779...46H}
{Hughes} A.,  et~al., 2013, \mn@doi [\apj] {10.1088/0004-637X/779/1/46}, \href {https://ui.adsabs.harvard.edu/abs/2013ApJ...779...46H} {779, 46}

\bibitem[\protect\citeauthoryear{{Hunter} et~al.,}{{Hunter} et~al.}{2012}]{Hunter2012}
{Hunter} D.~A.,  et~al., 2012, \mn@doi [\aj] {10.1088/0004-6256/144/5/134}, \href {https://ui.adsabs.harvard.edu/abs/2012AJ....144..134H} {144, 134}

\bibitem[\protect\citeauthoryear{{Imara}, {De Looze}, {Faesi}  \& {Cormier}}{{Imara} et~al.}{2020}]{2020ApJ...895...21I}
{Imara} N.,  {De Looze} I.,  {Faesi} C.~M.,   {Cormier} D.,  2020, \mn@doi [\apj] {10.3847/1538-4357/ab8883}, \href {https://ui.adsabs.harvard.edu/abs/2020ApJ...895...21I} {895, 21}

\bibitem[\protect\citeauthoryear{{Jeffreson}, {Keller}, {Winter}, {Chevance}, {Kruijssen}  et~al.}{{Jeffreson} et~al.}{2021}]{Jeffreson2021}
{Jeffreson} S. M.~R.,  {Keller} B.~W.,  {Winter} A.~J.,  {Chevance} M.,  {Kruijssen} J.~M.~D.,   et~al., 2021, \mn@doi [\mnras] {10.1093/mnras/stab1293}, \href {https://ui.adsabs.harvard.edu/abs/2021MNRAS.505.1678J} {505, 1678}

\bibitem[\protect\citeauthoryear{{Jeffreson}, {Semenov}  \& {Krumholz}}{{Jeffreson} et~al.}{2023}]{Jeffreson2023}
{Jeffreson} S. M.~R.,  {Semenov} V.~A.,   {Krumholz} M.~R.,  2023, \mn@doi [arXiv e-prints] {10.48550/arXiv.2301.10251}, \href {https://ui.adsabs.harvard.edu/abs/2023arXiv230110251J} {p. arXiv:2301.10251}

\bibitem[\protect\citeauthoryear{{Johnson}, {Leitherer}, {Vacca}  \& {Conti}}{{Johnson} et~al.}{2000}]{Johnson2000}
{Johnson} K.~E.,  {Leitherer} C.,  {Vacca} W.~D.,   {Conti} P.~S.,  2000, \mn@doi [\aj] {10.1086/301541}, \href {https://ui.adsabs.harvard.edu/abs/2000AJ....120.1273J} {120, 1273}

\bibitem[\protect\citeauthoryear{{Johnson}, {Leroy}, {Indebetouw}, {Brogan}, {Whitmore}  et~al.}{{Johnson} et~al.}{2015}]{Johnson2015}
{Johnson} K.~E.,  {Leroy} A.~K.,  {Indebetouw} R.,  {Brogan} C.~L.,  {Whitmore} B.~C.,   et~al., 2015, \mn@doi [\apj] {10.1088/0004-637x/806/1/35}, \href {https://ui.adsabs.harvard.edu/abs/2015ApJ...806...35J} {806, 35}

\bibitem[\protect\citeauthoryear{{Karakas}}{{Karakas}}{2010}]{Karakas2010}
{Karakas} A.~I.,  2010, \mn@doi [\mnras] {10.1111/j.1365-2966.2009.16198.x}, \href {https://ui.adsabs.harvard.edu/abs/2010MNRAS.403.1413K} {403, 1413}

\bibitem[\protect\citeauthoryear{{Kennicutt}}{{Kennicutt}}{1998}]{1998ARA&A..36..189K}
{Kennicutt} Robert~C. J.,  1998, \mn@doi [\araa] {10.1146/annurev.astro.36.1.189}, \href {https://ui.adsabs.harvard.edu/abs/1998ARA&A..36..189K} {36, 189}

\bibitem[\protect\citeauthoryear{{Kennicutt} \& {Evans}}{{Kennicutt} \& {Evans}}{2012}]{Kennicutt2012}
{Kennicutt} R.~C.,  {Evans} N.~J.,  2012, \mn@doi [\araa] {10.1146/annurev-astro-081811-125610}, \href {https://ui.adsabs.harvard.edu/abs/2012ARA\&A..50..531K} {50, 531}

\bibitem[\protect\citeauthoryear{{Kepley}, {Leroy}, {Johnson}, {Sandstrom}  \& {Chen}}{{Kepley} et~al.}{2016}]{Kepley2016}
{Kepley} A.~A.,  {Leroy} A.~K.,  {Johnson} K.~E.,  {Sandstrom} K.,   {Chen} C. H.~R.,  2016, \mn@doi [\apj] {10.3847/0004-637x/828/1/50}, \href {https://ui.adsabs.harvard.edu/abs/2016ApJ...828...50K} {828, 50}

\bibitem[\protect\citeauthoryear{{Kirby}, {Cohen}, {Guhathakurta}, {Cheng}, {Bullock}  \& {Gallazzi}}{{Kirby} et~al.}{2013}]{2013ApJ...779..102K}
{Kirby} E.~N.,  {Cohen} J.~G.,  {Guhathakurta} P.,  {Cheng} L.,  {Bullock} J.~S.,   {Gallazzi} A.,  2013, \mn@doi [\apj] {10.1088/0004-637X/779/2/102}, \href {https://ui.adsabs.harvard.edu/abs/2013ApJ...779..102K} {779, 102}

\bibitem[\protect\citeauthoryear{{Kroupa}}{{Kroupa}}{2001}]{Kroupa2001}
{Kroupa} P.,  2001, \mn@doi [\mnras] {10.1046/j.1365-8711.2001.04022.x}, \href {https://ui.adsabs.harvard.edu/abs/2001MNRAS.322..231K} {322, 231}

\bibitem[\protect\citeauthoryear{{Kruijssen} et~al.,}{{Kruijssen} et~al.}{2019a}]{2019Natur.569..519K}
{Kruijssen} J.~M.~D.,  et~al., 2019a, \mn@doi [\nat] {10.1038/s41586-019-1194-3}, \href {https://ui.adsabs.harvard.edu/abs/2019Natur.569..519K} {569, 519}

\bibitem[\protect\citeauthoryear{{Kruijssen} et~al.,}{{Kruijssen} et~al.}{2019b}]{Kruijssen2019}
{Kruijssen} J.~M.~D.,  et~al., 2019b, \mn@doi [\nat] {10.1038/s41586-019-1194-3}, \href {https://ui.adsabs.harvard.edu/abs/2019Natur.569..519K} {569, 519}

\bibitem[\protect\citeauthoryear{{Krumholz}}{{Krumholz}}{2012}]{Krumholz2012}
{Krumholz} M.~R.,  2012, \mn@doi [\apj] {10.1088/0004-637X/759/1/9}, \href {https://ui.adsabs.harvard.edu/abs/2012ApJ...759....9K} {759, 9}

\bibitem[\protect\citeauthoryear{{Lada} \& {Lada}}{{Lada} \& {Lada}}{2003}]{Lada2003}
{Lada} C.~J.,  {Lada} E.~A.,  2003, \mn@doi [\araa] {10.1146/annurev.astro.41.011802.094844}, \href {https://ui.adsabs.harvard.edu/abs/2003ARA\&A..41...57L} {41, 57}

\bibitem[\protect\citeauthoryear{{Lah{\'e}n}, {Naab}, {Johansson}, {Elmegreen}, {Hu}  et~al.}{{Lah{\'e}n} et~al.}{2019}]{Lahen2019}
{Lah{\'e}n} N.,  {Naab} T.,  {Johansson} P.~H.,  {Elmegreen} B.,  {Hu} C.-Y.,   et~al., 2019, \mn@doi [\apjl] {10.3847/2041-8213/ab2a13}, \href {https://ui.adsabs.harvard.edu/abs/2019ApJ...879L..18L} {879, L18}

\bibitem[\protect\citeauthoryear{{Lah{\'e}n}, {Naab}, {Johansson}, {Elmegreen}, {Hu}  et~al.}{{Lah{\'e}n} et~al.}{2020a}]{Lahen2020a}
{Lah{\'e}n} N.,  {Naab} T.,  {Johansson} P.~H.,  {Elmegreen} B.,  {Hu} C.-Y.,   et~al., 2020a, \mn@doi [\apj] {10.3847/1538-4357/ab7190}, \href {https://ui.adsabs.harvard.edu/abs/2020ApJ...891....2L} {891, 2}

\bibitem[\protect\citeauthoryear{{Lah{\'e}n}, {Naab}, {Johansson}, {Elmegreen}, {Hu}  et~al.}{{Lah{\'e}n} et~al.}{2020b}]{Lahen2020b}
{Lah{\'e}n} N.,  {Naab} T.,  {Johansson} P.~H.,  {Elmegreen} B.,  {Hu} C.-Y.,   et~al., 2020b, \mn@doi [\apj] {10.3847/1538-4357/abc001}, \href {https://ui.adsabs.harvard.edu/abs/2020ApJ...904...71L} {904, 71}

\bibitem[\protect\citeauthoryear{Larson}{Larson}{1981}]{Larson1981}
Larson R.~B.,  1981, \mn@doi [Monthly Notices of the Royal Astronomical Society] {10.1093/mnras/194.4.809}, 194, 809

\bibitem[\protect\citeauthoryear{{Leroy}, {Walter}, {Brinks}, {Bigiel}, {de Blok}, {Madore}  \& {Thornley}}{{Leroy} et~al.}{2008}]{2008AJ....136.2782L}
{Leroy} A.~K.,  {Walter} F.,  {Brinks} E.,  {Bigiel} F.,  {de Blok} W.~J.~G.,  {Madore} B.,   {Thornley} M.~D.,  2008, \mn@doi [\aj] {10.1088/0004-6256/136/6/2782}, \href {https://ui.adsabs.harvard.edu/abs/2008AJ....136.2782L} {136, 2782}

\bibitem[\protect\citeauthoryear{{Leroy} et~al.,}{{Leroy} et~al.}{2015}]{Leroy2015}
{Leroy} A.~K.,  et~al., 2015, \mn@doi [\apj] {10.1088/0004-637X/801/1/25}, \href {https://ui.adsabs.harvard.edu/abs/2015ApJ...801...25L} {801, 25}

\bibitem[\protect\citeauthoryear{{Leroy}, {Bolatto}, {Ostriker}, {Walter}, {Gorski}  et~al.}{{Leroy} et~al.}{2018}]{Leroy2018}
{Leroy} A.~K.,  {Bolatto} A.~D.,  {Ostriker} E.~C.,  {Walter} F.,  {Gorski} M.,   et~al., 2018, \mn@doi [\apj] {10.3847/1538-4357/aaecd1}, \href {https://ui.adsabs.harvard.edu/abs/2018ApJ...869..126L} {869, 126}

\bibitem[\protect\citeauthoryear{{Li}, {Tan}, {Christie}, {Bisbas}  \& {Wu}}{{Li} et~al.}{2018}]{2018PASJ...70S..56L}
{Li} Q.,  {Tan} J.~C.,  {Christie} D.,  {Bisbas} T.~G.,   {Wu} B.,  2018, \mn@doi [\pasj] {10.1093/pasj/psx136}, \href {https://ui.adsabs.harvard.edu/abs/2018PASJ...70S..56L} {70, S56}

\bibitem[\protect\citeauthoryear{{Li}, {Vogelsberger}, {Marinacci}, {Sales}  \& {Torrey}}{{Li} et~al.}{2020}]{2020MNRAS.499.5862L}
{Li} H.,  {Vogelsberger} M.,  {Marinacci} F.,  {Sales} L.~V.,   {Torrey} P.,  2020, \mn@doi [\mnras] {10.1093/mnras/staa3122}, \href {https://ui.adsabs.harvard.edu/abs/2020MNRAS.499.5862L} {499, 5862}

\bibitem[\protect\citeauthoryear{{Lombardi}, {Alves}  \& {Lada}}{{Lombardi} et~al.}{2010}]{2010A&A...519L...7L}
{Lombardi} M.,  {Alves} J.,   {Lada} C.~J.,  2010, \mn@doi [\aap] {10.1051/0004-6361/201015282}, \href {https://ui.adsabs.harvard.edu/abs/2010A&A...519L...7L} {519, L7}

\bibitem[\protect\citeauthoryear{{Mac Low}}{{Mac Low}}{1999}]{MacLow1999}
{Mac Low} M.-M.,  1999, \mn@doi [\apj] {10.1086/307784}, \href {https://ui.adsabs.harvard.edu/abs/1999ApJ...524..169M} {524, 169}

\bibitem[\protect\citeauthoryear{{Mac Low} \& {Klessen}}{{Mac Low} \& {Klessen}}{2004}]{MacLow2004}
{Mac Low} M.-M.,  {Klessen} R.~S.,  2004, \mn@doi [Reviews of Modern Physics] {10.1103/RevModPhys.76.125}, \href {https://ui.adsabs.harvard.edu/abs/2004RvMP...76..125M} {76, 125}

\bibitem[\protect\citeauthoryear{{McKee} \& {Tan}}{{McKee} \& {Tan}}{2003}]{McKee2003}
{McKee} C.~F.,  {Tan} J.~C.,  2003, \mn@doi [\apj] {10.1086/346149}, \href {https://ui.adsabs.harvard.edu/abs/2003ApJ...585..850M} {585, 850}

\bibitem[\protect\citeauthoryear{{McKee}, {Li}  \& {Klein}}{{McKee} et~al.}{2010}]{McKee2010}
{McKee} C.~F.,  {Li} P.~S.,   {Klein} R.~I.,  2010, \mn@doi [\apj] {10.1088/0004-637x/720/2/1612}, \href {https://ui.adsabs.harvard.edu/abs/2010ApJ...720.1612M} {720, 1612}

\bibitem[\protect\citeauthoryear{{Meidt} et~al.,}{{Meidt} et~al.}{2015}]{2015ApJ...806...72M}
{Meidt} S.~E.,  et~al., 2015, \mn@doi [\apj] {10.1088/0004-637X/806/1/72}, \href {https://ui.adsabs.harvard.edu/abs/2015ApJ...806...72M} {806, 72}

\bibitem[\protect\citeauthoryear{{Micha{\l}owski} et~al.,}{{Micha{\l}owski} et~al.}{2015}]{Michalowski2015}
{Micha{\l}owski} M.~J.,  et~al., 2015, \mn@doi [\aap] {10.1051/0004-6361/201526542}, \href {https://ui.adsabs.harvard.edu/abs/2015A&A...582A..78M} {582, A78}

\bibitem[\protect\citeauthoryear{{Myers}}{{Myers}}{1985}]{Myers1985}
{Myers} P.~C.,  1985, in {Black} D.~C.,  {Matthews} M.~S.,  eds, Protostars and Planets II. pp 81--103

\bibitem[\protect\citeauthoryear{{Naab} \& {Ostriker}}{{Naab} \& {Ostriker}}{2017}]{Naab2017}
{Naab} T.,  {Ostriker} J.~P.,  2017, \mn@doi [\araa] {10.1146/annurev-astro-081913-040019}, \href {https://ui.adsabs.harvard.edu/abs/2017ARA\&A..55...59N} {55, 59}

\bibitem[\protect\citeauthoryear{{Navarro}, {Frenk}  \& {White}}{{Navarro} et~al.}{1997}]{1997ApJ...490..493N}
{Navarro} J.~F.,  {Frenk} C.~S.,   {White} S. D.~M.,  1997, \mn@doi [\apj] {10.1086/304888}, \href {https://ui.adsabs.harvard.edu/abs/1997ApJ...490..493N} {490, 493}

\bibitem[\protect\citeauthoryear{{Nelson} \& {Langer}}{{Nelson} \& {Langer}}{1997}]{Nelson1997}
{Nelson} R.~P.,  {Langer} W.~D.,  1997, \mn@doi [\apj] {10.1086/304167}, \href {https://ui.adsabs.harvard.edu/abs/1997ApJ...482..796N} {482, 796}

\bibitem[\protect\citeauthoryear{{Nickerson}, {Teyssier}  \& {Rosdahl}}{{Nickerson} et~al.}{2019}]{2019MNRAS.484.1238N}
{Nickerson} S.,  {Teyssier} R.,   {Rosdahl} J.,  2019, \mn@doi [\mnras] {10.1093/mnras/stz048}, \href {https://ui.adsabs.harvard.edu/abs/2019MNRAS.484.1238N} {484, 1238}

\bibitem[\protect\citeauthoryear{{Ochsendorf}, {Zinnecker}, {Nayak}, {Bally}, {Meixner}  et~al.}{{Ochsendorf} et~al.}{2017}]{Ochsendorf2017}
{Ochsendorf} B.~B.,  {Zinnecker} H.,  {Nayak} O.,  {Bally} J.,  {Meixner} M.,   et~al., 2017, \mn@doi [Nature Astronomy] {10.1038/s41550-017-0268-0}, \href {https://ui.adsabs.harvard.edu/abs/2017NatAs...1..784O} {1, 784}

\bibitem[\protect\citeauthoryear{{Oka}, {Hasegawa}, {Sato}, {Tsuboi}, {Miyazaki}  \& {Sugimoto}}{{Oka} et~al.}{2001}]{Oka2001}
{Oka} T.,  {Hasegawa} T.,  {Sato} F.,  {Tsuboi} M.,  {Miyazaki} A.,   {Sugimoto} M.,  2001, \mn@doi [\apj] {10.1086/322976}, \href {https://ui.adsabs.harvard.edu/abs/2001ApJ...562..348O} {562, 348}

\bibitem[\protect\citeauthoryear{{Pettitt}, {Egusa}, {Dobbs}, {Tasker}, {Fujimoto}  \& {Habe}}{{Pettitt} et~al.}{2018}]{2018MNRAS.480.3356P}
{Pettitt} A.~R.,  {Egusa} F.,  {Dobbs} C.~L.,  {Tasker} E.~J.,  {Fujimoto} Y.,   {Habe} A.,  2018, \mn@doi [\mnras] {10.1093/mnras/sty2040}, \href {https://ui.adsabs.harvard.edu/abs/2018MNRAS.480.3356P} {480, 3356}

\bibitem[\protect\citeauthoryear{{Press} \& {Schechter}}{{Press} \& {Schechter}}{1974}]{1974ApJ...187..425P}
{Press} W.~H.,  {Schechter} P.,  1974, \mn@doi [\apj] {10.1086/152650}, \href {https://ui.adsabs.harvard.edu/abs/1974ApJ...187..425P} {187, 425}

\bibitem[\protect\citeauthoryear{{Rathjen} et~al.,}{{Rathjen} et~al.}{2021}]{Rathjen2021}
{Rathjen} T.-E.,  et~al., 2021, \mn@doi [\mnras] {10.1093/mnras/stab900}, \href {https://ui.adsabs.harvard.edu/abs/2021MNRAS.504.1039R} {504, 1039}

\bibitem[\protect\citeauthoryear{{Rosolowsky}}{{Rosolowsky}}{2005}]{Rosolowsky2005}
{Rosolowsky} E.,  2005, \mn@doi [\pasp] {10.1086/497582}, \href {https://ui.adsabs.harvard.edu/abs/2005PASP..117.1403R} {117, 1403}

\bibitem[\protect\citeauthoryear{{R{\"o}ttgers}, {Naab}, {Cernetic}, {Dav{\'e}}, {Kauffmann}  et~al.}{{R{\"o}ttgers} et~al.}{2020}]{Rottgers2020}
{R{\"o}ttgers} B.,  {Naab} T.,  {Cernetic} M.,  {Dav{\'e}} R.,  {Kauffmann} G.,   et~al., 2020, \mn@doi [\mnras] {10.1093/mnras/staa1490}, \href {https://ui.adsabs.harvard.edu/abs/2020MNRAS.496..152R} {496, 152}

\bibitem[\protect\citeauthoryear{{Schmidt}}{{Schmidt}}{1959}]{Schmidt1959}
{Schmidt} M.,  1959, \mn@doi [\apj] {10.1086/146614}, \href {https://ui.adsabs.harvard.edu/abs/1959ApJ...129..243S} {129, 243}

\bibitem[\protect\citeauthoryear{{Schruba} et~al.,}{{Schruba} et~al.}{2012}]{Schruba2012}
{Schruba} A.,  et~al., 2012, \mn@doi [\aj] {10.1088/0004-6256/143/6/138}, \href {https://ui.adsabs.harvard.edu/abs/2012AJ....143..138S} {143, 138}

\bibitem[\protect\citeauthoryear{{Scoville} \& {Hersh}}{{Scoville} \& {Hersh}}{1979}]{1979ApJ...229..578S}
{Scoville} N.~Z.,  {Hersh} K.,  1979, \mn@doi [\apj] {10.1086/156991}, \href {https://ui.adsabs.harvard.edu/abs/1979ApJ...229..578S} {229, 578}

\bibitem[\protect\citeauthoryear{{Scoville} \& {Solomon}}{{Scoville} \& {Solomon}}{1975}]{1975ApJ...199L.105S}
{Scoville} N.~Z.,  {Solomon} P.~M.,  1975, \mn@doi [\apjl] {10.1086/181859}, \href {https://ui.adsabs.harvard.edu/abs/1975ApJ...199L.105S} {199, L105}

\bibitem[\protect\citeauthoryear{{Skarbinski}, {Jeffreson}  \& {Goodman}}{{Skarbinski} et~al.}{2023}]{2023MNRAS.519.1887S}
{Skarbinski} M.,  {Jeffreson} S. M.~R.,   {Goodman} A.~A.,  2023, \mn@doi [\mnras] {10.1093/mnras/stac3627}, \href {https://ui.adsabs.harvard.edu/abs/2023MNRAS.519.1887S} {519, 1887}

\bibitem[\protect\citeauthoryear{{Smith} et~al.,}{{Smith} et~al.}{2020}]{2020MNRAS.492.1594S}
{Smith} R.~J.,  et~al., 2020, \mn@doi [\mnras] {10.1093/mnras/stz3328}, \href {https://ui.adsabs.harvard.edu/abs/2020MNRAS.492.1594S} {492, 1594}

\bibitem[\protect\citeauthoryear{{Smith}, {Bryan}, {Somerville}, {Hu}, {Teyssier}, {Burkhart}  \& {Hernquist}}{{Smith} et~al.}{2021}]{Smith2021}
{Smith} M.~C.,  {Bryan} G.~L.,  {Somerville} R.~S.,  {Hu} C.-Y.,  {Teyssier} R.,  {Burkhart} B.,   {Hernquist} L.,  2021, \mn@doi [\mnras] {10.1093/mnras/stab1896}, \href {https://ui.adsabs.harvard.edu/abs/2021MNRAS.506.3882S} {506, 3882}

\bibitem[\protect\citeauthoryear{{Solomon}, {Rivolo}, {Barrett}  \& {Yahil}}{{Solomon} et~al.}{1987}]{Solomon1987}
{Solomon} P.~M.,  {Rivolo} A.~R.,  {Barrett} J.,   {Yahil} A.,  1987, \mn@doi [\apj] {10.1086/165493}, \href {https://ui.adsabs.harvard.edu/abs/1987ApJ...319..730S} {319, 730}

\bibitem[\protect\citeauthoryear{{Springel}}{{Springel}}{2005}]{Springel2005}
{Springel} V.,  2005, \mn@doi [\mnras] {10.1111/j.1365-2966.2005.09655.x}, \href {https://ui.adsabs.harvard.edu/abs/2005MNRAS.364.1105S} {364, 1105}

\bibitem[\protect\citeauthoryear{{Steinwandel}, {Moster}, {Naab}, {Hu}  \& {Walch}}{{Steinwandel} et~al.}{2020}]{Steinwandel2020}
{Steinwandel} U.~P.,  {Moster} B.~P.,  {Naab} T.,  {Hu} C.-Y.,   {Walch} S.,  2020, \mn@doi [\mnras] {10.1093/mnras/staa821}, \href {https://ui.adsabs.harvard.edu/abs/2020MNRAS.495.1035S} {495, 1035}

\bibitem[\protect\citeauthoryear{{Stutzki}, {Bensch}, {Heithausen}, {Ossenkopf}  \& {Zielinsky}}{{Stutzki} et~al.}{1998}]{1998A&A...336..697S}
{Stutzki} J.,  {Bensch} F.,  {Heithausen} A.,  {Ossenkopf} V.,   {Zielinsky} M.,  1998, \aap, \href {https://ui.adsabs.harvard.edu/abs/1998A&A...336..697S} {336, 697}

\bibitem[\protect\citeauthoryear{{Szakacs}, {P{\'e}roux}, {Zwaan}, {Nelson}, {Schinnerer}, {Lah{\'e}n}, {Weng}  \& {Fresco}}{{Szakacs} et~al.}{2022}]{Szakacs2022}
{Szakacs} R.,  {P{\'e}roux} C.,  {Zwaan} M.~A.,  {Nelson} D.,  {Schinnerer} E.,  {Lah{\'e}n} N.,  {Weng} S.,   {Fresco} A.~Y.,  2022, \mn@doi [\mnras] {10.1093/mnras/stac510}, \href {https://ui.adsabs.harvard.edu/abs/2022MNRAS.512.4736S} {512, 4736}

\bibitem[\protect\citeauthoryear{{Tasker}}{{Tasker}}{2011}]{2011ApJ...730...11T}
{Tasker} E.~J.,  2011, \mn@doi [\apj] {10.1088/0004-637X/730/1/11}, \href {https://ui.adsabs.harvard.edu/abs/2011ApJ...730...11T} {730, 11}

\bibitem[\protect\citeauthoryear{{Tasker} \& {Tan}}{{Tasker} \& {Tan}}{2009}]{2009ApJ...700..358T}
{Tasker} E.~J.,  {Tan} J.~C.,  2009, \mn@doi [\apj] {10.1088/0004-637X/700/1/358}, \href {https://ui.adsabs.harvard.edu/abs/2009ApJ...700..358T} {700, 358}

\bibitem[\protect\citeauthoryear{{Tremonti} et~al.,}{{Tremonti} et~al.}{2004}]{Tremonti2004}
{Tremonti} C.~A.,  et~al., 2004, \mn@doi [\apj] {10.1086/423264}, \href {https://ui.adsabs.harvard.edu/abs/2004ApJ...613..898T} {613, 898}

\bibitem[\protect\citeauthoryear{{Tre{\ss}}, {Sormani}, {Smith}, {Glover}, {Klessen}, {Mac Low}, {Clark}  \& {Duarte-Cabral}}{{Tre{\ss}} et~al.}{2021}]{2021MNRAS.505.5438T}
{Tre{\ss}} R.~G.,  {Sormani} M.~C.,  {Smith} R.~J.,  {Glover} S. C.~O.,  {Klessen} R.~S.,  {Mac Low} M.-M.,  {Clark} P.,   {Duarte-Cabral} A.,  2021, \mn@doi [\mnras] {10.1093/mnras/stab1683}, \href {https://ui.adsabs.harvard.edu/abs/2021MNRAS.505.5438T} {505, 5438}

\bibitem[\protect\citeauthoryear{Walch \& Naab}{Walch \& Naab}{2015}]{Walch2015a}
Walch S.,  Naab T.,  2015, \mn@doi [Mon. Not. R. Astron. Soc.] {10.1093/mnras/stv1155}, 451, 2757

\bibitem[\protect\citeauthoryear{{Walch} et~al.,}{{Walch} et~al.}{2015}]{Walch2015}
{Walch} S.,  et~al., 2015, \mn@doi [\mnras] {10.1093/mnras/stv1975}, \href {https://ui.adsabs.harvard.edu/abs/2015MNRAS.454..238W} {454, 238}

\bibitem[\protect\citeauthoryear{{Westera}, {Lejeune}, {Buser}, {Cuisinier}  \& {Bruzual}}{{Westera} et~al.}{2002}]{Westera2002}
{Westera} P.,  {Lejeune} T.,  {Buser} R.,  {Cuisinier} F.,   {Bruzual} G.,  2002, \mn@doi [\aap] {10.1051/0004-6361:20011493}, \href {https://ui.adsabs.harvard.edu/abs/2002A\&A...381..524W} {381, 524}

\bibitem[\protect\citeauthoryear{{Whitmore}, {Chandar}, {Schweizer}, {Rothberg}, {Leitherer}  et~al.}{{Whitmore} et~al.}{2010}]{Whitmore2010}
{Whitmore} B.~C.,  {Chandar} R.,  {Schweizer} F.,  {Rothberg} B.,  {Leitherer} C.,   et~al., 2010, \mn@doi [\aj] {10.1088/0004-6256/140/1/75}, \href {https://ui.adsabs.harvard.edu/abs/2010AJ....140...75W} {140, 75}

\bibitem[\protect\citeauthoryear{{Wiersma}, {Schaye}  \& {Smith}}{{Wiersma} et~al.}{2009}]{Wiersma2009}
{Wiersma} R. P.~C.,  {Schaye} J.,   {Smith} B.~D.,  2009, \mn@doi [\mnras] {10.1111/j.1365-2966.2008.14191.x}, \href {https://ui.adsabs.harvard.edu/abs/2009MNRAS.393...99W} {393, 99}

\bibitem[\protect\citeauthoryear{{Williamson}, {Thacker}, {Wurster}  \& {Gibson}}{{Williamson} et~al.}{2014}]{2014MNRAS.442.3674W}
{Williamson} D.~J.,  {Thacker} R.~J.,  {Wurster} J.,   {Gibson} B.~K.,  2014, \mn@doi [\mnras] {10.1093/mnras/stu1121}, \href {https://ui.adsabs.harvard.edu/abs/2014MNRAS.442.3674W} {442, 3674}

\makeatother
\end{thebibliography}

%%%%%%%%%%%%%%%%% APPENDICES %%%%%%%%%%%%%%%%%%%%%
\appendix
\section{Linking lenght study} \label{appendixA}
\begin{figure}
\includegraphics[width=1\columnwidth]{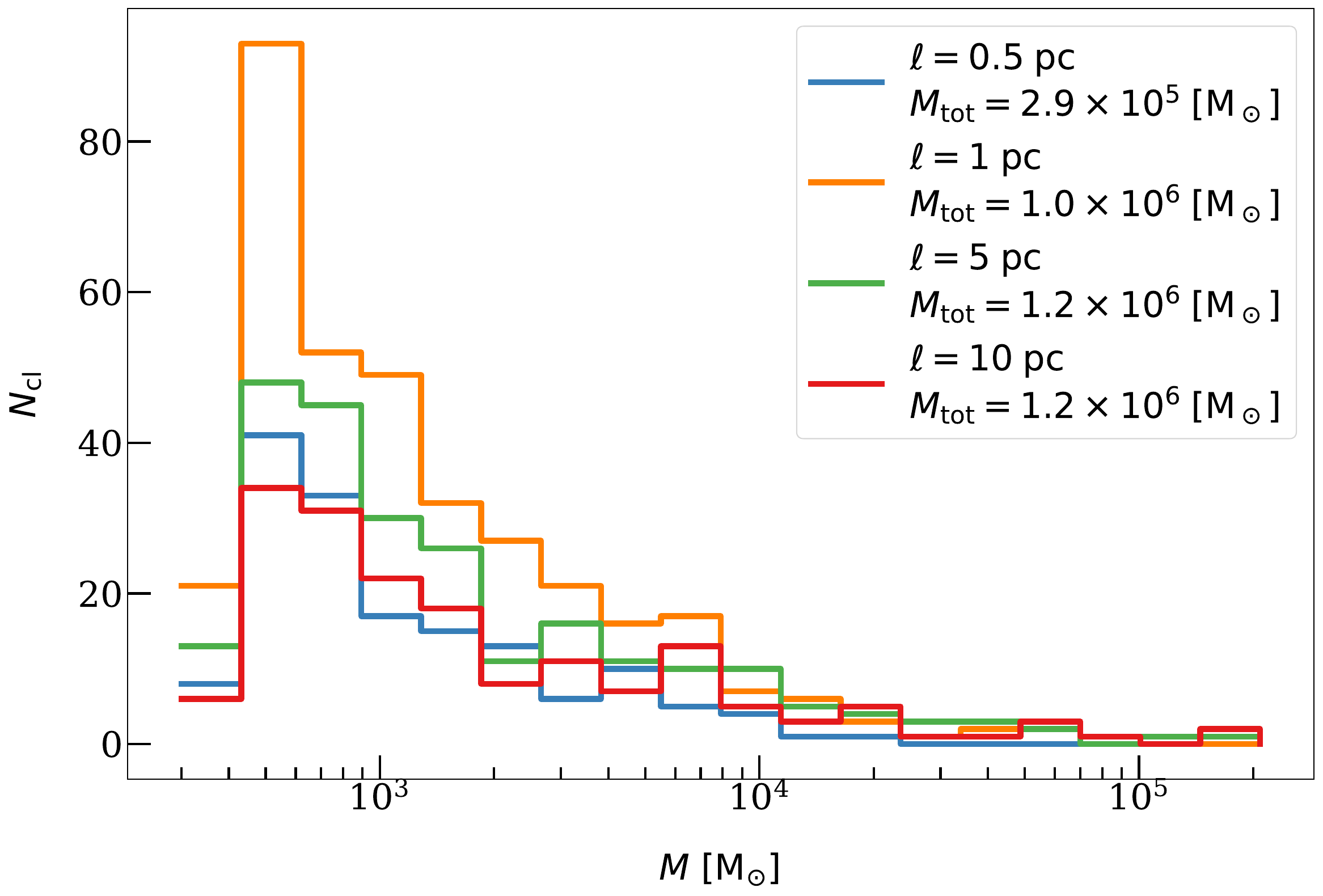}
\caption{Histograms of the masses of the identified clouds for different linking lengths used in the FoF algorithm: $\mathbf{\ell}$ = 0.5 pc (blue), 1 pc (orange), 5 pc (green) and 10 pc (red).}
\label{different_linking_lengths}
\end{figure}

In the paper we use a linking length of \mbox{1 pc} to identify cold clouds. In Fig. \ref{different_linking_lengths} we show the histogram of cold cloud masses for a representative snapshot identified with linking lengths of \mbox{0.5 pc} (blue), \mbox{1 pc} (orange), \mbox{5 pc} (green) and \mbox{10 pc} (red). A linking length of \mbox{1 pc} allows for the best identification of low mass clouds ($\lesssim 10^3 M_\odot$) while the total mass detected in cold clouds ($M_\mathrm{tot}$) changes very little, mainly for larger linking lengths for which we loose the information about low mass clouds again. Therefore we consider a linking length if \mbox{1 pc} to be the best choice.  

\section{Cloud lifetime study}
\begin{figure}
\includegraphics[width=1\columnwidth]{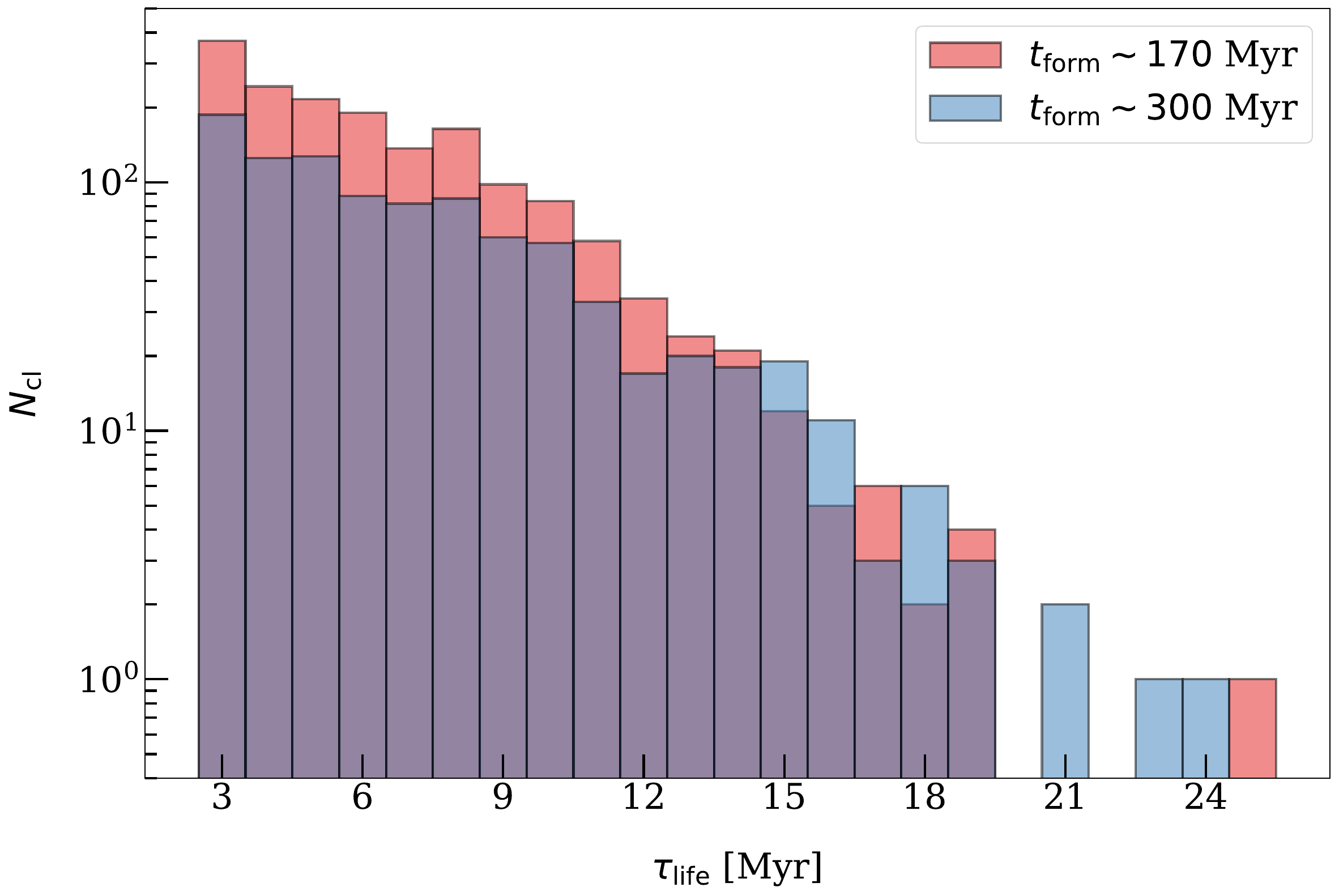}
\caption{Histograms of the lifetimes of cold clouds forming during the merger: $160 \; \mathrm{Myr} < t_\mathrm{form} < 180 \; \mathrm{Myr}$ (red) and forming during a more quiescent later stage of the evolution of the system: $290 \; \mathrm{Myr} < t_\mathrm{form} < 310 \; \mathrm{Myr}$ (blue).
}
\label{tracer_lifespan_comparison}
\end{figure}

In Fig. \ref{tracer_lifespan_comparison} we show a comparison of the cloud lifetime distributions at the peak starburst, when we would expect the strongest impact of the environment and 130 Myr later when the system is in a more relaxed state. We do not find evidence for a change in the lifetime distribution with respect to the environment.

%%%%%%%%%%%%%%%%%%%%%%%%%%%%%%%%%%%%%%%%%%%%%%%%%%

% Don't change these lines
\bsp	% typesetting comment
\label{lastpage}
\end{document}